\pdfoutput=1

\documentclass[11pt,letterpaper,hyperref,dvipsnames]{cernyrep}
\usepackage[utf8]{inputenc}

\DeclareUnicodeCharacter{2212}{-}
\DeclareUnicodeCharacter{202F}{\,}

\hypersetup{
  colorlinks=true,
  linkcolor=red,
  urlcolor=blue,
  citecolor=darkgreen,
  pdfauthor={LH21 participants},
  pdftitle={Les Houches 2021 Proceedings}
}

\usepackage{cite}
\usepackage{amsmath,amssymb,graphicx,xspace,subfigure}
\usepackage{mathtools}
\usepackage{wasysym}
\usepackage{bigints}
\usepackage{wrapfig}

\usepackage[T1]{fontenc}
\usepackage{lmodern}
\usepackage{rotating}
\usepackage{xcolor}
\usepackage{booktabs}
\usepackage{arydshln}
\usepackage{lscape}
\usepackage{floatpag}
\floatpagestyle{plain}
\usepackage{lineno}
\usepackage{graphpap}
\usepackage[super]{nth}

\usepackage{import}
\usepackage{enumitem}
\usepackage{caption}
\usepackage{xspace}
\usepackage{adjustbox}
\usepackage{tikz}
\usetikzlibrary{positioning,arrows}
\usetikzlibrary{decorations.pathmorphing}
\usetikzlibrary{decorations.markings}

\newcommand{\eV}{\ensuremath{\text{e\kern-0.15ex{}V}}\xspace}

\definecolor{darkred}{rgb}{.8, 0.1, 0.1}
\definecolor{darkyellow}{rgb}{0.45, 0.45, 0.}
\definecolor{violet}{rgb}{1.0, 0.0, 1.0}
\definecolor{darkgreen}{rgb}{0.15, .8, 0.15}

\makeatletter
\renewcommand{\thechapter}{\@Roman\c@chapter}
\makeatother

\begin{document}

\begin{flushright}
CERN-TH-2025-071, FR-PHENO-2025-005, IPPP/25/19, TIF-UNIMI-2025-9
\end{flushright}

{\centering{\LARGE{\bf{Les Houches 2023 - Physics at TeV Colliders:\\ Report on the Standard Model Precision Wishlist \par }}}}

\pagenumbering{roman}

\vspace{0.7cm}

Alexander~Huss$^{1\,}$, Joey~Huston$^{2\,}$, Stephen~Jones$^{3\,}$, Mathieu~Pellen$^{4\,}$,
Raoul~R\"ontsch$^{5\,}$

\vspace{0.7cm}
 
\noindent{\small\it $^1$Theoretical Physics Department, CERN,} \\ %
{\small\it 1211 Geneva 23, Switzerland}\\[3mm]
\noindent{\small\it $^2$Department of Physics and Astronomy, Michigan State University,} \\ %
{\small\it East Lansing, MI 48824, USA}\\[3mm]
\noindent{\small\it $^3$Institute for Particle Physics Phenomenology, Durham University,} \\ %
{\small\it Durham DH1 3LE, United Kingdom}\\[3mm]
\noindent{\small\it $^4$Albert-Ludwigs-Universit\"at Freiburg, Physikalisches Institut,} \\ %
{\small\it Hermann-Herder-Stra\ss e 3, D-79104 Freiburg, Germany}\\[3mm]
\noindent{\small\it $^5$University of Milan and INFN Milan,} \\ %
{\small\it Via Celoria 20133, Milan, Italy}\\[3mm]

{\leftline{\bf{Abstract}}}

\vspace{0.5cm}

Les Houches returned to an in-person format in 2023 and the bi-yearly tradition of updating the standard model precision wishlist has continued. 
In this work we review recent progress (since Les Houches 2021) in fixed-order computations for LHC applications.
In addition, necessary ingredients for such calculations such as parton distribution functions, amplitudes, and subtraction methods are discussed.
Finally, we indicate processes and missing higher-order corrections that are required to reach the theoretical accuracy that matches the anticipated experimental precision.

\newpage
\pagenumbering{arabic}
\setcounter{footnote}{0}


\setcounter{tocdepth}{3}
\tableofcontents


\newpage




\newcommand{\alphas}{\ensuremath{\alpha_\text{s}}\xspace}

\newcommand{\NLLgen}[1]{N${}^{#1}$LL\xspace}

\newcommand{\NLL}[1]{N${}^{#1}$LL\xspace}
\newcommand{\NLLone}{NLL\xspace}

\newcommand{\LL}{LL\xspace}
\newcommand{\NLLp}{NLL'\xspace}
\newcommand{\NNLL}{NNLL\xspace}
\newcommand{\NNLLp}{NNLL'\xspace}
\newcommand{\NNNLL}{\NLLgen3\xspace}
\newcommand{\NNNLLp}{\NLLgen3'\xspace}

\newcommand{\NLO}[1]{N${}^{#1}$LO\xspace}
\newcommand{\NLOone}{NLO\xspace}
\newcommand{\NLOgen}{NLO\xspace}
\newcommand{\NNLOgen}{NNLO\xspace}
\newcommand{\NNNLOgen}{\NLO3}

\newcommand{\NLOH}[1]{N${}^{#1}$LO${}_{\rm HTL}$\xspace}
\newcommand{\NLOHone}{NLO${}_{\rm HTL}$\xspace}
\newcommand{\NLOHTL}{NLO${}_{\rm HTL}$\xspace}
\newcommand{\NNLOHTL}{NNLO${}_{\rm HTL}$\xspace}
\newcommand{\NNNLOHTL}{\NLOH3}

\newcommand{\NLOQ}[1]{N${}^{#1}$LO${}_{\rm QCD}$\xspace}
\newcommand{\LOQ}{LO${}_{\rm QCD}$\xspace}
\newcommand{\NLOQone}{NLO${}_{\rm QCD}$\xspace}
\newcommand{\LOQCD}{LO${}_{\rm QCD}$\xspace}
\newcommand{\NLOQCD}{NLO${}_{\rm QCD}$\xspace}
\newcommand{\NNLOQCD}{NNLO${}_{\rm QCD}$\xspace}
\newcommand{\NNNLOQCD}{\NLOQ3}

\newcommand{\NLOE}[1]{N${}^{#1}$LO${}_{\rm EW}$\xspace}
\newcommand{\NLOEone}{NLO${}_{\rm EW}$\xspace}
\newcommand{\LOEW}{LO${}_{\rm EW}$\xspace}
\newcommand{\NLOEW}{NLO${}_{\rm EW}$\xspace}
\newcommand{\NNLOEW}{NNLO${}_{\rm EW}$\xspace}

\newcommand{\NLOD}[1]{N${}^{#1}$LO${}_{\rm QED}$\xspace}
\newcommand{\NLODone}{NLO${}_{\rm QED}$\xspace}
\newcommand{\NLOQED}{NLO${}_{\rm QED}$\xspace}
\newcommand{\NNLOQED}{NNLO${}_{\rm QED}$\xspace}

\newcommand{\NLOSM}{NLO${}_{\rm SM}$\xspace}

\newcommand{\NLOQE}[2]{N${}^{(#1,#2)}$LO${}_{{\rm QCD}\otimes{\rm EW}}$\xspace}

\newcommand{\NLOHE}[2]{N${}^{(#1,#2)}$LO${}^{\rm (HTL)}_{{\rm QCD}\otimes{\rm EW}}$\xspace}

\newcommand{\NLOmixQED}[2]{N${}^{(#1,#2)}$LO${}_{{\rm QCD}\otimes{\rm QED}}$\xspace}

\newcommand{\NLOQmtsix}[1]{N${}^{#1}$LO${}_{\rm QCD}^{(1/{m_t^8})}$\xspace}
\newcommand{\NLOQzzero}[1]{N${}^{#1}$LO${}_{\rm QCD}^{(z\to0)}$\xspace}
\newcommand{\NLOQVBF}[1]{N${}^{#1}$LO${}_{\rm QCD}^{(\rm VBF)}$\xspace}
\newcommand{\NLOQoneVBF}{NLO${}_{\rm QCD}^{(\rm VBF)}$\xspace}
\newcommand{\NLOQCDVBF}{NLO${}_{\rm QCD}^{(\rm VBF)}$\xspace}
\newcommand{\NNLOQCDVBF}{NNLO${}_{\rm QCD}^{(\rm VBF)}$\xspace}
\newcommand{\NLOQoneDIS}{NLO${}_{\rm QCD}^{(\rm DIS)}$\xspace}
\newcommand{\NLOQDIS}[1]{N${}^{#1}$LO${}_{\rm QCD}^{(\rm DIS)}$\xspace}
\newcommand{\NLOEoneVBF}{NLO${}_{\rm EW}^{(\rm VBF)}$\xspace}
\newcommand{\NLOEWVBF}{NLO${}_{\rm EW}^{(\rm VBF)}$\xspace}

\newcommand{\NLOQoneVBFstar}{NLO${}_{\rm QCD}^{(\rm VBF^{*})}$\xspace}
\newcommand{\NLOQVBFstar}[1]{N${}^{#1}$LO${}_{\rm QCD}^{(\rm VBF^{*})}$\xspace}
\newcommand{\NLOQCDVBFstar}{NLO${}_{\rm QCD}^{(\rm VBF^{*})}$\xspace}
\newcommand{\NNLOQCDVBFstar}{NNLO${}_{\rm QCD}^{(\rm VBF^{*})}$\xspace}
\newcommand{\NNNLOQCDVBFstar}{\NLOQVBFstar3}

\newcommand{\NLOEoneVBFstar}{NLO${}_{\rm EW}^{(\rm VBF^{*})}$\xspace}
\newcommand{\NLOggHVtb}[1]{N${}^{#1}$LO${}_{gg\to HZ}^{(t,b)}$\xspace}
\newcommand{\NNLOQCDT}{NNLO${}_{\rm QCD}^{(t)}$\xspace}
\newcommand{\NNLOQCDBC}{NNLO${}_{\rm QCD}^{(b,c)}$\xspace}
\newcommand{\NNLOQCDTTXB}{NNLO${}_{\rm QCD}^{(t,t \times b)}$\xspace}

\newcommand{\xs}{$\sigma$}
\newcommand{\tb}{\bar{t}}
\newcommand{\bb}{\bar{b}}
\newcommand{\qb}{\bar{q}}

\newcommand{\wodecay}{(w/o decay)}
\newcommand{\wdecay}{}
\newcommand{\wodecays}{(w/o decays)}
\newcommand{\wdecays}{}
\newcommand{\wleptdecays}{}

\newcommand{\MadgraphaMCatNLO}{\textsc{Madgraph5}\_a\textsc{MC@NLO}\xspace}
\newcommand{\OpenLoops}{O\protect\scalebox{0.8}{PENLOOPS}\xspace}
\newcommand{\Recola}{R\protect\scalebox{0.8}{ECOLA}\xspace}
\newcommand{\GoSam}{G\protect\scalebox{0.8}{O}S\protect\scalebox{0.8}{AM}\xspace}
\newcommand{\MadLoop}{M\protect\scalebox{0.8}{AD}L\protect\scalebox{0.8}{OOP}\xspace}
\newcommand{\Powheg}{P\protect\scalebox{0.8}{OWHEG}\xspace}
\newcommand{\Powhegboxres}{P\protect\scalebox{0.8}{OWHEG-BOX-RES}\xspace}
\newcommand{\PowhegboxVtwo}{P\protect\scalebox{0.8}{OWHEG-BOX-V2}\xspace}
\newcommand{\Herwig}{H\protect\scalebox{0.8}{ERWIG}\xspace}
\newcommand{\Matrix}{M\protect\scalebox{0.8}{ATRIX}\xspace}
\newcommand{\Munich}{M\protect\scalebox{0.8}{UNICH}\xspace}
\newcommand{\Geneva}{G\protect\scalebox{0.8}{ENEVA}\xspace}
\newcommand{\Sherpa}{S\protect\scalebox{0.8}{HERPA}\xspace}
\newcommand{\NNLOjet}{NNLO\protect\scalebox{0.8}{JET}\xspace}
\newcommand{\MiNLO}{M\protect\scalebox{0.8}{iNLO}\xspace}
\newcommand{\MiNNLO}{M\protect\scalebox{0.8}{iNNLO}\xspace}
\newcommand{\MiNNLOPS}{M\protect\scalebox{0.8}{iNNLOPS}\xspace}
\newcommand{\NLOX}{NLOX\xspace}
\newcommand{\MCFM}{MCFM\xspace}
\newcommand{\Pythia}{P\protect\scalebox{0.8}{YTHIA}\xspace}
\newcommand{\Vincia}{V\protect\scalebox{0.8}{INCIA}\xspace}
\newcommand{\eps}{\epsilon}

\section{Introduction}
\label{sec:SM_wishlist}

The advancement of our understanding of fundamental physics at high energies necessarily relies on a detailed comparison between experimental measurements and theoretical predictions based on first-principles quantum field theory.
At the Large Hadron Collider (LHC), this approach has consistently demonstrated its utility and efficiency over the years.
It is therefore essential to recognize that advancements in fundamental physics at the LHC can only be achieved through the simultaneous improvement of experimental measurements and the development of precision computations.
For the latter, it has proven particularly beneficial to systematically monitor the level of precision required to fully exploit the available experimental data.
In this context, the so-called \emph{Les Houches wishlist}, motivated by the bi-annual workshops at Les Houches on physics at TeV colliders, has been invaluable over the years.

In the first part of the document, some selected topics related to fixed-order techniques and calculations as well as related phenomenological studies are briefly highlighted.
This is followed by what constitutes the main part of the document, the precision wishlist of Standard Model calculations.
The present edition builds on the previous ones and in particular the one of the 2021 edition~\cite{Huss:2022ful}.
For each process, the state of the art as of Ref.~\cite{Huss:2022ful} is briefly summarised, followed by an overview of the progress that has been made since then.
Given the rapid and continuous progress in the field of precision calculations, this summary is bound to be incomplete and we apologize for any omissions.%
\footnote{The knowledge cutoff for this wishlist is \nth{31} December 2024, we also remind the reader of the Les Houches Disclaimer: \emph{never attribute to malice that which is adequately explained by incompetence}.}
While the whishlist has served as a useful resource for both theorists and experimentalists as a summary of the current stat-of-the-art calculations, it does not constitute a comprehensive review on the topic of precision calculations.
We instead refer to dedicated reviews~\cite{Tricoli:2020uxr,Heinrich:2020ybq,Covarelli:2021gyz,Jakobs:2023fxh,Jones:2023uzh} for in-depths discussions.

\section{Higher-order techniques}
\label{sec:HOT}

While the years before the Les Houches 2021 report~\cite{Huss:2022ful} had been marked
by significant progress in the production of \NNLOgen results in an almost industrial manner
with most useful $2\to2$ processes having been calculated,
the last two years have seen a saturation due to the unavailability of 2-loop amplitudes
beyond $2\rightarrow2$ scattering.
However, remarkable progress was achieved in this direction by several groups and approaches
culminating in the first $2\rightarrow3$ calculations of a hadron collider process.
Closely related is the huge progress in the calculation of 2-loop 5-point amplitudes,
as well as 2-loop amplitudes for $2\to2$ processes involving internal masses.
For a review of some recent developments see also Ref.~\cite{Heinrich:2020ybq}.

However, it is not only the amplitude community that has seen impressive development. There have also
been significant steps forward on the side of subtraction schemes, and there are in the meanwhile several subtraction and slicing methods available to deal (in principle) with higher-multiplicity processes at \NNLOgen (see below).

On the parton shower side, NLO QCD matched results and matrix element improved multi-jet merging techniques have become a standard level of theoretical precision.
The automation of full SM corrections including NLO electroweak predictions has also seen major improvements.

Another challenge is to make the \NNLOgen $2\to2$ predictions or complex NLO predictions publicly available to experimental analyses, and there has been major progress to achieve this goal. {\sc Root nTuples} have been a useful tool for complicated final states at NLO and
allow for very flexible re-weighting and analysis.
More recently a similar approach was put forward at NNLO dubbed HighTEA~\cite{Czakon:2023hls}.
The cost for these approaches is
the large disk space required to store the event information.

Finally, the application of APPLgrid~\cite{Carli:2010rw}, fastNLO~\cite{Kluge:2006xs}, and PineAPPL~\cite{Carrazza:2020gss} interpolation libraries to higher-order calculations offers a convenient method to distribute precision predictions.
To this end, the Ploughshare project\footnote{\url{https://ploughshare.web.cern.ch}} provides a central location to distribute such grids.
Although the number of publicly available grids is still limited, steady progress is being made with interfaces to various parton-level Monte Carlo tools being implemented to make the production of such grids accessible to the general public.

Below, we discuss some aspects of higher-order computations.

\subsection{Parton distribution functions}

One of the key elements in improving the accuracy of theoretical predictions at the LHC lies in the determination of parton distribution functions (PDFs). PDFs are most commonly determined by global fits to experimental data, taking into account the experimental errors in the data. The standard now is for the PDFs to be determined at NNLO QCD, although fits at NLO QCD and LO are still available. 
It is encouraged to use NLO QCD PDFs for predictions at both next-to-leading order and leading-order accuracy. There are large NLO corrections for the deep-inelastic scattering process that are not reflected in most LHC cross sections, that distort the leading order PDFs. Differences between NLO and NNLO PDFs tend to be smaller, but there is some indication that the use of NNLO PDFs with NLO predictions leads to better agreement with fully NNLO predictions~\cite{Campbell:2017hsr}.
The results of the global fits are central values for each flavor PDF, along with an estimate of the PDF uncertainty, dominated by the input experimental errors for the data included in the fit. The formalism used in the fit can either be Hessian~\cite{Hou:2019efy,Bailey:2020ooq} or based on Monte Carlo replicas~\cite{NNPDF:2021njg}. The number of data points included in the global PDF fits is typically of the order of 4000--5000 from a wide range of processes. Diagnostic tools, such as the $L_2$ sensitivity~\cite{Jing:2023isu}, have been developed to allow a detailed examination of how the interplay between the different data sets used in global PDF fits determine both the PDFs and their uncertainties.  Lattice gauge theory has reached a level of precision where information from such calculations has provided useful input for PDF determination, especially at large $x$~\cite{Constantinou:2022yye}. This will continue to improve.

In 2021-22, a benchmarking exercise was conducted using the CT18~\cite{Hou:2019efy}, MSHT20~\cite{Bailey:2020ooq}, NNPDF3.1~\cite{NNPDF:2017mvq}/4.0~\cite{NNPDF:2021njg} PDFs, and a combination (PDF4LHC21~\cite{Ball:2022hsh}) was formed, using Monte Carlo replicas generated from each of the three PDF sets. As the benchmarking exercise continued over the transition from NNPDF3.1 to NNPDF4.0, an updated version of 3.1 was used which utilized some of the key new data sets added to 4.0 (and already present in CT18 and MSHT20).  PDF4LHC21 PDF sets are available either in a 40 member Hessian format, or a 100 member Monte Carlo replica format. The PDF4LHC21 PDFs show a reduction in uncertainty from the combined PDFs determined in 2015, but perhaps not to the extent that may have been expected through the introduction of a variety of new LHC data. This is partially due to the central values of the three input PDFs not coinciding exactly, and partially because the tensions between the data sets that limit the resultant possible uncertainty.%
\footnote{Ref.~\cite{Courtoy:2022ocu} points out one problem that PDF fits may face is the bias that results from improper sampling in very large data spaces. The bias can not only result in an underestimate of the true uncertainty, but also an incorrect central PDF. An alternative perspective is provided by Ref.~\cite{Ball:2022uon}}
The PDF4LHC21 PDF sets are appropriate for use in general predictions for state-of-the-art calculations, and indeed the prior PDF4LHC15 PDFs have been used in just that way.

More recently, the ABMP PDFs were updated,  with a emphasis on the impact of LHC top quark data~\cite{Alekhin:2024bhs}.

Many differential cross section measurements from the LHC have been included in the PDF determination.
This was made possible by the \NNLOQCD calculations of the relevant $2\rightarrow2$ matrix elements that have been discussed in past iterations of the wishlist.
For use in calculations at \NNNLOgen, several of which are discussed here, nominally \NNNLOgen PDFs would be needed. As they are not yet available, NNLO PDFs are used in their stead with an unknown uncertainty introduced into the predictions as a result, as for example in the Higgs Cross Section Working Group (see the discussion on approximate  \NNNLOgen PDFs below). This has a non-negligible impact on the Higgs cross section at \NNNLOgen through gluon--gluon fusion, for example. Indeed, this mis-match in order leads to a notable contribution to the uncertainty for predictions for gluon--gluon fusion  Higgs boson production. There are efforts to estimate the theoretical uncertainties due to (missing) higher order terms. These would be in addition to the (dominant) experimental uncertainties from the data included in the PDF fits. One method to estimate the theory uncertainties would be by variations of the renormalization and factorization scales that are used to evaluate the matrix elements at \NNLOgen. Considering separate scales of each type for each data-set calculation would add too many degrees of freedom and remove much of the constraining power of the PDF fit. Connecting the renormalization or factorization scales, even for similar processes, may be treating those scales as more physical than they deserve. Perhaps there is more justification for treating the factorization scale in this manner than the renormalization scale. There is also the issue of whether introducing additional uncertainties in the PDFs through scale variations, and then in addition, performing scale variations in the predictions in the nominal manner, may lead to an over-counting of the uncertainty. Ref.~\cite{Harland-Lang:2018bxd}  used a physical basis (for example structure functions or similar observables) rather than the PDFs themselves to demonstrate that performing factorization scale variations in cross sections results in double-counting of uncertainties. Considering correlated factorization scale variations in the PDF fit, and not in the resultant predictions, may not be ideal but an acceptable solution for certain specific physical quantities. See also Refs.~\cite{Ball:2021icz,Kassabov:2022orn} for further discussion.

Ref.~\cite{NNPDF:2021njg} proposes taking into account the missing higher order uncertainties in the cross sections included in the PDF fits by adding a theory uncertainty to the experimental covariance matrix. Since the theory uncertainties are uncorrelated with the experimental ones, the two uncertainties can be added in quadrature in the covariance matrix. The global fit processes are divided into five separate types (DIS NC, DIS CC, Drell--Yan, jets and top), with a hypothesis that calculations within a given type will be likely to
have similar structures of higher-order corrections. An assumption is made that the  renormalization scale
is only correlated within a single type of process, while the factorization scale is fully correlated across all processes. Resultant fits to the NNPDF4.0 data set  do not substantially change the PDF uncertainties, but may have a non-negligible effect on PDF central values. One drawback to this method is the need to impose a higher value of $Q^2$ cut on the data, in order to be able to perform a scale uncertainty variation both above and below the central value. This has the impact of removing some of the low-x HERA data commonly used in the PDF fits. 

MSHT~\cite{McGowan:2022nag}
has carried out an exercise of parametrising the higher order effects with nuisance parameters based on a prior probability distribution (using the information currently available regarding \NNNLOgen matrix elements and the approximate splitting functions). Where not explicitly available, the \NNNLOgen/\NNLOgen K-factors are parametrised as a superposition of both \NLOgen and \NNLOgen K-factors, allowing the fit to determine the combination of shapes and an overall magnitude. The result is a reduction in $\chi^2$ for the global fit greater than that expected by the extra degrees of freedom.

In order to fully determine PDFs at \NNNLOgen, a number of contributing items have to be known:

\begin{itemize}
    \item parton splitting functions at 4 loops to evolve the PDFs in $x$ and $Q^2$
    \item transition matrix elements at 3 loops to change the number of PDF flavors at heavy quark mass thresholds
    \item coefficient functions for DIS at 3 loops
    \item hadronic cross sections at \NNNLOgen
\end{itemize}

Recently, additional moments have been calculated allowing a better determination of the necessary 4-loop splitting functions (see the benchmarking exercise and references therein in Ref.~\cite{Cooper-Sarkar:2024crx}; or more recent updates, see Ref.~\cite{Falcioni:2024qpd}). Full information is known for the 3-loop transition matrix elements,  and the 3-loop light flavor coefficient functions are known for DIS, with approximations for the heavy flavor coefficient functions related to the transition matrix elements (although there has been recent progress on this front~\cite{Ablinger:2024qxg}). There is limited information, however, at \NNNLOgen for the hadronic cross sections that enter into the PDF fits, hence the need for the nuisance parameters described above. Most of the discrimination power for the global PDF fits arises from differential data from processes such as DIS, DY, inclusive jet and $t\bar{t}$ production. Their full use in \NNNLOgen PDF fits requires the availability of differential predictions at that level.  Such predictions exist for Drell--Yan but are very CPU-intensive (see the discussion in the Drell--Yan section of the wishlist) and thus not yet at a stage to enable their use in global PDF analyses. It will be some time before such differential predictions are available at \NNNLOgen for inclusive jet and $t\bar{t}$ production, and even then the computing resources needed may be prohibitive.

Using information from  this list, there have been two approximate \NNNLOgen PDF fits, first by MSHT~\cite{McGowan:2022nag} and second by NNPDF~\cite{NNPDF:2024nan}. The Higgs Cross Section Working group 
has allotted a theory uncertainty for the use of \NNLOgen PDFs with \NNNLOgen matrix elements of the order of one percent. Nominally, the determination of PDFs at this (approximate) order would allow the retirement of the uncertainty for those cross sections known to \NNNLOgen due to the use of \NNLOgen PDFs; however, as mentioned earlier, very little information is known about the relevant hadronic cross sections at that order~\footnote{It is not surprising that the gluon changes from \NNLOgen to \NNNLOgen, as there are large new logs that appear in the splitting functions at the latter order (there were accidental zeros at \NLOgen and \NNLOgen.)}. In addition, the differences between the gluon distributions for the two approximate \NNNLOgen PDFs may result in an uncertainty for the ggF Higgs boson cross section larger than that observed at the previous order (and much larger than observed for the similar $q\bar{q}$ PDF luminosity comparison), primarily due to differing impacts from the \NNLOgen to approximate \NNNLOgen transition.

\begin{figure}[h]
	\includegraphics[width=0.49\linewidth]{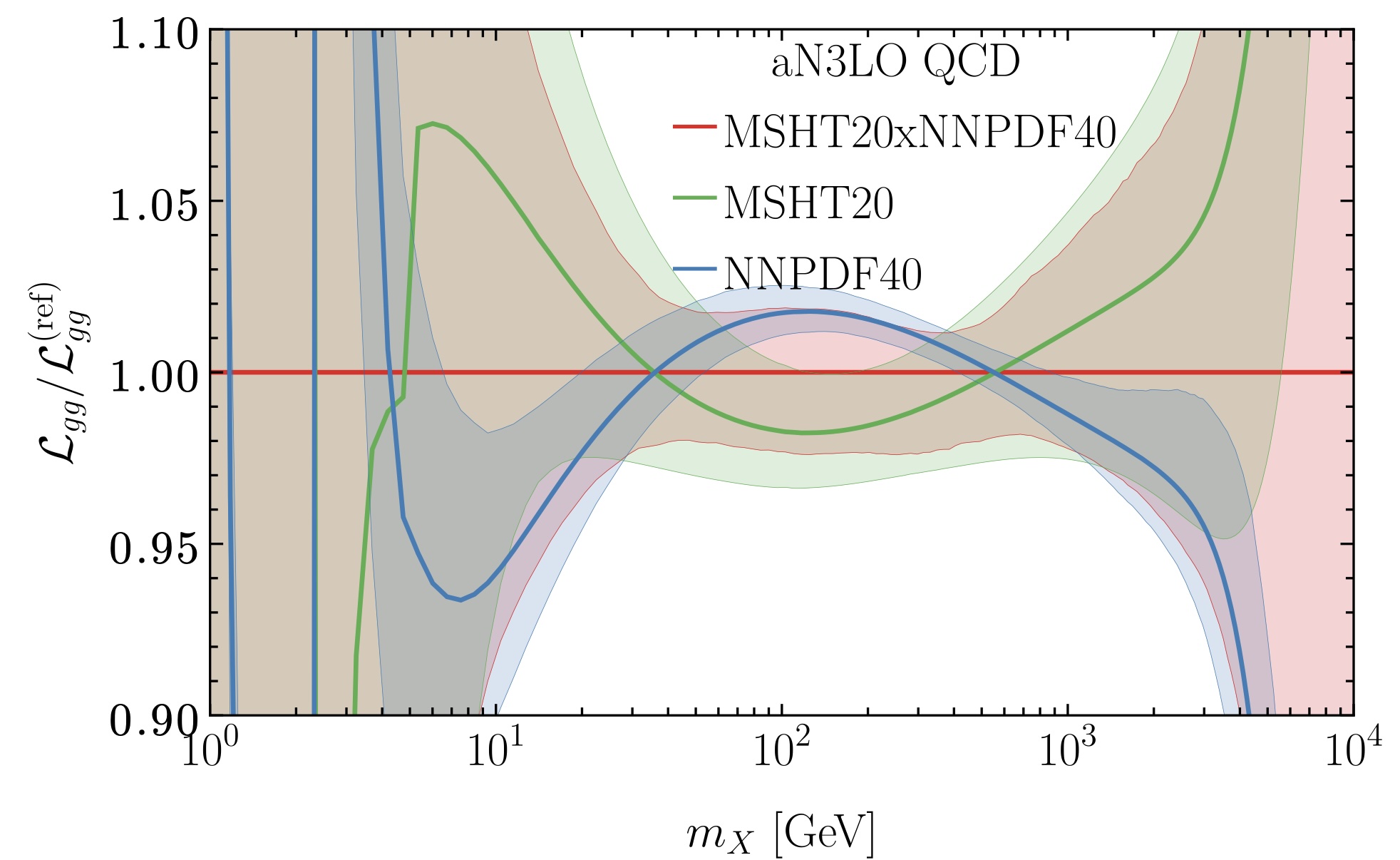}
	\includegraphics[width=0.49\linewidth]{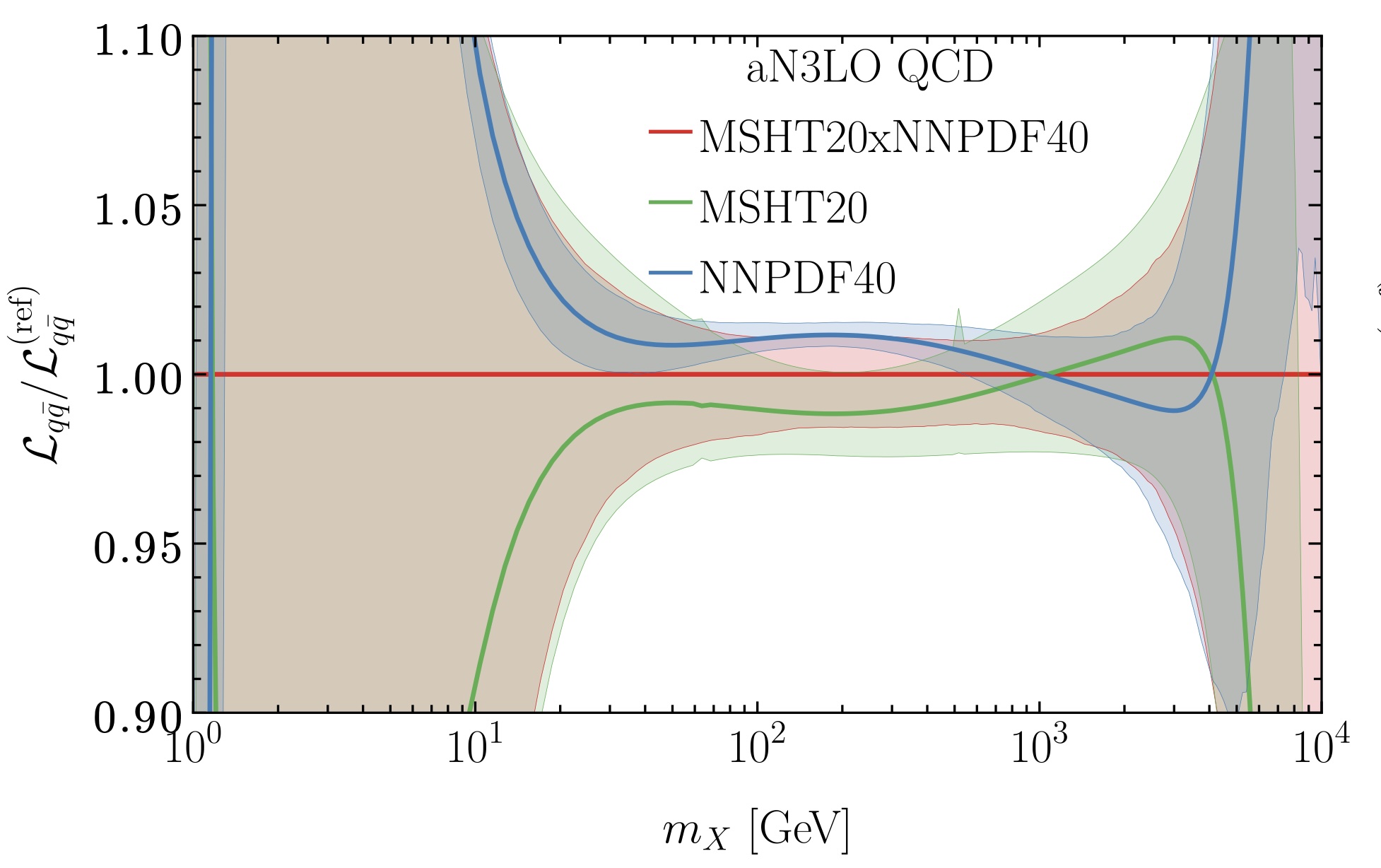}
	\caption{A comparison of the aN3LO PDF luminosities for MSHT20 and NNPDF4.0 to their combination (MSHT20xNNPDF40) for $gg$ (left) and $q\bar{q}$ (right).
 }
\label{an3lo_pdf_lum}
\end{figure}

\begin{figure}[h]
	\includegraphics[width=0.49\linewidth]{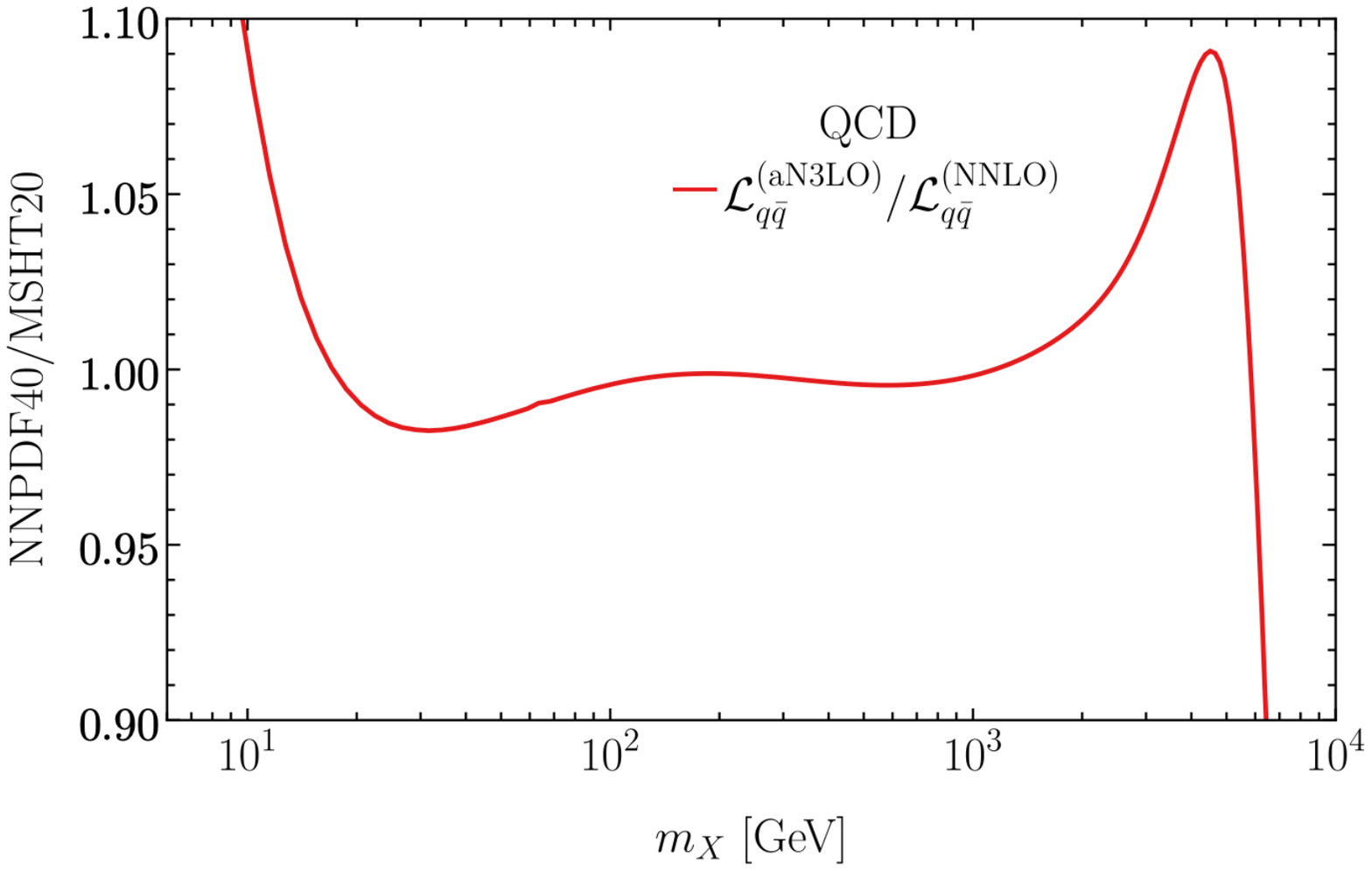}
    \includegraphics[width=0.49\linewidth]{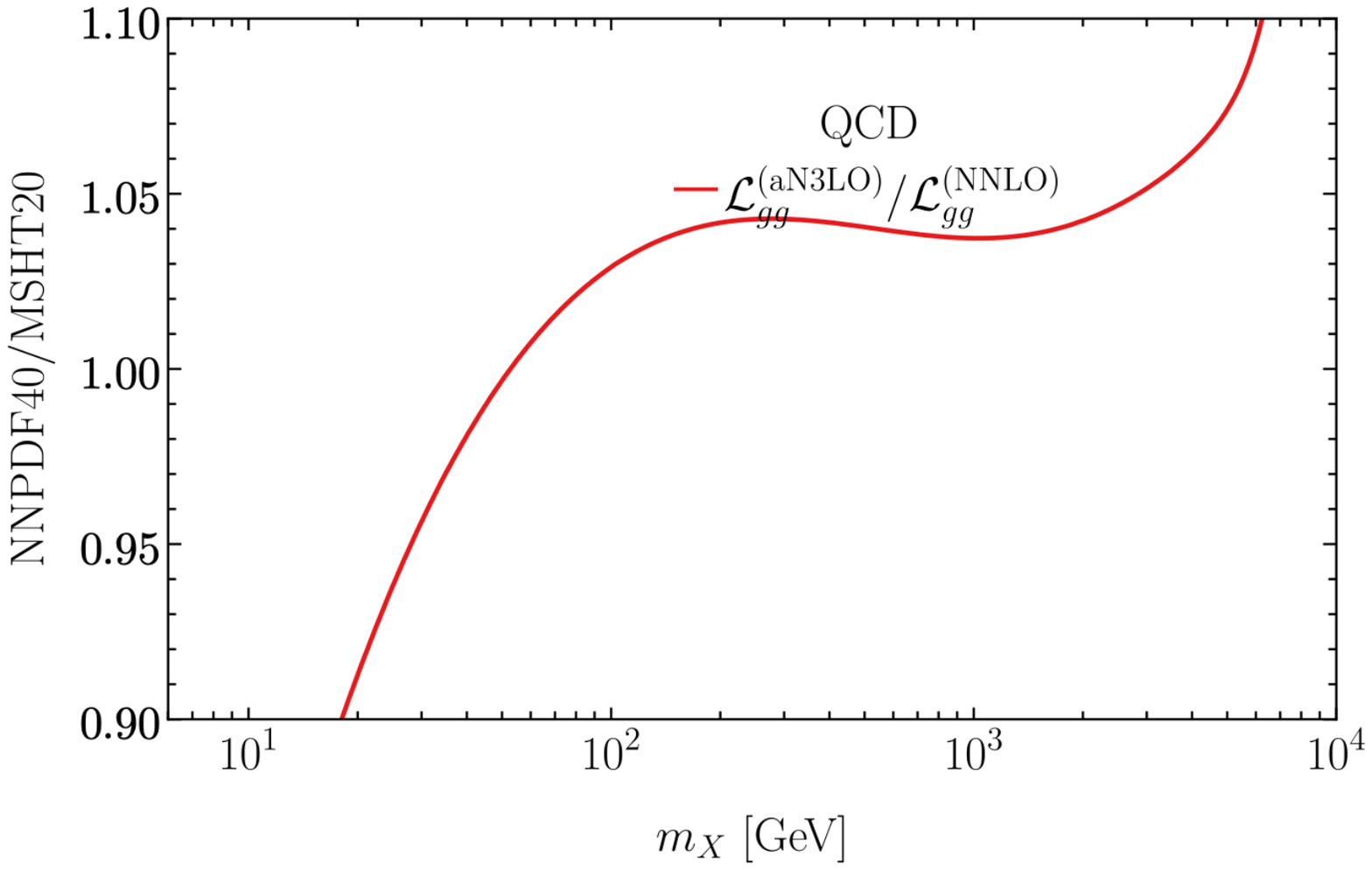}
 \caption{The ratio of the aN3LO PDF luminosities to the NNLO PDF luminosities are determined separately, for MSHT20 and NNPDF4.0, and the ratio of the two ratios is plotted for $q\bar{q}$ (left) and $gg$  (right).
 }
\label{an3lo_pdf_lum_ratios}
\end{figure}

A combination of the two \NNNLOgen PDFs has been carried out, named MSHT20xNNPDF40~\cite{MSHT:2024tdn}. In Figure~\ref{an3lo_pdf_lum}, the $gg$ PDF luminosity (left) and the $q\bar{q}$ PDF luminosity (right) at 13.6 TeV is shown for MSHT20 and NNPDF4.0 and MSHT20xNNPDF40 for approximate \NNNLOgen. 
The $gg$ PDF luminosities for MSHT20 and NNPDF4.0 are in greater disagreement than the corresponding $q\bar{q}$ PDF luminosities.
Some of the differences in the $gg$ PDF luminosities may build upon existing variations in analysis and technique already existing at \NNLOgen. This can be tested by taking the ratio of the  approximate \NNNLOgen $gg$ and $q\bar{q}$ PDF luminosities to the corresponding \NNLOgen PDF luminosities  for MSHT20 and NNPDF4.0, and then to calculate the ratio of the two ratios.  This is shown in Figure~\ref{an3lo_pdf_lum_ratios}.  This ratio of ratios examines differences arising from the inclusion/treatment of the approximate \NNNLOgen information. For the $q\bar{q}$ luminosities (left), the MSHT20 and NNPDF4.0 PDF a\NNNLOgen/\NNLOgen  ratio of ratios is within a percent or so of unity at a mass scale of 100 GeV, indicating that the impact of the approximate \NNNLOgen corrections is very similar for the two groups. For the $gg$ PDF luminosities (right), the corresponding ratio of ratios at the Higgs boson mass shows a deviation from unity of  more than 3\%.  The benchmarking exercise referred to earlier~\cite{Cooper-Sarkar:2024crx}, demonstrated that, when using a common toy PDF, both the MSHT and NNPDF approaches produce similar evolution results, indicating that the evolution at \NNNLOgen does not seem to be the cause of the observed differences. One of the two a\NNNLOgen PDFs was determined earlier than other,  so some of the difference observed may be due to a lower number of moments included in the gluon splitting functions. Given the importance of the a\NNNLOgen gluon distribution, this will continue to be investigated.

It will be some time before the \NNNLOgen information for the other processes becomes available, as discussed above, but this may indicate the need for some additional understanding/benchmarking, as was done for PDF4LHC21 (and before that PDF4LHC15). QED effects have been included in global PDF fits for some time~\cite{Xie:2021equ,Cridge:2021pxm,NNPDF:2024djq,Cridge:2023ryv,Barontini:2024eii}, and the impact is non-negligible for the level of precision demanded at the LHC. The differences between the QED versions of the a\NNNLOgen PDFs is somewhat smaller than for the pure QCD case~\cite{Cridge:2023ryv}. However, this presents another puzzle in understanding as to why the QED corrections are different, given that the underlying base of the determination is the same. This again suggests a needed benchmarking to further understand. This is currently in progress by the PDF4LHC working group. 
It may be necessary to include a fraction of the differences observed between 
NNLO and a\NNNLOgen PDFs as an additional source of uncertainty.

The impact of NLL small-$x$ resummation corrections on the PDFs, especially those of the gluon, may notably alter the
low-$x$ behavior in kinematic regions where \NNNLOgen effects may also be important (and may contain contributions that may overlap with \NNNLOgen, but go beyond them).

\subsection{Development in amplitude and loop integral techniques}

Computing fixed-order amplitudes for scattering processes remains a key obstacle to producing precise predictions for the LHC and HL-LHC.
For ease of presentation, we divide the computation of multi-loop amplitudes into two broad categories:
\begin{enumerate}
\item Obtaining the amplitudes and simplifying (\textit{reducing}) them,
\item Calculating the integrals which appear in the amplitudes.
\end{enumerate}
In the previous wishlist~\cite{Huss:2022ful}, to which we refer the interested reader, we described the state of the art of each of these categories in some detail.
Here we only briefly highlight a selection of the most interesting recent advances in this area since the last wishlist.
Thorough reviews of formal developments in the calculation of scattering amplitudes can be found in Ref.~\cite{Travaglini:2022uwo}.
A modern introduction to techniques for computing multi-loop Feynman integrals can be found in Ref.~\cite{Weinzierl:2022eaz}. Further details on recent developments can be found in the SAGEX review~\cite{Abreu:2022mfk,Blumlein:2022zkr} and Snowmass White Paper~\cite{Cordero:2022gsh}.

The use of integration-by-parts (IBP) identities~\cite{Tkachov:1981wb,Chetyrkin:1981qh,Laporta:2000dsw}, Lorenz invariance (LI)~\cite{Gehrmann:1999as}, and dimension shift relations~\cite{Tarasov:1996br,Lee:2009dh} remains a critically important technique in modern loop calculations, but also presents a major bottleneck.
Several efficient codes exist to facilitate their use, including:
{\sc Air}~\cite{Anastasiou:2004vj},
{\sc Fire}~\cite{Smirnov:2008iw,Smirnov:2013dia,Smirnov:2014hma,Smirnov:2019qkx} (recently updated in Ref.~\cite{Smirnov:2023yhb}),
{\sc LiteRed}~\cite{Lee:2012cn,Lee:2013mka},
{\sc Reduze}~\cite{Studerus:2009ye,vonManteuffel:2012np}, and
{\sc Kira}~\cite{Maierhofer:2017gsa,Maierhofer:2018gpa,Klappert:2020nbg}.
The {\sc Blade} reduction package~\cite{Guan:2024byi} aims to reduce the total time to obtain a reduction by generating block-triangular IBP systems, which can be orders of magnitude smaller than traditional tools.
The {\sc NeatIBP} tool~\cite{Wu:2023upw} uses syzygy and module intersection techniques to provide IBP systems in which the propagator degrees are limited.
The use of finite field techniques, as implemented in {\sc FireFly}~\cite{Klappert:2019emp,Klappert:2020aqs}, {\sc FiniteFlow}~\cite{Peraro:2019svx} and various private codes~\cite{vonManteuffel:2014ixa}, has widely been adopted to accelerate the reduction to master integrals.
The {\sc RatRacer} package~\cite{Magerya:2022hvj} can be used to further speed up the use of finite fields by separating the construction of expressions, tracing, and their subsequent evaluation during rational reconstruction, replaying the existing trace with different inputs.
Recent algorithmic improvements in the reconstruction of rational functions are presented in Refs.~\cite{Liu:2023cgs,Maier:2024djk}.
In principle, the need for IBP reduction can be side-stepped using techniques from intersection theory, for an introduction see Ref.~\cite{Frellesvig:2020qot}, several advances in this direction were presented in Refs.~\cite{Fontana:2023amt,Brunello:2023rpq,Brunello:2023fef,Crisanti:2024onv,Brunello:2024tqf}.
Significant developments in the methods and tools used for simplifying the resulting reduced expressions have also been achieved.
The {\sc FUEL} package provides routines for the manipulation of rational functions,  a tool for partial fractioning such expressions was described in Ref.~\cite{Heller:2021qkz}.
In Refs.~\cite{DeLaurentis:2022otd,Chawdhry:2023yyx}, advances in techniques for directly obtaining simplified expressions using $p$-adic numbers were presented.

The methods used to calculate Feynman integrals continue to evolve.
Several new ideas and methods have been presented in the literature and existing techniques have been refined and applied in new contexts.
The use of canonical differential equations, for an introduction see Refs.~\cite{Argeri:2007up,Henn:2014qga}, remains an essential technique.
Previously, all integrals required for $2 \rightarrow 3$ massless scattering had been computed and expressed in terms of (analytic) pentagon functions~\cite{Gehrmann:2015bfy,Papadopoulos:2015jft,Abreu:2018rcw,Abreu:2018aqd,Chicherin:2018old,Abreu:2020jxa,Chicherin:2020oor}.
Recently, the master integrals required for five-point one-mass scattering have also been obtained analytically~\cite{Chicherin:2021dyp,Abreu:2023rco} using the differential equations method.
In Ref.~\cite{Henn:2024ngj}, a family of planar two-loop massless six-point master integrals relevant for $2 \rightarrow 4$ scattering were obtained using the technique.
As the number of loops, scattered particles and internal/external masses increases, it is increasingly common to encounter functions beyond multiple polylogarithms (MPLs), for a review of the various developments see Ref.~\cite{Bourjaily:2022bwx}.
Very significant advances have occurred in this area in recent years, stemming from joint research by both the phenomenology and amplitude communities.
These advances have helped to clarify the analytic properties of integrals beyond MPLs and is enabling their numeric evaluation, see \eg Refs.~\cite{Duhr:2022dxb,Lairez:2022zkj,Wilhelm:2022wow,Hannesdottir:2022xki,Loebbert:2022nfu,Gong:2022erh,Duhr:2023eld,Marzucca:2023gto,McLeod:2023qdf,Fevola:2023kaw,Fevola:2023fzn,Doran:2023yzu,DHoker:2023vax,Duhr:2024hjf,Jockers:2024uan,Britto:2024mna
}.
When a fully analytic solution of the differential equations cannot be obtained, the use of generalised series expansions as implemented in
\textsc{DiffExp}~\cite{Hidding:2020ytt} and the recent \textsc{SeaSyde} package~\cite{Armadillo:2022ugh} remain indispensable.
The method of Auxiliary Mass Flow~\cite{Liu:2017jxz,Liu:2020kpc,Liu:2021wks}, as implemented in \textsc{AMFlow}~\cite{Liu:2022chg}, is also used in many cutting-edge calculations either to directly evaluate the relevant master integrals or for obtaining high-precision numerical boundary values for differential equations.

Methods to evaluate integrals directly in parameter space, either analytically as implemented in e.g., {\sc HyperInt}~\cite{Panzer:2014caa}, or numerically as implemented in {\sc Fiesta}~\cite{Smirnov:2015mct,Smirnov:2021rhf} or {\sc pySecDec}~\cite{Borowka:2017idc,Borowka:2018goh,Heinrich:2021dbf,Heinrich:2023til}, continue to be developed and used in modern calculations.
A procedure to efficiently evaluate parameter integrals based on tropical Monte Carlo quadrature~\cite{Borinsky:2020rqs} has been implemented in the public tool {\sc feyntrop}~\cite{Borinsky:2023jdv} and applied also to integrals in the Minkowski regime.
The analytic and numeric computation of Feynman integrals via their Mellin-Barnes representation provides another avenue of research, for a recent introducion and review see Ref.~\cite{Dubovyk:2022obc}.
Feynman integrals satisfy a Gel\'fand-Kapranov-Zelevinsky (GKZ) system of partial differential equations, an automated package, {\sc FeynGKZ}, to derive the associated GKZ system and solve it in terms of hypergeometric functions was presented in Ref.~\cite{Ananthanarayan:2022ntm}.
Various new approaches have also been developed in the last few years.
In Ref.~\cite{Zeng:2023jek},
A method of evaluating Euclidean integrals via positivity constraints was derived.
In Ref.~\cite{Huang:2024nij}, a procedure for reformulating Feynman integrals as integrals over a small set of parameters was proposed.

Loop–Tree Duality provides a framework for treating real and virtual corrections simultaneously, this
can help to avoid having to separately treat the IR divergences arising in and then cancelling
between the amplitudes. Progress continues to be made in this direction, some recent advances were presented in \eg~\cite{Sterman:2023xdj,Ramirez-Uribe:2024rjg,Rios-Sanchez:2024xtv,LTD:2024yrb,Kermanschah:2021wbk,Kermanschah:2024utt,deLejarza:2024pgk}.
We also point the reader to the reviews of Refs.~\cite{deJesusAguilera-Verdugo:2021mvg,Sborlini:2024uhh}.

A regularly updated review of the various applications of machine learning in high-energy physics including for the computation, simplification and approximation of scattering amplitudes and Feynman integrals can be found in Ref.~\cite{Feickert:2021ajf}. We refrain from reviewing this very active area.

\subsection{Infrared subtraction methods for differential cross sections}

Fully differential higher-order calculations must retain the complete information on the final-state kinematics, which includes regions of the real-emission phase space that are associated with soft and/or collinear configurations and thus where the Matrix Elements can develop singularities.
While such infrared (IR) singularities must cancel with the explicit poles in the virtual amplitudes for any IR-safe observable, this entails some level of integration of the unresolved emission to expose the singularity.
IR subtraction methods facilitate the explicit cancellation of singularities to obtain finite cross sections,
\begin{equation}
  d\sigma_{2\to n} \text{\NLO{k}} = {\rm IR}_k(A^{k}_{2\to n}, A^{k-1}_{2\to n+1},\cdots, A^{0}_{2\to n+k})\,,
\end{equation}
where the function ${\rm IR}_k$ represents an infrared subtraction technique that leaves the kinematic information for each particle multiplicity intact, and $A^k_{2 \to N}$ denotes the amplitude for a $2 \to N$ particle process with $k$ loops.

While full automation of \NLOgen subtractions has been achieved, this is not yet the case at \NNLOgen.
Nonetheless, tremendous progress has been made in differential \NNLOgen calculations, essentially completing all relevant $2 \to 1$ and $2 \to 2$ processes as well as several important $2 \to 3$ processes.
Nevertheless, the substantial computing times required for these results have motivated the re-appraisal of subtraction schemes at \NNLOgen, with the aims of streamlining them, making them applicable to broader class of processes, and/or including previously ignored sub-leading effects. At the same time, there are ongoing efforts to revisit prior approximations that could potentially limit the interpretation of theory--data comparisons (\eg combination of production and decay subprocesses, flavoured jet definition, photon-jet separation and hadron fragmentation, on-shell vs.\ off-shell, etc.).
Lastly, we have observed remarkable progress in the area of differential \NNNLOgen calculations with results being available for $2 \to 1$ benchmark processes.

\begin{itemize}
\item Antenna subtraction~\cite{Gehrmann-DeRidder:2005btv,Currie:2013vh}:\\
  Applicable to processes with hadronic initial and final states with analytically integrated counterterms.
  An almost completely local subtraction up to angular correlations that are removed through the averaging over azimuthal angles.
  Applied to processes in \(\mathrm{e}^+\mathrm{e}^-\), deep-inelastic scattering (DIS), and hadron–hadron collisions:
  \(\mathrm{e}^+\mathrm{e}^-\to 3j\)~\cite{Gehrmann-DeRidder:2014hxk,Gehrmann:2017xfb},
  (di-)jets in DIS~\cite{Currie:2017tpe,Niehues:2018was},
  \(pp\to \text{(di)-jets}\)~\cite{Currie:2016bfm,Currie:2017eqf},
  \(pp\to \gamma\gamma\)~\cite{Gehrmann:2020oec},
  \(pp\to \gamma+j/X\)~\cite{Chen:2019zmr},
  \(pp\to V+j\)~\cite{Gehrmann-DeRidder:2015wbt,Gehrmann-DeRidder:2016cdi,Gehrmann-DeRidder:2017mvr},
  \(pp\to H+j\)~\cite{Chen:2016zka},
  \(pp\to VH(+\mathrm{jet})\)~\cite{Gauld:2019yng,Gauld:2020ced,Gauld:2021ule},
  and Higgs production in VBF~\cite{Cruz-Martinez:2018rod}.
  Extensions to cope with identified jet flavours~\cite{Gauld:2019yng,Gauld:2020deh,Gauld:2023zlv,Gehrmann-DeRidder:2023gdl}, the photon fragmentation function~\cite{Gehrmann:2022cih,Chen:2022gpk} and hadron fragmentation~\cite{Bonino:2024adk,Caletti:2024xaw}.

Recent refinements have focused on  streamlining the construction of antenna functions by reducing the number of spurious divergences~\cite{Braun-White:2023sgd,Braun-White:2023zwd,Fox:2023bma,Fox:2024bfp} as well as the formulation of the method in color space~\cite{Gehrmann:2023dxm,Chen:2022ktf} allowing high-multiplicity processes to be computed beyond the leading-color approximation in a semi-automated manner. Extensions to accommodate fragmentation functions for identified hadrons have also been considered~\cite{Bonino:2024adk}.

\item Sector-improved residue subtraction~\cite{Czakon:2010td,Czakon:2011ve,Boughezal:2011jf}:\\
  Capable of treating hadronic initial and final states through a fully local subtraction that incorporates ideas of the FKS approach at NLO~\cite{Frixione:1995ms,Frederix:2009yq} and a sector decomposition~\cite{Binoth:2000ps} approach
  for real radiation singularities~\cite{Heinrich:2002rc,Anastasiou:2003gr,Binoth:2004jv}.
  Counterterms obtained numerically with improvements using a four-dimensional formulation~\cite{Czakon:2014oma}.
  Applied to
  top-quark processes~\cite{Czakon:2013goa,Czakon:2014xsa,Czakon:2015owf,Czakon:2016ckf,Brucherseifer:2013iv,Brucherseifer:2014ama}, to $pp\to H+j$~\cite{Boughezal:2015dra,Caola:2015wna}, inclusive jet production~\cite{Czakon:2019tmo}, $pp\to3\gamma$~\cite{Chawdhry:2019bji},
  $pp\to2\gamma+j$~\cite{Chawdhry:2021hkp},
  $pp\to\gamma+2j$~\cite{Badger:2023mgf},
  $pp\to W+j$~\cite{Pellen:2022fom}, and $pp\to3j$~\cite{Czakon:2021mjy}, the latter being the most complicated process from the point of view of infrared divergences that has been computed to date.
  Extensions to deal with flavoured jets~\cite{Czakon:2020coa,Czakon:2022khx} and $B$-hadron production~\cite{Czakon:2021ohs,Czakon:2022pyz,Czakon:2024tjr}.

\item $q_T$-subtraction~\cite{Catani:2007vq}:\\
  A slicing approach for processes with a colourless final state and/or a pair of massive coloured particles.
  Applied to
  $H$~\cite{Catani:2007vq,Grazzini:2008tf},
  $V$~\cite{Catani:2009sm,Catani:2010en}
  and $VV'$ production processes~\cite{Catani:2011qz,Grazzini:2013bna,Gehrmann:2014fva,Cascioli:2014yka,Grazzini:2015nwa,Grazzini:2015hta,Grazzini:2016swo,Grazzini:2016ctr,Grazzini:2017ckn,Catani:2018krb,Kallweit:2018nyv}, which are available in the \Matrix program~\cite{Grazzini:2017mhc}.
  Predictions at \NNLOQCD for $H$, $V$, $VH$, $V\gamma$, $\gamma\gamma$, and $VV'$ available in the MCFM program~\cite{Campbell:2022gdq}.
  Further applications at \NNLOQCD include $VH$~\cite{Ferrera:2011bk,Ferrera:2013yga,Ferrera:2014lca}, $HH$~\cite{deFlorian:2016uhr,Grazzini:2018bsd}, $VHH$~\cite{Li:2016nrr,Li:2017lbf}.
  Extended to cope with a pair of massive coloured particles~\cite{Bonciani:2015sha,Angeles-Martinez:2018mqh} and applied to top-pair production~\cite{Catani:2019iny,Catani:2019hip} and $b\bar{b}$ production~\cite{Catani:2020kkl}; more recently extended to processes beyond the Born back-to-back configuration and applied to $t\tb H$~\cite{Catani:2022mfv,Devoto:2024nhl} and $t\tb W$~\cite{Buonocore:2023ljm}.
  The same developments allowed the mixed QCD--EW corrections to Drell--Yan with massive leptons to be tackled~\cite{Buonocore:2021rxx,Bonciani:2021zzf,Armadillo:2024ncf}.
  Method extended to \NNNLOQCD~\cite{Gehrmann:2010ue,Li:2016ctv,Luo:2019szz,Ebert:2020yqt} with applications to Higgs production~\cite{Cieri:2018oms,Billis:2021ecs} and Drell--Yan production~\cite{Chen:2021vtu,Camarda:2021ict,Camarda:2021jsw,Chen:2022cgv,Chen:2022lwc,Campbell:2023lcy}.

  Adding sub-leading power corrections, computed to a given logarithmic accuracy, can improve the numerical accuracy of $q_T$ subtraction. These have been studied in Refs.~\cite{Ebert:2018gsn,Ebert:2019zkb}. At NLO accuracy, a method to compute all-order power corrections was recently presented, and used to compute next-to-next-to-leading power corrections to Higgs production~\cite{Ferrera:2023vsw}.
  Lastly, special types of linear power corrections arise from fiducial cuts~\cite{Frixione:1997ks,Ebert:2019zkb,Alekhin:2021xcu}, which can be eliminated from the $q_T$ slicing calculation through a simple recoil prescription~\cite{Catani:2015vma,Ebert:2020dfc} or alternatively through the adjustment of cuts~\cite{Salam:2021tbm}.

\item $N$-jettiness~\cite{Boughezal:2015eha,Boughezal:2015dva,Gaunt:2015pea}:\\
  A slicing approach based on the resolution variable $\tau_N$ ($N$-jettiness) that is suited for processes beyond the scope of the $q_T$ method, i.e.\ involving final-state jets. The N-jettiness soft function has now been calculated for arbitrarily many jets~\cite{Bell:2023yso,Agarwal:2024gws}. 
  Applied to $V(+j)$~\cite{Boughezal:2015dva,Boughezal:2015aha,Boughezal:2015ded,Boughezal:2016isb,Boughezal:2016yfp,Boughezal:2016dtm,Campbell:2016jau,Campbell:2016lzl,Campbell:2017aul} and
  $H+j$~\cite{Campbell:2019gmd}.
  Colourless final state production available in the MCFM program~\cite{Boughezal:2016wmq,Campbell:2019dru}.
  Same technique also used in the calculation of top decay~\cite{Gao:2012ja} and $t$-channel single top production~\cite{Berger:2016oht}.
  Important progress towards the extension for \NNNLOQCD calculations have been made~\cite{Melnikov:2018jxb,Melnikov:2019pdm,Behring:2019quf,Billis:2019vxg,Baranowski:2020xlp,Ebert:2020unb,Baranowski:2022khd,Baranowski:2022vcn,Bell:2023yso,Agarwal:2024gws,Baranowski:2024ene,Baranowski:2024vxg,Baranowski:2024ysi} with all ingredients now known for zero-jettiness slicing at \NNNLOgen.

  Including sub-leading power corrections, computed to a given logarithmic accuracy, can improve the numerical performance of the $N$-jettiness method. The leading power corrections are known  for color singlet production to \NNNLOgen, computed in Soft--Collinear Effective Theory (SCET)~\cite{Vita:2024ypr} and to next-to-leading power accuracy for color singlet production to NLO, both using SCET~\cite{Ebert:2018lzn} and direct QCD~\cite{Boughezal:2018mvf}.   The next-to-leading power  corrections to $V+j$ productions at NLO were computed in Ref.~\cite{Boughezal:2019ggi}. The impact of fiducial and isolation cuts on power corrections in both the $N$-jettiness and $q_T$ subtractions was analyzed in~Ref.~\cite{Ebert:2019zkb}. Recently, a procedure to improve both the $N$-jettiness slicing method using projection-to-Born correction factors was proposed, and exhibits an improved numerical behavior for Higgs, Drell--Yan and diphoton production~\cite{Vita:2024ypr,Campbell:2024hjq}.

\item ColorFul subtraction~\cite{DelDuca:2015zqa}:\\
  Fully local subtraction extending the ideas of the Catani--Seymour dipole method at NLO~\cite{Catani:1996vz}.
  Analytically integrated counter-terms for the infrared poles, numerical integration for finite parts.
  Fully worked out for processes with hadronic final states and applied to $H\to b\bb$~\cite{DelDuca:2015zqa} and $e^+e^-\to$ 3 jets~\cite{DelDuca:2016csb,DelDuca:2016ily,Tulipant:2017ybb}.
  Extended to the case of colourless final states in hadron collisions~\cite{VanThurenhout:2024hmd} with a public implementation for $H$ production in the code {\tt NNLOCAL}~\cite{DelDuca:2024ovc}.

\item Nested soft--collinear subtraction~\cite{Caola:2017dug,Caola:2018pxp,Delto:2019asp}:\\
  Fully local subtraction with analytic results for integrated subtraction counterterms.
  Worked out for processes with hadronic initial and final states~\cite{Caola:2019nzf, Caola:2019pfz,Asteriadis:2019dte}.
  Applied to compute \NNLOQCD corrections to VH~\cite{Caola:2017xuq} and VBF~\cite{Asteriadis:2021gpd}, as well as mixed QCD--EW corrections to the Drell--Yan process~\cite{Delto:2019ewv,Buccioni:2020cfi,Behring:2020cqi}. The first step towards a generalization of this method was taken in Ref.~\cite{Devoto:2023rpv}, where the analytical cancellation of IR singularities in the production of arbitrarily many gluons in quark--antiquark annihilation was demonstrated. Further development for additional partonic channels is underway.

\item Local analytic sector subtraction~\cite{Magnea:2018hab, Magnea:2018ebr,Magnea:2020trj}:\\
  Local subtraction with analytic integration of the counterterms aiming to combine the respective advantages from two NLO approaches of FKS subtraction~\cite{Frixione:1995ms,Frederix:2009yq} and dipole subtraction~\cite{Catani:1996vz}.
  First proof-of-principle results for $e^+e^-\to 2$\,jets were presented in~\cite{Magnea:2018hab}.

  The analytic pole cancellation in fully differential observables in the production of arbitrarily many massless partons in $e^-e^+$ collisions was demonstrated in Ref.~\cite{Bertolotti:2022aih}. Progress towards hadronic initial states is underway~\cite{Bertolotti:2022ohq}. The first steps towards an extension to \NNNLOQCD were taken in Ref.~\cite{Magnea:2024jqg}, where the architecture of infrared subtraction in full generality and the organisation of relevant counterterms was presented.

\item Projection to Born~\cite{Cacciari:2015jma}:\\
  Requires the knowledge of inclusive calculations that retain the full differential information with respect to Born kinematics.
  With the necessary ingredients in place, generalisable to any  order.
  Applied at \NNLOQCD to VBF~\cite{Cacciari:2015jma}, Higgs-pair production via VBF~\cite{Dreyer:2018rfu}, and $t$-channel single top production~\cite{Berger:2016oht,Campbell:2020fhf}.
  Fully differential \NNNLOQCD predictions obtained for jet production in DIS~\cite{Currie:2018fgr,Gehrmann:2018odt}, $H\to b\bar{b}$~\cite{Mondini:2019gid}, and Higgs production in gluon fusion~\cite{Chen:2021isd}.
\end{itemize}


\section{Update on the precision Standard Model wish list}
\label{sec:SM_wishlist:precision_wish_list}

This section is divided in four parts which comprise: Higgs-boson associated processes, jet final states, vector-boson associated processes, and top-quark associated processes.

The terms of the expansion are defined with respect to the Born contribution and expanded in the QCD and electroweak couplings as:
\begin{equation}
  d\sigma_X = d\sigma_X^{\rm LO} \left(1 +
      \sum_{k=1} \alphas^k d\sigma_X^{\delta \text{\NLOQ{k}}}
    + \sum_{k=1} \alpha^k d\sigma_X^{\delta \text{\NLOE{k}}}
    + \sum_{k,l=1} \alphas^k \alpha^l d\sigma_X^{\delta \text{\NLOQE{k}{l}}}
    \right).
  \label{eq:SM_wishlist:dsigmapertexp}
\end{equation}
Note that Eq.~\eqref{eq:SM_wishlist:dsigmapertexp} only applies to cases where the leading-order process is uniquely defined through the powers of the respective couplings.
In the following, the notation \NLOSM is used to denote NLO calculations that include the full Standard Model corrections, \ie all QCD and EW corrections to all leading-order contributions.

Given that the fields of resummation and parton showers have seen tremendous progress in the past years, we feel that it warrants a specific document.%
\footnote{A wish of the Les Houches wishlist.}
The interested reader may consult Ref.~\cite{Campbell:2022qmc} for an overview.
Nonetheless, where relevant, we provide the recent developments in parton shower and resummation that are relevant for the given process.

Below, an overview of the current status of fixed-order calculations within the Standard Model is provided.
The references mainly focus on the state of the art at the time of writing.
In particular, superseded computations can be found in the respective process categories of prior wishlists.
In detail, we provide a short overview of the status of theory predictions as documented in the previous wishlist (LH21), followed by a description of the progress since then.
Before moving to the actual wishlist, several aspects and highlights of the recent years of fixed-order calculations are discussed.

\paragraph*{Electroweak corrections}

Given the present and anticipated experimental precision from Run~3 of the LHC and its future HL-LHC upgrade, EW radiative corrections have become essential to be included in the analysis of many SM processes alongside higher-order QCD corrections.
The increase in experimental precision further demands the inclusion of mixed QCD--EW corrections for some key processes such as the Drell--Yan like production of electroweak gauge bosons.

Generally, EW corrections can receive sizeable enhancements in two scenarios:
First, in the vicinity of resonances and shoulders where photon emission (in QED) induce large corrections that can further be enhanced in the case of non-collinear safe observable (such as bare-lepton observables).
Second, in the high-energy limit where Sudakov logarithms (in the weak theory) can become large.
These effects have been studied for a plethora of processes and are well understood; the interested reader can consult the comprehensive review article~\cite{Denner:2019vbn} for further details.
Nonetheless, in contrast to QCD predictions where the scale variation offers a convenient approach to estimate the impact of missing higher orders, this is typically not the case of EW corrections as they are renormalised at physical points.
The issue of assessing the uncertainties on EW corrections is thus more subtle with first steps in this direction taken in Ref.~\cite{Lindert:2017olm} and continued in~\cite{Andersen:2024czj}.

One-loop Matrix Elements for EW corrections are readily available from many one-loop providers:
\OpenLoops~\cite{Cascioli:2011va,Buccioni:2019sur}, \GoSam~\cite{Cullen:2011ac,Cullen:2014yla,Chiesa:2017gqx}, \Recola~\cite{Actis:2012qn,Actis:2016mpe,Denner:2017wsf}, \MadLoop~\cite{Alwall:2014hca,Frederix:2018nkq}, and \NLOX~\cite{Honeywell:2018fcl}
are publicly available and incorporated in various public and private Monte Carlo programs capable of performing NLO calculations.
The highest multiplicity achieved at \NLOEW so far is for a $2\to8$ scattering process, the associated-top production~\cite{Denner:2021hqi,Denner:2023eti} (off-shell $t\tb W$ and $t\tb Z$).

Electroweak Sudakov logarithms have received renewed interest in the recent years with their incorporation into different automated tools~\cite{Chiesa:2013yma,Bothmann:2020sxm,Pagani:2021vyk,Lindert:2023fcu} based on the original work of Ref.~\cite{Denner:2000jv}.
Isolating the enhanced Sudakov corrections allows to incorporate dominant effects in certain phase-space regions while avoiding the additional complexity that a full EW calculation entails, in particular from IR singularities induced by QED corrections.
Moreover, they serve as a convenient starting point for QCD parton shower matching~\cite{Kallweit:2017khh,Gutschow:2018tuk,Brauer:2020kfv,Bothmann:2021led,Pagani:2023wgc} and the resummation of Sudakov logarithms~\cite{Denner:2024yut}.
Lastly, their impact in the context of new physics has been studied in the context of Effective Field Theories~\cite{ElFaham:2024egs}.

\paragraph*{On-shell and off-shell descriptions}

The resonance of intermediate unstable particles admits various approximations that allow to reduce the complexity of the calculation.
Among the most common are the Narrow-Width Approximation (NWA) and the Pole Approximation (PA).
The NWA is valid for narrow resonances in which case the intermediate particle can be approximately treated as stable, effectively replacing the internal propagator by an on-shell delta distribution and thus only retaining resonant diagrams.
The PA instead performs a consistent expansion around the resonance, retaining all leading terms in which the resonant propagators are kept intact while their residues are evaluated on-shell.
This approximation includes resonant diagrams as well as non-factorizable contributions that arise from soft gluon or photon exchange.

In order to describe non-resonant effects of a process, a full off-shell calculation is required.
In this case, the complete final state after the decay of the unstable particle must be considered, including all contributions that may or may not include the resonant state.
This however comes at an additional cost in the complexity of the calculation (larger number of Feynman diagrams with more complex expressions) that in turn reflects in an increase in computing time.
The current frontier calculations have achieved a multiplicity of $2\to8$ scattering at \NLOgen, while the multiplicity frontier at \NNLOQCD is currently at $2\to3$ processes.
Nonetheless, all $2\to3$ calculations at \NNLOQCD involve either massless final-state particles (photons or jets) or unstable particles treated as stable, meaning that off-shell and non-resonant contributions are not yet accounted for at \NNLOQCD accuracy.

When reviewing the status of the calculations for the wishlist below, off-shell effects are assumed to be included.
For QCD corrections to processes featuring a purely EW decay, the different treatments of the resonances does not give rise to additional complications.
This is for instance not the case for EW corrections and processes featuring top quarks.
In the latter case of top quarks, we explicitly indicate if off-shell effects are included in a calculation.

\paragraph*{Jet algorithms, identified final states, and fragmentation}

\NNLOgen predictions are necessary to achieve the highest precision for $2\to 2$ (and $2\to 3$) processes at the LHC.
The presence of one or more jets in the final state requires the application of a jet algorithm, almost universally the anti-$k_t$ algorithm as they give rise to geometrically regular jets.
However, there can be accidental cancellations that can result in artificially small scale uncertainties, especially close to jet radii of $R=0.4$.
A more realistic estimate of the uncertainty can be obtained by the use of a larger radius jet ($R=0.6$--$0.7$), or by alternate estimates for uncertainties from missing higher orders~\cite{Bellm:2019yyh,Rauch:2017cfu,Buckley:2021gfw}.

Increasingly, many of the precision LHC measurements involve the presence of heavy quarks in the final state, e.g.\ V+c/b (see later discussion in the vector boson section). If the heavy quark is treated as massless, any calculation at \NNLOgen requires the application of an IR-collinear safe jet algorithm, to reduce the sensitivity to log-enhanced terms [proportional to $\alphas^n \log^m(m_q/p_t$)], such as with the
flavour-$k_t$ algorithm~\cite{Banfi:2006hf}.
The experimental approach is to first reconstruct the jet using the anti-$k_t$ jet algorithm, and then afterwards to look for the presence of heavy flavour tag within that jet. The transverse momentum requirement for the heavy flavour tag is typically much less than the transverse momentum of the jet itself. This can lead to many jets being tagged as heavy-flavour due to gluon splittings into a (relatively soft) quark--antiquark pair, an indication of the log-enhancement described above for the theory calculation.

The mis-match between experimental and theoretical algorithms can result in an error of the order of 10\%, potentially larger than the other sources of uncertainty in the measurement/prediction.
A computation based on massive heavy quarks  (see e.g.\ Ref.~\cite{Behring:2020uzq} for a comparison against flavour-$k_t$ in $WH$ production), or with the inclusion of the fragmentation contribution at \NNLOgen (see e.g.\ Ref.~\cite{Czakon:2021ohs} for \NNLOQCD predictions for $B$-hadron production in $t\bar{t}$) can reduce the theory uncertainty.
Alternatively, new jet-tagging algorithms compatible with the anti-$k_t$ definition~\cite{Caletti:2022glq,Caletti:2022hnc,Czakon:2022wam,Gauld:2022lem,Caola:2023wpj} can be used for the same purpose.
Also, the mismatch between the \emph{particle-level} observables and \emph{parton-level} observables, which are particularly important for measurements with jets is an important effect which has been partially discussed in follow-up studies of the Les Houches workshop 2023~\cite{Andersen:2024czj,Behring:2025ilo}.

The flavour-tagging algorithms referenced above require a complete knowledge of the heavy flavour content of the event, something that is difficult to obtain in any experimental measurement, especially if it involves the tagging of charm quarks. It is currently not well known (1) the efficiency with which LHC experiments can reconstruct gluon splitting into heavy quark pairs and (2) how well the parton shower Monte Carlos estimate the rate of this splitting. The latter was a well-known problem at the Fermilab Tevatron~\cite{Campbell:2004gj}. A recent workshop\footnote{\url{https://conference.ippp.dur.ac.uk/event/1301/}} discussed these issues, and will lead to experimental studies which hopefully provide a better understanding of the situation, such that the \NNLOgen predictions can be used to their fullest extent.

A similar issue with a mismatch between experiment and theory arises in the case of identified photons that require an isolation procedure to distinguish the prompt production from the overwhelming background.
Differences in a fixed-cone isolation versus a smooth-cone isolation~\cite{Frixione:1998jh,Siegert:2016bre} have been the subject of many studies which assessed the impact to be at the few-percent level~\cite{Andersen:2014efa,Andersen:2016qtm,Catani:2018krb,Catani:2013oma,Amoroso:2020lgh}.
Precision phenomenology based on processes with external photons thus demands for an extension of the fragmentation contribution to \NNLOgen that has been achieved recently~\cite{Gehrmann:2022cih,Chen:2022gpk}.

\paragraph*{Polarised predictions for gauge-boson production}

The increased experimental precision not only enables a detailed study of the gauge-boson production processes through cross sections and differential distributions, but also the access to the polarization states of the gauge bosons.
To this end, the longitudinal polarization is of particular interest due to its intimate connection to the mechanism of electroweak symmetry breaking and how weak gauge bosons acquire their masses.
As such, the study of the longitudinal component of massive gauge bosons not only allows to scrutinize the Standard Model at a deeper level, but also may reveal hints for new physics that lies beyond.

In this context, the past few years has seen great progresses in polarized predictions for a plethora of LHC processes.
While most work has focused on \NLOQCD corrections for di-boson production~\cite{Denner:2020eck,Denner:2022riz}, significant progress on the respective \NLOEW corrections have been made recently~\cite{Denner:2021csi,Le:2022lrp,Denner:2023ehn,Dao:2023kwc,Le:2022ppa}.
Electroweak corrections entail significant complications due to the need for a consistent isolation of the resonant (on-shell) parts only for which polarizations are properly defined.
Such calculations thus rely on pole approximations with power corrections that can e.g.\ be induced by the details of the mappings to project onto on-shell states.
These efforts at NLO accuracy recently culminated with the \NLOQCD+\NLOEW corrections to vector-boson scattering in the same-sign $WW$ channel~\cite{Denner:2024tlu}.
\NNLOQCD corrections are so far limited to a handful of processes: diboson production~\cite{Poncelet:2021jmj} and $W+j$~\cite{Pellen:2021vpi}.
More recently, also \NLOQCD corrections matched to a parton shower have became available for all di-boson processes~\cite{Pelliccioli:2023zpd}.

Finally, while these calculations have been exclusively obtained with private Monte Carlo codes, there have been efforts in enabling such calculations within general-purpose Monte Carlo programs.
Progress was made at LO including parton-shower corrections within the {\sc MadGraph5\_aMC@NLO} framework~\cite{BuarqueFranzosi:2019boy} as well as within the {\sc Sherpa} framework, where also approximate \NLOQCD corrections can be incorporated consistently with the shower~\cite{Hoppe:2023uux}.

\subsection{Higgs boson associated processes}

An overview of the status of Higgs boson associated processes for decay and production is given in Tables \ref{tab:decays} and \ref{tab:SM_wishlist:wlH}, respectively.
In the following, the acronym \emph{Heavy Top limit} (HTL) is used to denote the effective field theory in the $m_t\to\infty$ limit.
In this limit, the Higgs bosons couple directly to gluons via the following effective Lagrangian
\begin{equation}
\mathcal{L}_{\rm eff} = - \frac{1}{4} G^a_{\mu \nu} G_a^{\mu \nu} \left(C_H \frac{H}{v} - C_{HH} \frac{H^2}{2 v^2} + C_{HHH} \frac{H^3}{3 v^3} + \ldots\right)\,,
\end{equation}
whose matching coefficients are known up to fourth order in $\alphas$~\cite{Chetyrkin:1997iv,Chetyrkin:2005ia,Kramer:1996iq,Schroder:2005hy,Djouadi:1991tka,Grigo:2014jma,Spira:2016zna,Gerlach:2018hen}.
The HTL results are often used to correct complete QCD results available at a lower perturbative order.
We will generically indicate this combination of HTL and QCD results using the notation $\text{\NLOH{x}} \otimes \text{\NLOQ{y}}$.
One strategy used for this combination is to compute a multiplicative $K$-factor in the HTL that is then applied to the complete QCD result.
Alternatively, the HTL $K$-factor can be used to correct only unknown parts of the QCD results, for example the virtual part of a calculation, which are then combined with exact real corrections.
The latter procedure is generally preferred, especially where a differential description is required.

\subsubsection{Decays}

\begin{table}[h!]
\begin{center}
\begin{tabular}{ll}
\hline
Partial Width             & known              \\ \hline
$b\overline{b}/c\overline{c}$               & \begin{tabular}[c]{@{}l@{}}$\text{\NLOH4}\otimes\text{\NNLOQCD}$\\ $\text{\NLOEW}$\end{tabular} \\ \hline
$WW/ZZ$   & \begin{tabular}[c]{@{}l@{}}$\text{\NLOQCD}$\\ $\text{\NLOEW}$\end{tabular}         \\ \hline
$\tau^+\tau^-/\mu^+\mu^-$ & \begin{tabular}[c]{@{}l@{}} - \\ $\text{\NLOEW}$\end{tabular}        \\ \hline
$gg$                      & \begin{tabular}[c]{@{}l@{}}$\text{\NLOH{4}} \otimes \text{\NNLOQCD}$\\ $\text{\NLOEW}$\end{tabular}          \\ \hline
$\gamma \gamma$           & \begin{tabular}[c]{@{}l@{}}$\text{\NLOH{3}} \otimes \text{\NNLOQCD}$\\ $\text{\NLOEW}$\end{tabular}           \\ \hline
$Z\gamma$                 & \begin{tabular}[c]{@{}l@{}}$\text{\NLOQCD}$\\ $\text{\NLOEW}$ \end{tabular}      \\ \hline
\end{tabular}
\caption{
Available theory results for Higgs boson decay, adapted from Refs.~\cite{Proceedings:2019vxr,Spira:2016ztx}.
}
\label{tab:decays}
\end{center}
\end{table}

In Table~\ref{tab:decays}, we list the Standard Model Higgs boson decays, as well as the available theory precision for each channel.
Below we briefly summarise the key theoretical works contributing to the known precision; for more detailed reviews we refer the reader to Refs.~\cite{LHCHiggsCrossSectionWorkingGroup:2016ypw,Spira:2016ztx,Proceedings:2019vxr,Jones:2023uzh}.
This is the first time that a review of Higgs decays is included in the Les Houches wishlist.

The $H \rightarrow b\overline{b}$ and $H \rightarrow c\overline{c}$ partial widths are known at
\NNLOQCD~\cite{Bernreuther:2018ynm,Primo:2018zby,Behring:2019oci,Somogyi:2020mmk,Wang:2023xud}
including quark mass effects.
The \NNNLOQCD correction for the $y_t$ induced contribution, including the mass of the bottom quark, was recently computed~\cite{Wang:2024ilc}.
In the heavy top limit and neglecting the bottom quark mass, results are known to order \NLOH4~\cite{Herzog:2017dtz}
and fully differentially at \NLOH3~\cite{Mondini:2019gid}.
The \NLOEW and $\mathcal{O}(\alpha\alpha_s)$ mixed QCD-EW corrections were obtained in Refs.~\cite{Djouadi:1997rj,Kataev:1997cq,Kniehl:1994ju,Kwiatkowski:1994cu,Chetyrkin:1996wr,Mihaila:2015lwa} (with two-loop master integrals in Ref.~\cite{Chaubey:2019lum}), these calculations also represent the state of the art for $H \rightarrow \tau^+ \tau^-$ and $H \rightarrow \mu^+ \mu^-$ decays.

The $H \rightarrow WW$ and $H \rightarrow ZZ$ widths are known at \NLOQCD~\cite{Rizzo:1980gz,Keung:1984hn,Cahn:1988ru}
and \NLOEW~\cite{Fleischer:1980ub,Kniehl:1990mq,Bardin:1991dp,Bredenstein:2006rh,Bredenstein:2006ha}. Recently, also $\mathcal{O}(\alpha\alpha_s)$ mixed QCD-EW corrections have been calculated~\cite{Kaur:2023eyv}.
The $H \rightarrow gg$ decay is known at \NNLOQCD~\cite{Harlander:2019ioe,Czakon:2020vql} including quark mass effects.
The corrections in the heavy top limit are known at \NLOH4~\cite{Baikov:2006ch,Herzog:2017dtz}.
The \NLOEW corrections were computed in Refs.~\cite{Djouadi:1994ge,Aglietti:2004nj,Degrassi:2004mx,Aglietti:2006yd,Actis:2008ug,Actis:2008ts}.
The $H \rightarrow \gamma \gamma$ decay is known at \NNLOQCD including the exact mass dependence~\cite{Maierhofer:2012vv,Niggetiedt:2020sbf}
and at \NLOH{3} in the large mass approximation~\cite{Davies:2021zbx}
the \NLOEW corrections are also known~\cite{Actis:2008ts,Djouadi:1997rj,Degrassi:2005mc,Passarino:2007fp}.

The $H \rightarrow Z \gamma$ decay is measured as part of the Dalitz decays $H \rightarrow f \overline{f} \gamma$, which receive contributions from $H \rightarrow \gamma^* \gamma \rightarrow f \overline{f} \gamma$, $H \rightarrow Z^* \gamma \rightarrow f \overline{f} \gamma$ and direct/non-resonant $H \rightarrow f \overline{f} \gamma$.
The relevance of each channel depends on the experimental cuts and reconstruction strategy.
For the loop-induced $H \rightarrow Z \gamma$ decay, \NLOQCD results are known~\cite{Spira:1991tj,Bonciani:2015eua,Gehrmann:2015dua}, the \NLOEW corrections were obtained very recently~\cite{Chen:2024vyn,Sang:2024vqk} and found to be small, and unable to explain the discrepancy between theory predictions and experimental measurements.
For direct $H \rightarrow f \overline{f} \gamma$ results are known at \NLOQCD and \NLOEW accuracy~\cite{Abbasabadi:1996ze,Abbasabadi:2006dd,Dicus:2013ycd,Chen:2012ju,Passarino:2013nka,Sun:2013rqa}.

Significant effort has recently been invested in the description of hadronic Higgs decays through both the $H \rightarrow b \overline{b}$ and $H \rightarrow gg$ channels.
In Ref.~\cite{Chen:2023fba} results at \NLOH{3} were presented.
Four jet event shapes have been studied at the \NLOQCD level~\cite{Gehrmann-DeRidder:2023uld} and three jet event shapes at \NLOQCD with \NLL{} resummation for hadronic Higgs decays are known~\cite{Coloretti:2022jcl,Gehrmann-DeRidder:2024avt}.
A study of flavour-sensitive observables at \NLOQCD for up to three jet hadronic Higgs decays was presented in Ref.~\cite{Aveleira:2024dcx}.

Many of the dominant QCD and EW corrections for the major Higgs boson decay modes are available in the Prophecy4f~\cite{Bredenstein:2006rh,Bredenstein:2006ha,Denner:2019fcr}, HDECAY~\cite{Djouadi:1997yw,Djouadi:2018xqq} and Hto4L~\cite{Boselli:2015aha} programs, which are widely used.

\subsubsection{Production}

\begin{table}[h!]
  \renewcommand{\arraystretch}{1.5}
\setlength{\tabcolsep}{5pt}
  \begin{center}
  \begin{tabular}{lll}
    \hline
    \multicolumn{1}{c}{process} & \multicolumn{1}{c}{known} &
    \multicolumn{1}{c}{desired} \\
    \hline
    $pp\to H$ &
    \begin{tabular}{l}
      \NNNLOHTL \ \\
      \NNLOQCDTTXB \ \\
      \NLOHE11 \\
      \NLOQCD
    \end{tabular} &
    \begin{tabular}{l}
      \NLOH4 (incl.) \\
    \end{tabular} \\
    \hline
    $pp\to H+j$ &
    \begin{tabular}{l}
      \NNLOHTL \\
      \NLOQCD \\
      \NLOQE11 \\ 
    \end{tabular} &
    \begin{tabular}{l}
      \NNLOHTL$\!\otimes\,$\NLOQCD\!+\,\NLOEW \\
      \NNNLOHTL \\
      \NNLOQCD \
    \end{tabular} \\
    \hline
    $pp\to H+2j$ &
    \begin{tabular}{l}
      \NLOHone$\!\otimes\,$\LOQCD \\
      \NNNLOQCDVBFstar (incl.) \\
      \NNLOQCDVBFstar  \\
      \NLOEWVBF
    \end{tabular} &
    \begin{tabular}{l}
      \NNLOHTL$\!\otimes\,$\NLOQCD\!+\,\NLOEW\\
      \NNNLOQCDVBFstar \\
      \NNLOQCDVBF \\
      \NLOQCD
    \end{tabular} \\
    \hline
    $pp\to H+3j$ &
    \begin{tabular}{l}
      \NLOHone \\
      \NLOQCDVBF
    \end{tabular} &
    \begin{tabular}{l}
      \NLOQCD\!+\,\NLOEW \\
      \NNLOQCDVBFstar
    \end{tabular} \\
    \hline
    $pp\to VH$ &
    \begin{tabular}{l}
      \NNNLOQCD\!(incl.)+\,\NLOEW \\
      \NLOggHVtb{} \\
    \end{tabular} &
    \begin{tabular}{l}
    \NNNLOQCD \\
    \NLOQE11 \\
    \end{tabular} \\
    \hline
    $pp\to VH + j$ &
    \begin{tabular}{l}
      \NNLOQCD \ \\
      \NLOQCD\!+\,\NLOEW \\
    \end{tabular} &
    \begin{tabular}{l}
    \end{tabular} \\
    \hline
    $pp\to HH$ &
    \begin{tabular}{l}
      \NNNLOHTL$\!\otimes\,$\NLOQCD \\
      \NLOEW \\
    \end{tabular} &
    \begin{tabular}{l}
      \NNLOQCD \\
    \end{tabular} \\
    \hline
    $pp\to HH + 2j$ &
    \begin{tabular}{l}
      \NNNLOQCDVBFstar (incl.) \\
      \NNLOQCDVBFstar  \\
      \NLOEWVBF
    \end{tabular} &
    \begin{tabular}{l}
    \NLOQCD \\
    \end{tabular} \\
    \hline
    $pp\to HHH$ &
    \begin{tabular}{l}
      \NNLOHTL \\
    \end{tabular} &
    \begin{tabular}{l}
    \NLOQCD
      \\
    \end{tabular} \\
    \hline
    $pp\to H+t\tb$ &
    \begin{tabular}{l}
      \NLOQCD\!+\,\NLOEW\\
      \NNLOQCD (approx.)
    \end{tabular} &
    \begin{tabular}{l}
     \NNLOQCD
    \end{tabular}  \\
    \hline
    $pp\to H+t/\tb$ &
    \begin{tabular}{l}
      \NLOQCD\!+\,\NLOEW\\
    \end{tabular} &
    \begin{tabular}{l}
      \NNLOQCD
    \end{tabular} \\
    \hline
  \end{tabular}
  \caption{Precision wish list: Higgs boson final states. \NLOQVBFstar{x} means a
   calculation using the structure function approximation. $V=W,Z$.}
  \label{tab:SM_wishlist:wlH}
  \end{center}
\renewcommand{\arraystretch}{1.0}
\end{table}

\subsubsection{$H$}

\textit{LH21 status}
Results at \NNLOHTL known for two decades~\cite{Harlander:2002wh,Anastasiou:2002yz,Ravindran:2003um,Catani:2007vq,Grazzini:2008tf}.
Inclusive \NNNLOHTL results computed in \cite{Anastasiou:2015vya,Anastasiou:2016cez,Mistlberger:2018etf} and available exactly in the programs {\sc iHixs 2}~\cite{Dulat:2018rbf} and in an expansion around the Higgs production threshold in {\sc SusHi}~\cite{Harlander:2016hcx}.
Differential results at \NNNLOHTL were presented in Ref.~\cite{Dulat:2017brz,Dulat:2017prg,Dulat:2018bfe,Cieri:2018oms,Chen:2021isd} and the transverse momentum spectrum of the Higgs boson has been studied at \NNLOgen\!+\,\NNNLLp~\cite{Chen:2018pzu,Bizon:2018foh,Re:2021con} and at \NNNLOHTL\!+\,\NNNLLp~\cite{Billis:2021ecs}. \NNLOgen+\NNLL predictions for $gg \rightarrow H(\rightarrow \gamma \gamma)$ are publicly available through \textsc{Hturbo}~\cite{Camarda:2022wti}.
The $m_t$-dependence is known at 3-loops for the virtual piece~\cite{Davies:2019nhm,Czakon:2020vql,Harlander:2019ioe} and at 4-loops in a large-$m_t$ expansion~\cite{Davies:2019wmk}.
Complete \NLOQCD corrections are known for arbitrary quark masses~\cite{Dawson:1990zj,Djouadi:1991tka,Graudenz:1992pv,Spira:1995rr,Harlander:2005rq,Anastasiou:2006hc,Aglietti:2006tp,Anastasiou:2009kn}, while the top mass dependence is known at \NNLOQCD~\cite{Czakon:2021yub}. The top-quark mass  renormalisation scheme uncertainty for offshell Higgs production has been studied at the same order~\cite{Mazzitelli:2022scc}.
Bottom quark effects have been studied for intermediate Higgs transverse momentum $m_b \lesssim p_T \lesssim m_t$ at \NLOgen\!+\,\NNLL~\cite{Caola:2018zye}.
Power-suppressed logarithms  of the form $ y_q m_q \alphas^n \ln^{2n-1}(\frac{m_H}{m_q})$, where $y_q$ is the Yukawa coupling, arise for small virtual quark masses in $gg \to H$. These have been resummed for next-to-leading power $\mathcal{O}(m_q^2)$ corrections at \NLL accuracy~\cite{Liu:2017vkm,Liu:2017axv,Liu:2018czl,Anastasiou:2020vkr} for next-to-next-to-leading power at double-logarithmic accuracy~\cite{Liu:2021chn}.
Mixed QCD--EW corrections, \NLOHE11, are known in the limit of small electroweak gauge boson mass~\cite{Bonetti:2018ukf,Anastasiou:2018adr}, and the dominant light-quark contribution to the \NLOgen mixed QCD-EW corrections have been computed including the exact EW-boson mass dependence~\cite{Becchetti:2020wof}.
\medskip

\noindent \textit{Progress}
Inclusive results to \NNNLOHTL are now available in the public program {\sc n3loxs}~\cite{Baglio:2022wzu}.
In Ref.~\cite{Campbell:2023cha}, jet-veto resummation was implemented to \NNLOQCD matched to \NNNLLp in the public code \MCFM, and theoretical predictions at this order were compared with ATLAS and CMS data. A substantial reduction of theoretical uncertainties relative to the \NNLL accuracy was observed.

There has been important recent progress on the impact of virtual quark masses. Interference effects between amplitudes with top and bottom loops were studied in Ref.~\cite{Czakon:2023kqm} with \NNLOQCD accuracy, where the bottom and top quark masses are renormalized onshell. The perturbative convergence is observed to be quite bad, with the $\mathcal{O}(\alphas^3)$ and $\mathcal{O}(\alphas^4)$ being almost identical numerically. Using the $\overline{\rm MS}$ renormalization scheme for both the bottom mass and the bottom Yukawa improves the convergence dramatically~\cite{Czakon:2024ywb}. This reference also presents comparisons between results computed using the 4FS and 5FS. The calculation of Ref.~\cite{Czakon:2021yub} was matched to parton showers using the \MiNNLOPS formalism in Ref.~\cite{Niggetiedt:2024nmp}.

There has also been recent progress in advancing the precision of predictions for Higgs-interference processes.
The interference between amplitudes for Higgs-mediated signal and prompt background processes in the diboson decay channel ($gg \to  H \to VV$ and $gg \to VV$, respectively) are known to be significant for offshell Higgs production, which comprises around 10\% of the events in this channel~\cite{Kauer:2012hd}, and can be used to place a stringent bound on the Higgs width~\cite{Caola:2013yja,Campbell:2013una,Campbell:2013wga}. For many years, the computation of the exact \NLOQCD corrections was hampered by the difficulty in computing two-loop $gg \to VV$ amplitudes with a massive quark loop. Approximate \NLOQCD corrections were computed using the heavy-top expansion~\cite{Melnikov:2015laa,Campbell:2016ivq,Caola:2016trd}, the high-energy expansion~\cite{Davies:2020lpf}, and a combination of these two approximations and the threshold approximation was presented in Ref.~\cite{Grober:2017uho,Grober:2019kuf}. \NLOQCD corrections are also available with a reweighting procedure for the massive two-loop $gg \to VV$ amplitudes~\cite{Grazzini:2018owa,Grazzini:2021iae}. \NLOQCD corrections matched to parton showers in the \Powheg approach were presented in Ref.~\cite{Alioli:2021wpn}. The two-loop amplitudes for $gg \to VV$ processes with massive quark loops were computed in Refs.~\cite{Agarwal:2020dye,Bronnum-Hansen:2021olh,Bronnum-Hansen:2020mzk}. Recently, the amplitudes of Ref.~\cite{Agarwal:2020dye} were used to compute the exact \NLOQCD corrections to $gg \to VV$, including the signal and background processes and their interference~\cite{Agarwal:2024pod}. It is found that the reweighting approach provides an extremely accurate approximation to the full two-loop amplitudes.

The interference between the Higgs-mediated $gg \to H \to \gamma \gamma$ process and the prompt background $gg \to \gamma \gamma$ process results in a shift in the Higgs peak in the diphoton invariant mass distribution, which can be used to constrain the Higgs width~\cite{Martin:2012xc}. Although it is expected that such bounds are about 5-30 times the Standard Model Higgs width value, and therefore less constraining than measurements in using off-shell Higgs production, they do not suffer from the same model dependence, and in that sense are complementary. The LO analysis of Ref.~\cite{Martin:2012xc} was refined to include all partonic channels~\cite{deFlorian:2013psa} and the emission of one ~\cite{Martin:2013ula} and two~\cite{Coradeschi:2015tna} hard jets. \NLOQCD corrections, first presented  in Ref.~\cite{Dixon:2013haa} and  later confirmed in Ref.~\cite{Campbell:2017rke}, reduce the shift of the mass peak by around 40\%. Small $p_T$ resummation was performed in Ref.~\cite{Cieri:2017kpq}. Recently, the \NNLOQCD corrections in the soft-virtual limit (i.e.\ neglecting hard emissions) were presented~\cite{Bargiela:2022dla}, using three-loop helicity amplitudes for $gg \to \gamma \gamma$~\cite{Bargiela:2021wuy} and two-loop amplitudes for $\gamma \gamma j$ production~\cite{Badger:2021imn, Agarwal:2021vdh}. These corrections reduce the mass shift by a further 30\%.

Interference effects between the Higgs-mediated process $gg \to H \to Z \gamma$ and the prompt background $gg \to Z\gamma$ have been computed in Ref.~\cite{Buccioni:2023qnt} to \NLOQCD, employing the soft-virtual approximation. It was found that the interference effects amount to around $-3\%$ of the signal $gg \to H \to Z\gamma$ cross section, and that the \NLOQCD corrections are small, below the
current experimental accuracy.

\medskip \noindent \textit{Experimental status}
The experimental uncertainty on the total Higgs boson cross section is currently
of the order of 8\%~\cite{ATLAS:2019mju,CMS:2025ihj}
based on a data sample of 138-139\,fb$^{-1}$,
and is expected to reduce to the order of 3\% or less with a data sample
of 3000\,fb$^{-1}$~\cite{Campbell:2017hsr}. Most Higgs boson couplings will be known to 2-5\%~\cite{Cepeda:2019klc}.
To achieve the desired theoretical uncertainty, it may be necessary to also consider the
finite-mass effects at \NNLOQCD from $b$ and $c$ quarks, combined with fully differential \NNNLOHTL corrections.

Sometimes the form of experimental cuts can affect the perturbative stability of the theoretical prediction through linear fiducial power corrections. This is the case, for example, for the traditional cuts applied to the decay photons in Higgs boson diphoton events. The traditional cuts require that each of the two photons have a transverse momentum greater than a given fraction of the Higgs boson mass (typically 0.35 for the leading and 0.25 for the sub-leading photon). This leads to an instability of the perturbative convergence of the prediction and increased scale uncertainties, most noticeable at N3LO. This issue can be avoided by a re-design of the cuts~\cite{Salam:2021tbm} and is currently being investigated by the LHC experiments.
A brief summary of the issues associated with fiducial cuts was given in the 2021 document~\cite{Huss:2022ful}.

\subsubsection{$H+j$}

\textit{LH21 status}
Known at \NNLOHTL~\cite{Chen:2014gva,Chen:2016zka,Boughezal:2015dra,Boughezal:2015aha,Caola:2015wna,Campbell:2019gmd} and at \NLOQCD including both top-quark and bottom-quark mass effects~\cite{Jones:2018hbb,Lindert:2018iug,Neumann:2018bsx,Bonciani:2019jyb,Frellesvig:2019byn,Bonciani:2022jmb}; top--bottom interference effects are also known~\cite{Melnikov:2016qoc,Lindert:2017pky}.
Fiducial cross sections for the four-lepton decay mode were calculated in Ref.~\cite{Chen:2019wxf}.
The Higgs $p_T$ spectrum with finite quark mass effects is known beyond LO using high-energy resummation techniques at \LL accuracy~\cite{Caola:2016upw} and in the "High-energy jets" framework~\cite{Andersen:2009nu,Andersen:2009he,Andersen:2011hs,Andersen:2017kfc,Andersen:2018tnm,Andersen:2018kjg}; parton shower predictions including finite mass effects available in various approximations~\cite{Frederix:2016cnl,Neumann:2016dny,Hamilton:2015nsa,Buschmann:2014sia}.
The transverse momentum spectrum has also been studied at \NLOgen\!+\,\NNLL in the case a jet veto, $p_t^j \ll p_t^{j,v}$, is applied~\cite{Monni:2019yyr,ATLAS:2022fnp}.
The leading EW effects for the $qg$ and $q\bar{q}$ channels were computed some time ago~\cite{Mrenna:1995cf,Keung:2009bs} and amplitudes for the leading mixed QCD-EW corrections are known~\cite{Becchetti:2018xsk,Bonetti:2020hqh,Becchetti:2021axs,Bonetti:2022lrk}. The $b\bar{b} \rightarrow H + j$ process is known differentially at \NNLOQCD~\cite{Mondini:2021nck}, $H+c$ is known at \NLOQCD~\cite{Bizon:2021nvf}.

\medskip

\noindent \textit{Progress}
Efforts to compute $H+j$ production at \NNNLOHTL are ongoing, with results for many of the relevant Feynman integrals (including all required for the leading colour approximation) now known~\cite{Henn:2023vbd,Gehrmann:2024tds}.
In Ref.~\cite{Gehrmann:2023etk}, the two-loop (\NNLOHTL) helicity amplitudes amplitudes were presented to higher orders in the dimensional regulator, this is an ingredient required for the \NNNLOQCD corrections.
Electroweak corrections involving the trilinear Higgs self-coupling are also now available fully differentially, including the exact quark mass dependence~\cite{Haisch:2024nzv}.

In Ref.~\cite{Liu:2024tkc}, Higgs plus jet production was studied in the small quark mass limit, namely $m_q^2 \ll p_T^2 \ll s, m_H^2$.
Using the leading logarithmic approximation, a \NNLOQCD prediction for the bottom-quark correction was presented.
This work provides an ingredient for understanding the quark mass renormalisation uncertainty at the LHC.
The $\mathcal{O}(y_b^0)$ bottom mass corrections in the kinematic region $q_T \sim m_b \ll m_H$ were computed in Ref.~\cite{Pietrulewicz:2023dxt}.
In the high-energy regime, \LL-accurate results matched to \NLOQCD fixed-order results were presented in Refs~\cite{Andersen:2022zte}, with the resummed results producing a harder transverse momentum spectrum compared to the fixed-order ones. These calculations were made public through the {\tt HEJ-2.2} code~\cite{Andersen:2023kuj}.
In Ref.~\cite{Cal:2024yjz}, a jet-veto resummation was performed at \NNLOgen + a\NNLL$^\prime$ for exclusive $H+1\ \mathrm{jet}$ production.
This provides an important input for the STXS in the regime $p_T^\mathrm{cut} \ll p_T^H \sim m_H^2$.

The impact of anomalous Higgs couplings is now known at \NLOQCD retaining the quark mass effects~\cite{Aveleira:2024byi}, which are relevant at high Higgs $p_T$.

\medskip \noindent \textit{Experimental status}
The current experimental uncertainty on the Higgs + $\ge$ 1 jet differential cross section is of the order of $10-15\%$, dominated by the statistical error, for example, the fit statistical errors for the case of the combined $H\rightarrow \gamma \gamma$ and $H\rightarrow 4\ell$ analyses~\cite{ATLAS:2022fnp,ATLAS:2020wny,CMS:2022wpo,CMS:2023gjz}.
With a sample of
3000 fb$^{-1}$ of data, the statistical error will nominally decrease by about a factor of 5, resulting in a statistical error of
the order of $2.5\%$.  If the remaining systematic errors
(dominated for the diphoton analysis by the spurious signal systematic error) remain the same,
the resultant systematic error would be of the order of 9\%, leading to a
total error of approximately $9.5\%$.  This is similar enough to the current theoretical uncertainty that it may motivate
improvements on the $H+j$ cross section calculation. Of course, any improvements in the systematic errors would reduce the experimental uncertainty further.
Improvements in the theory could entail a combination of the \NNLOHTL results with the
full \NLOQCD results,
similar to the reweighting procedure that has been done one perturbative order lower.

Theoretical precision is not only needed for the full Higgs boson cross section, but specifically for production at high transverse momentum.
High $p_T$ Higgs boson production is of great interest, as it allows the probe of any new particles that may enter into the top quark loop, or indeed of any other new physics that might become evident at high $p_T$. The best foreseeable precision requires the $H+j$ calculation at N3LO.
Traditionally, the boosted $H\rightarrow b\bar{b}$ channel has been viewed as the most efficient way to examine high transverse momentum Higgs boson production, given the large branching ratio  into that final state. For ATLAS, for example, with the full Run 2 data sample at 13 TeV for $H\rightarrow b\bar{b}$, the 95\% confidence-level upper limit on the cross section for Higgs boson production with transverse momentum above 450 GeV is 115 fb with an uncertainty of 128 fb~\cite{ATLAS:2021tbi}.
Above 1 TeV it is 9.6 fb with an uncertainty of 17 fb.
The Standard Model cross section predictions for a Higgs boson with a mass of 125 GeV in the same kinematic regions are 18.4 fb and 0.13 fb, respectively.
Both results are consistent with the Standard Model, but also allow for a possible interesting excess at high $p_T$.
The Run 2 CMS extraction of the gluon-fusion contribution to $H\rightarrow b\overline{b}$ with Higgs transverse momentum between 450--650 GeV is 26 fb with an uncertainty of 27 fb. For transverse momentum $>650$ GeV, CMS extract 4 fb with an uncertainty of 6 fb~\cite{CMS:2024ddc}. Both uncertainties are completely dominated by statistics.

Fiducial measurements of the Higgs diphoton decay channel at CMS are currently available with a binning up to 450-550 GeV, yielding a cross section of $14 \pm 21$\,fb, in agreement with the SM prediction~\cite{CMS:2022wpo}.
The ATLAS Higgs to diphoton channel has allowed for measurements of the Higgs boson cross section of $38\pm19$\,fb from 450-650 GeV and $5.4\pm7.6$\,fb for 650-1300 GeV, both in agreement with the SM prediction~\cite{ATLAS:2022fnp}. The result is dominated by statistical errors for both bins. The 95\% CL upper limits on the ratio of the observed cross section to Standard Model prediction are 3.1 and 5.8 for the 450-650 GeV and 650-1300 GeV bins, respectively, a substantial improvement on the limits provided by the $b\bar{b}$ channel. The cross section times branching ratio is more limited than for $b\bar{b}$ decays, but the channel benefits from the rising signal-to-background ratio (due primarily to the $2\rightarrow3$ nature of the diphoton background process compared to $2\rightarrow2$  for Higgs boson production) for high Higgs boson transverse momenta. As with Sherlock Holmes' dog that didn't bark (The Memoirs of Sherlock Holmes, Arthur Conan Doyle, 1892), the presence of no Higgs diphoton events at very high $p_T$ can serve as a useful limit on the possible Higgs boson cross section in that kinematic region.

\subsubsection{$H+\geq 2j$}

\textit{LH21 status}
VBF production known at \NNNLOQCD accuracy for the total cross section~\cite{Dreyer:2016oyx} and at \NNLOQCD accuracy differentially~\cite{Cacciari:2015jma,Cruz-Martinez:2018rod} in the ``DIS'' approximation~\cite{Han:1992hr}. LO Higgs decays $H\to WW^*$ and $H \to b \bar b$ were included to the \NNLOQCD description of the VBF production process in Ref.~\cite{Asteriadis:2021gpd}. The double-virtual contributions to non-factorizable corrections are known in the eikonal approximation~\cite{Liu:2019tuy, Dreyer:2020urf}.
Full \NLOQCD corrections for $H + 3j$ in the VBF channel available~\cite{Campanario:2013fsa,Campanario:2018ppz}.
$H + \le 3j$ in the gluon fusion channel was studied in Ref.~\cite{Greiner:2016awe} and an assessment of the mass dependence of the various jet multiplicities was made in Ref.~\cite{Greiner:2015jha}; the impact of the top-quark mass in $H+1,2$ jets was studied in Ref.~\cite{Chen:2021azt};
\NLOEW corrections to stable Higgs boson production in VBF calculated~\cite{Ciccolini:2007jr} and available in {\sc Hawk}~\cite{Denner:2014cla}.
Mass effects in $H+2j$ at large energy are known within the ``High Energy Jets'' framework\cite{Andersen:2009nu,Andersen:2009he,Andersen:2011hs,Andersen:2017kfc,Andersen:2018tnm,Andersen:2018kjg}. Parton shower and matching uncertainties for VBF Higgs productions have been studied in detail using \Pythia and \Herwig matched to \MadgraphaMCatNLO and \Powheg in Ref.~\cite{Jager:2020hkz}; the \Pythia and \Vincia parton showers were compared in Ref.~\cite{Hoche:2021mkv}.
A comparative study of VBF Higgs production at fixed order and with parton shower Monte Carlos has been carried out in Ref.~\cite{Buckley:2021gfw}, as an outgrowth of Les Houches 2019.
\medskip

\noindent \textit{Progress}
VBF production at \NNLOQCD with the inclusion
of both the \NLOQCD and  the \NNLOQCD corrections to the Higgs decay $H \to b\bar b$ was presented in Ref.~\cite{Asteriadis:2024nbg}. These effects are substantial, amounting to a decrease of $7\%$ at both \NLOQCD and \NNLOQCD, largely due to the interplay between the radiation off the $b$-quarks and the kinematic cut placed on the $b$-tagged jets.

It is known~\cite{Liu:2019tuy,Dreyer:2020urf} that double-virtual contributions to non-factorizable QCD corrections beyond the DIS approximation are an order of magnitude smaller than the \NNLOQCD corrections in the factorized approximation, although with a large scale uncertainty of around $20\%-30\%$ (due to the fact that they appear at \NNLOQCD for this first time). In Ref.~\cite{Asteriadis:2023nyl}, the double-virtual corrections were combined with real-real and real-virtual corrections for the non-factorizable contributions, and it was observed that the double-virtual contributions are completely dominant.
In Ref.~\cite{Long:2023mvc}, the next-to-leading eikonal contributions were shown to modify the non-factorizable corrections by approximately $30\%$.
In Ref.~\cite{Bronnum-Hansen:2023vzh} the $\mathcal{O}(\beta_0 \alphas^3 )$ corrections were computed, and were shown to reduce the scale uncertainty associated with the non-factorizable corrections to around $5\%$.
Fully analytic expressions for the two-loop amplitude in the eikonal approximation were presented in Ref.~\cite{Gates:2023iiv}.

 Recently, \NLL{} accurate parton showers for VBF production have become available with the PanScales method, and these were matched to LO calculations in Ref.~\cite{vanBeekveld:2023chs}. For exclusive observables, such as those related to the third jet, the impact of the \NLL{} corrections can be as large as 15\%, and generally results in a softer spectrum of the third jet. \NLOQCD and \NLOEW corrections for electroweak Higgs production (i.e. including both the $VH$ and VBF processes) have been matched to parton showers in Ref.~\cite{Jager:2022acp}.

\medskip \noindent \textit{Experimental status}
The current experimental error on the $H+\geq 2j$ cross section is on the order of
25\%~\cite{ATLAS:2022fnp,CMS:2022wpo}, again dominated by statistical errors,
and again for the diphoton final state, by the fit statistical error. With the same assumptions
as above, for 3000 fb$^{-1}$, the statistical error will reduce to the order of 3.5\%.
If the systematic errors remain the same, at approximately 12\% (in this case the largest
systematic error is from the jet energy scale uncertainty and the jet energy resolution uncertainty),
a total uncertainty of approximately 12.5\% would result, less than the
current theoretical uncertainty.
To achieve a theoretical uncertainty less than this value would require the calculation
of $H+\geq 2j$ to \NNLOHTL$\!\otimes\,$\NLOQCD in the gluon fusion production mode.

\subsubsection{$VH$}
\textit{LH21 status}
The total cross section is known at \NNNLOQCD~\cite{Baglio:2022wzu}. Inclusive \NNLOQCD corrections are available in {\sc VH@NNLO}~\cite{Brein:2003wg,Brein:2011vx,Brein:2012ne}, and  soft-gluon resummation effects are known~\cite{Dawson:2012gs}.
\NNLOQCD differential results are known for $WH$~\cite{Ferrera:2011bk} and $ZH$~\cite{Ferrera:2014lca}; matched to parton shower using the MiNLO procedure in Ref.~\cite{Astill:2016hpa,Astill:2018ivh}; supplemented with \NNLLp resummation in the 0-jettiness variable and matched to a parton shower within the \Geneva Monte Carlo framework in Ref.~\cite{Alioli:2019qzz}. In this last reference, $H \to b\bb$ decays were included at LO through the parton shower.
\NLOQCD corrections to the $H\to b\bar{b}$ decay are available in MCFM~\cite{Campbell:2016jau}, using massive $b$-quarks. The consistent combination of \NNLOQCD corrections to $VH$ production and $H \to b\bar b$ decay were presented in Refs.~\cite{Ferrera:2017zex,Caola:2017xuq,Gauld:2019yng}, where the first of these studies considered $ZH$ and $W^+H$ production, the second considered $W^-H$ production, and the third considered all three processes $W^{\pm}H$ and $ZH$. All of these calculations employed massless $b$-quarks in the decay. Bottom quark mass effects in \NNLOQCD corrections to $pp \to W^+H(\to b\bar b)$ production were presented in Ref.~\cite{Behring:2020uzq}. \NNLOQCD predictions for $pp \to ZH(\to b\bar b)$ and $pp \to W^\pm H(\to b\bar b)$ were matched to a parton shower using the MiNNLO method in Ref.~\cite{Zanoli:2021iyp}, for massive $b$-quarks. \NLOEW corrections calculated~\cite{Ciccolini:2003jy,Denner:2011id,Obul:2018psx,Granata:2017iod} also including parton shower effects~\cite{Granata:2017iod}.
The process $b\bb\to ZH$ in the 5FS, but with a non-vanishing bottom-quark Yukawa coupling, was investigated in the soft-virtual approximation at \NNLOQCD~\cite{Ahmed:2019udm}.
The polarised $q\bar{q} \rightarrow ZH$ amplitudes were studied at \NNLOQCD in Ref.~\cite{Ahmed:2020kme}.
The loop-induced $gg \rightarrow ZH$ channel accounts for $\sim 10\%$ of the total cross section and contributes significantly to the $pp \rightarrow ZH$ theoretical uncertainty.
The \NLOHTL
results reweighted by the full LO cross section were presented in Ref.~\cite{Altenkamp:2012sx}; finite $m_t$ effects at \NLOQCD known in a $1/m_t$ expansion~\cite{Hasselhuhn:2016rqt}; threshold resummation calculated in Ref.~\cite{Harlander:2014wda}.
The NLO virtual amplitudes were computed in a small-$p_T$ expansion~\cite{Alasfar:2021ppe}, high-energy expansion~\cite{Davies:2020drs}, and numerically~\cite{Chen:2020gae}.
The complete NLO corrections were recently presented in Ref.~\cite{Wang:2021rxu} (based on a small-$m_H,m_Z$ expansion), in Ref.~\cite{Chen:2022rua} (based on a combination of the numerical results and high-energy expansion), and in Ref.~\cite{Degrassi:2022mro} (based on a combination of the small-$p_T$ and high-energy expansion~\cite{Bellafronte:2022jmo}).
\NLOQCD with dimension-six Standard Model Effective Field Theory (SMEFT) operators investigated~\cite{Degrande:2016dqg}, matched to a parton shower in the \MadgraphaMCatNLO framework.
Higgs pseudo-observables investigated at \NLOQCD~\cite{Greljo:2017spw}.
Anomalous $HVV$ couplings were studied at \NNLOQCD for $W^\pm H$ and $ZH$ in Ref.~\cite{Bizon:2021rww}.
In the SMEFT, a \NNLOQCD event generator for $pp \rightarrow Z(\rightarrow l\bar{l}) H(\rightarrow b\bar{b})$ was presented in Ref.~\cite{Haisch:2022nwz}.
\medskip

\noindent \textit{Progress}
The process $pp \to W^+H(\to W^+W^-)$ with a subsequent leptonic decay of the two $W^+$ bosons and the hadronic decay of the $W^-$ was considered in Ref.~\cite{Denner:2024ufg} as a contribution to the $\mu^+ \nu_{\mu} e^+ \nu_e j j$ final state. Full \NLOQCD+\NLOEW corrections have been calculated. In addition, the \NLOQCD corrections, matched to the \Sherpa parton shower, as well as virtual \NLOEW corrections, are presented in this reference.
The \NNLOQCD corrections to $VH$ production, matched to parton showers and including the complete set of SMEFT operators, were implemented in the \Powheg in Ref.~\cite{Gauld:2023gtb}. The inclusive cross section to \NNNLOQCD accuracy are now publicly available through the program {\sc n3loxs}~\cite{Baglio:2022wzu}.

\medskip \noindent \textit{Experimental status}
Published results for the $VH$ cross section are available for data samples up to 138--139\,fb$^{-1}$,
with uncertainties on the order of 20\%, equally divided between statistical and
systematic errors~\cite{ATLAS:2020fcp,CMS:2023vzh}. For 3000\,fb$^{-1}$, the statistical error will reduce
to 4--5\%, resulting in a measurement that is systematically limited, unless there are significant improvements to the systematic errors.
The general $VH$ process has been calculated to \NNLOQCD,
leading to a small scale uncertainty.
However, for the best description of the $ZH$ process, the exact NLO corrections to the $gg\rightarrow ZH$ sub-process, described above, should be included.

\medskip

\subsubsection{$VH+j$}
\textit{LH21 status} Known at \NLOQCD + PS~\cite{Luisoni:2013cuh} and \NLOSM + PS~\cite{Granata:2017iod}. Fully differential \NNLOQCD corrections known~\cite{Gauld:2020ced,Gauld:2021ule}.

\subsubsection{$HH$}

\textit{LH21 status}
\NNNLOHTL corrections are known in the infinite top mass limit~\cite{Chen:2019lzz,Banerjee:2018lfq} and have been reweighted by the \NLOQCD result~\cite{Chen:2019fhs}.
Finite $m_t$ effects are incorporated in \NNLOHTL calculation by reweighting and combined with full-$m_t$ double-real corrections in Ref.~\cite{Grazzini:2018bsd}.
\NLOQCD results including the full top-quark mass dependence are known numerically~\cite{Borowka:2016ehy,Borowka:2016ypz,Baglio:2018lrj,Baglio:2020ini} and matched to parton showers~\cite{Heinrich:2017kxx,Jones:2017giv};
exact numerical results have also been supplemented by results obtained in a small-$m_t$ expansion~\cite{Davies:2019dfy,Davies:2018qvx}; a Pad{\'e} approximated result based on the large-$m_t$ expansion and analytic results near the top threshold was presented in Ref.~\cite{Grober:2017uho}.
The top quark mass renormalisation scheme uncertainty is known at \NLOQCD~\cite{Baglio:2018lrj,Baglio:2020ini,Baglio:2020wgt}.
Threshold resummation was performed at \NLOHTL\!+\,\NNLL~\cite{Shao:2013bz} and \NNLOHTL\!+\,\NNLL~\cite{deFlorian:2015moa}.
\NLOHTL\!+\,\NLL{} resummation for the $p_T$ of the Higgs boson pair was presented in~\cite{Ferrera:2016prr}.
\NNLOQCD virtual and real-virtual corrections (involving three closed top-quark loops) known in a large-$m_t$ expansion~\cite{Grigo:2015dia,Davies:2019djw}.
Sensitivity of $HH$ production to the quartic self-coupling (which enters via EW corrections) was studied in Refs.~\cite{Liu:2018peg,Bizon:2018syu,Borowka:2018pxx}.
The $b\bb\to HH$ process is known at \NNLOQCD~\cite{Ajjath:2018ifl}, two-loop amplitudes for the quark annihilation contributions are known at \NNLOHTL~\cite{Ahmed:2021hrf}.
Results in the HEFT and SMEFT are known at \NLOQCD~\cite{Heinrich:2020ckp,Heinrich:2022idm} and \NNLOHTL{}~\cite{deFlorian:2021azd}.
\medskip

\noindent \textit{Progress}
Results at \NNNLOHTL matched to soft-gluon threshold resummation at \NNNLL and reweighted by the \NLOQCD result have been presented in Ref.~\cite{AH:2022elh}.
The central prediction is found to be compatible with the known \NNNLOHTL results within the percent-level remaining scale uncertainties.
By combining the small-$p_T$~\cite{Bonciani:2018omm} and high-energy~\cite{Davies:2018qvx} expansions, the \NLOQCD result has been matched to parton shower in the \Powheg-box framework including the dependence on the top quark mass, allowing the renormalisation scheme to be varied~\cite{Bagnaschi:2023rbx}.
In Ref.~\cite{Alioli:2022dkj} the \NNLOHTL result is matched to parton shower using the \Geneva framework, and zero-jettiness logarithms are resummed to \NNNLL.
Ref.~\cite{Campbell:2024tqg} presents compact analytic results for $pp \rightarrow HHj$ at \LOQCD (1-loop).
These results have been used to produce an improved MCFM implementation of $HH$ production at \NLOQCD.

In Ref.~\cite{Manzoni:2023qaf}, the $pp \rightarrow b\overline{b}H$ channel, a background to $pp \rightarrow H(\rightarrow b \overline{b})H$, has been studied at \NLOQCD matched to parton shower including the $y_b^2$ and $y_t^2$ (at \NLOHTL) contributions.
This work approximately halves the remaining theoretical scale/shower uncertainties originating from the $\mathcal{O}(y_t^2)$ contributions to this background process.
The 5FS \NNLOQCD+PS calculation is found to overestimate the background by a factor of 2.
Ref.~\cite{Li:2024ujf} presents $pp \rightarrow H(\rightarrow b \overline{b}) H(\rightarrow \gamma \gamma)$, treating both production and decay subprocesses at \NLOQCD in the narrow-width approximation.
The \NLOQCD corrections in the decay decrease the result by 19\% relative to \LOQCD.

The complete \NLOEW corrections to $HH$ production were presented in Ref.~\cite{Bi:2023bnq}.
The electroweak corrections are found to reduce the cross section by around $4\%$ and induce a large $15\%$ enhancement near the $HH$ production threshold.
Results for the \NLOEW corrections in the HTL were presented in Ref.~\cite{Davies:2023npk}.
In Ref.~\cite{Muhlleitner:2022ijf}, the $y_t$-enhanced piece of the EW corrections was studied in the HTL, and it is found that the corrections amount to around $0.2\%$, and are not well described by introducing an effective trilinear-Higgs coupling.
Ref.~\cite{Davies:2022ram} presents results for the $y_t^2$ corrections in a high-energy expansion.
Refs.~\cite{Li:2024iio} presents results for corrections proportional to the triple and quartic Higgs couplings and studies their impact in a modified-$\kappa$ framework.
Ref.~\cite{Heinrich:2024dnz} presents results for the $y_t$, triple and quartic Higgs corrections.
These corrections are found to enhance the cross section by $1\%$ with a large enhancement near to the $HH$ production threshold, similar to that present in the complete \NLOEW corrections of Ref.~\cite{Bi:2023bnq}.

Work towards the \NNLOQCD corrections has started, with results available in the large top quark mass expansion for the virtual~\cite{Davies:2019djw} and real corrections~\cite{Davies:2019xzc,Davies:2021kex}.
The light-fermion ($n_F$) piece is also known in a small-$t$ expansion~\cite{Davies:2023obx}, while the reducible contribution is known in an expansion about small gluon virtuality or small Higgs boson mass~\cite{Davies:2024znp}.

In Ref.~\cite{Jaskiewicz:2024xkd} the quark mass corrections to the $gg \rightarrow HH$ virtual amplitude were studied at high-energy, it was argued that the dominant part of the mass scheme uncertainty could be understood and removed.

Higher-order corrections have also been computed in both the HEFT and SMEFT frameworks, see Ref.~\cite{Alasfar:2023xpc} for a recent review.

\medskip \noindent \textit{Experimental status}
The experimental limits on $HH$ production are currently at the level of approximately three times the
SM cross section for the ATLAS combined analysis (with an expected limit of 2.4) based on a data sample of 120--140\,fb$^{-1}$~\cite{ATLAS:2024ish}. The observed (expected) constraints on the Higgs boson trilinear coupling modifier $\kappa_\lambda$ are determined to be $[−1.2,7.2]$ ($[−1.6,7.2]$) at 95\% confidence level, where the expected constraints on $\kappa_\lambda$ are obtained excluding $pp\to HH$ production from the background hypothesis. For CMS, a 95\% CL limit of 3.9 (expected 7.8) times the Standard Model has been obtained in the $b\overline{b}b\overline{b}$ channel~\cite{CMS:2022cpr} and 3.3 (expected 5.2) in the $b \overline{b} \tau \overline{\tau}$ channel~\cite{CMS:2022hgz}, for a data sample of 138\,fb$^{-1}$.
Constraints have also been set on the modifiers of the Higgs field self-coupling $\kappa_\lambda$ with this measurement in the range of $-2.3$ to 9.4,  with an expected range of $-5.0$ to 12.0.
With a data sample of 3000 fb$^{-1}$, it is projected that a limit of
$0.5 < \lambda_{hhh}/\lambda_{hhh,\mathrm{SM}} < 1.5$ can be achieved at the $68\%$ CL
for ATLAS and CMS combined~\cite{Cepeda:2019klc}.
The projected experimental precision at the HL-LHC, presented in Ref.~\cite{Cepeda:2019klc}, is comparable to the current theoretical uncertainty, motivating the desired improvement of the theoretical results. Any improvement in the experimental projection would very strongly motivate this and could require further theoretical precision. Given the sensitivity of this channel to the Higgs potential, it is also motivated to extend the desired theoretical precision to scenarios with modified Higgs boson couplings or general Effective Field Theory analyses.

\subsubsection{$HH + 2j$}

\textit{LH21 status}
\sloppy
Fully differential results for VBF $HH$ production are known at \NNLOQCDVBFstar~\cite{Dreyer:2018rfu} and at \NNNLOQCDVBFstar for the inclusive cross section~\cite{Dreyer:2018qbw}. The non-factorisable \NNLOQCD contributions~\cite{Dreyer:2020urf} and \NLOEW corrections are known~\cite{Dreyer:2020xaj} and have been combined.

\subsubsection{$HHH$}

\textit{LH21 status}
Known at \NNLOHTL~\cite{deFlorian:2016sit,deFlorian:2019app}, finite quark mass effects are included by reweighting with the full Born result.

\noindent \textit{Progress}
Triple production provides a direct handle on the quartic coupling of the Higgs boson, although very suppressed with respect to single and double Higgs production in the Standard Model, it can be enhanced in BSM scenarios.
It is therefore interesting to begin placing experimental constraints on this process.
For an overview of ongoing theoretical and experimental efforts we refer the reader to the HHH whitepaper~\cite{Abouabid:2024gms}.

\subsubsection{$t\bar{t}H$}

\textit{LH21 status}
\NLOQCD corrections for on-shell $t\tb H$ production known~\cite{Beenakker:2001rj,Reina:2001sf,Beenakker:2002nc,Dawson:2003zu}.
\NLOEW~corrections studied within the \MadgraphaMCatNLO
framework~\cite{Frixione:2014qaa,Frixione:2015zaa}.
Combined \NLOQCD and \NLOEW corrections with NWA top-quark decays computed in Ref.~\cite{Zhang:2014gcy}.
Corrections to $t\bar{t}H$ including top quark decays and full off-shell effects
computed at \NLOQCD~\cite{Denner:2015yca},
and combined with \NLOEW~\cite{Denner:2016wet}.
\NLOQCD results merged to parton showers~\cite{Garzelli:2011vp,Hartanto:2015uka} and \NLOgen\!+\,\NNLL resummation performed in
Refs.~\cite{Kulesza:2015vda,Broggio:2015lya,Broggio:2016lfj,Kulesza:2017ukk}.
The \NLOQCD corrections including off-shell effects were also presented in Ref.~\cite{Stremmer:2021bnk}, further considering the LO decays of the Higgs boson in the NWA.
\NLOQCD results in the SMEFT calculated~\cite{Maltoni:2016yxb}. The flavor off-diagonal channels were computed at \NNLOQCD in Ref.~\cite{Catani:2021cbl}. Fragmentation and splitting functions for the final-state transitions $t \rightarrow H$ and $g \rightarrow H$, are known at $\mathcal{O}(y_t^2 \alphas)$~\cite{Brancaccio:2021gcz}.
\medskip

\noindent \textit{Progress}
The \NNLOQCD calculation of Ref.~\cite{Catani:2021cbl} has been extended to encompass all partonic channels for the inclusive cross section~\cite{Catani:2022mfv} and differential distributions~\cite{Devoto:2024nhl}, using a soft-Higgs approximation for the as-yet unknown two-loop amplitudes. The \NNLOQCD effects increase the inclusive cross section by around $4\%$, and leave a residual perturbative uncertainty of $4\%$ for the cross section and around  $4\%-6\%$  for the Higgs transverse momentum spectrum.

There have been initial steps towards the evaluation of the two-loop amplitudes. These were computed in Ref.~\cite{Wang:2024pmv} in the boosted limit in which all quarks are massless. The master integrals for the leading-colour two loop amplitudes have been computed in Ref.~\cite{FebresCordero:2023pww},  the one-loop $gg \to t\bar t H$ amplitude was computed to $\mathcal{O}(\eps^2)$ in Ref.~\cite{Buccioni:2023okz}, and the $n_f$-dependent part of the two-loop amplitude was computed in Ref.~\cite{Agarwal:2024jyq}.

Results for an admixture of CP-even and CP-odd Higgs states have been computed to \NLOQCD, including offshell effects of the top decay~\cite{Hermann:2022vit}.

\medskip \noindent \textit{Experimental status}
The cross section for $t\bar{t}H$ has been measured with a data sample of 139\,fb$^{-1}$,
with a total uncertainty on the order of 20\%, dominated by the  statistical  error~\cite{ATLAS:2020ior,CMS:2020mpn}.
The statistical error will shrink to the order of 4--5\% for 3000\,fb$^{-1}$,
leaving a systematics-dominated measurement. Given that this calculation is currently known
only at \NLOQCD, with a corresponding scale uncertainty of the order of 10--15\%,
this warrants a calculation of the process to \NNLOQCD.

\subsubsection{$tH$}

\textit{LH19 status:}
\NLOQCD corrections known~\cite{Campbell:2013yla,Demartin:2015uha}. \NLOQCD and \NLOEW corrections known for on-shell top quarks, computed in both the four- and five-flavour schemes~\cite{Pagani:2020mov}.
\medskip

\subsubsection{$b\bar{b}H$ (including $H$ production in bottom quark fusion treated in 5FS)}

\textit{LH21 status}
\NNLOQCD predictions for $b\bar  b
\to H$ in the 5FS known for a long time, inclusively~\cite{Harlander:2003ai} and later differentially~\cite{Harlander:2011fx,Buhler:2012ytl}; resummed calculation at \NNLOgen\!+\,\NNLL available~\cite{Harlander:2014hya}. \NNNLOQCD in threshold approximation~\cite{Ahmed:2014cha,Ahmed:2014era} calculated; complete inclusive \NNNLOQCD corrections to $b\bar b
\to H$  in the 5FS presented in Ref.~\cite{Duhr:2019kwi} and matched to the 4FS in Ref.~\cite{Duhr:2020kzd}.
Threshold resummations up \NNNLL were combined with the \NNNLOgen results for the inclusive cross section in Ref.~\cite{Ajjath:2019neu}.
Massless 4-loop QCD corrections presented in Ref.~\cite{Chakraborty:2022yan}.
\NLOmixQED{1}{1} as well as \NNLOQED predictions were derived in Ref.~\cite{Ajjath:2019ixh}.
\NLOQCD corrections to $b\bar b H$ production in the 4FS known since long ago~\cite{Dittmaier:2003ej,Dawson:2003kb};
\NLOQCD (including the formally \NNLOHTL $y_t^2$ contributions) using the 4FS presented in Ref.~\cite{Deutschmann:2018avk}.
\NLOQCD matched to parton shower and compared to 5FS in Ref.~\cite{Wiesemann:2014ioa}; various methods proposed to combine 4FS and 5FS predictions~\cite{Harlander:2011aa,Bonvini:2015pxa,Forte:2015hba,Bonvini:2016fgf,Forte:2016sja}; \NLOEW corrections calculated~\cite{Zhang:2017mdz}. Complete predictions at $\mathcal{O}(\alphas^m \alpha^{n+1})$ with $m+n=2,3$ (i.e. including both QCD and EW corrections) for $b\bar{b}H$ production presented in Ref.~\cite{Pagani:2020rsg} in the 4FS. Two-loop leading colour planar helicity amplitudes in the 5FS computed in Ref.~\cite{Badger:2021ega}.
\medskip

\noindent \textit{Progress}
The \NNNLOQCD corrections to $b \bar b \to H$ are available in the public code {\sc n3loxs}~\cite{Baglio:2022wzu}. The \NNLOQCD corrections have been matched to parton shower using the \MiNNLOPS method in Ref.~\cite{Biello:2024vdh}, and the resulting transverse momentum distribution of the Higgs was compared to \NNLOQCD\!+ \NNLL results of Ref.~\cite{Harlander:2014hya}. Results for the Higgs transverse momentum spectrum resummed to \NNNLLp and matched to \NNLOQCD and approximate \NNNLOQCD were presented in Ref.~\cite{Cal:2023mib}. The perturbative uncertainties are small, opening the possibility of distinguishing different flavour production modes. Soft-virtual (SV) and next-to-soft-virtual (NSV) terms, resummed to \NNNLL accuracy, were presented in Ref.~\cite{Das:2024pac} for the inclusive cross section and Higgs rapidity distribution. The authors observe that the improvement in the perturbative precision when including the SV and NSV terms on top of the \NNNLOQCD results is quite small, indicating good convergence of the series.

In $b\bar b H$ production, the calculation of Ref.~\cite{Deutschmann:2018avk}, which included both $\mathcal{O}(y_b^2)$ and $\mathcal{O}(y_t^2)$ contributions to \NLOQCD in the 4FS, has been matched to PS in Ref.~\cite{Manzoni:2023qaf}.
\NNLOQCD results matched to parton shower were presented in Ref.~\cite{Biello:2024pgo}, using the \MiNNLOPS framework. The bottom quarks are treated as massive, with the two-loop amplitudes being evaluated using a small-mass expansion. The authors find that the \NNLOQCD corrections in the 4FS resolve tensions between this scheme and the 5FS.
Analytic expressions for the two-loop amplitudes for $b\bar{b}H$ production with massless $b$ quarks are now available for all color structures~\cite{Badger:2024awe}.

\subsection{Jet final states}
An overview of the status of jet final states is given in Table~\ref{tab:SM_wishlist:wljets}.

\begin{table}
  \renewcommand{\arraystretch}{1.5}
\setlength{\tabcolsep}{5pt}
  \centering
  \begin{tabular}{lll}
    \hline
    \multicolumn{1}{c}{process} & \multicolumn{1}{c}{known} & \multicolumn{1}{c}{desired} \\
    \hline
    $pp\to 2$\,jets &
    \begin{tabular}{l}
      \NNLOQCD \\
      \NLOQCD\!+\,\NLOEW
    \end{tabular} &
    \begin{tabular}{cl}
      \NNNLOQCD\!+\,\NLOEW \\
    \end{tabular} \\
    \hline
    $pp\to 3$\,jets &
    \begin{tabular}{l}
      \NNLOQCD\!+\,\NLOEW
    \end{tabular} &
    \begin{tabular}{l}
      \\
    \end{tabular} \\
    \hline
  \end{tabular}
  \caption{Precision wish list: jet final states.}
  \label{tab:SM_wishlist:wljets}
  \renewcommand{\arraystretch}{1.0}
\end{table}

\subsubsection{2j}

\textit{LH21 status} Differential \NNLOQCD corrections available from two independent groups using the antenna~\cite{Currie:2016bfm,Chen:2022tpk} and the sector-improved residue subtraction~\cite{Czakon:2019tmo} formalisms.
Complete NLO QCD+EW corrections available~\cite{Frederix:2016ost}.

\medskip

\noindent \textit{Progress}
\NNLOQCD interpolation grids were made available for this process in~\cite{Britzger:2022lbf}, which facilitate the inclusion of this process into PDF fits without $K$-factor approximations.
Such grids were used to perform an \alphas extraction based on LHC di-jet data in~\cite{Ahmadova:2024emn}.
Based on the \NNLOQCD corrections to massive bottom quark production~\cite{Catani:2020kkl},
NNLO+PS predictions were obtained for B-hadron production at the LHC in~\cite{Mazzitelli:2023znt}.
Ref.~\cite{Buonocore:2023rdw} explored new slicing variables at \NLOQCD for jet processes.
Jet angularities in a resummed and matched calculation were studied in Ref.~\cite{Reichelt:2021svh} that also took into account non-perturbative corrections from the underlying event and hadronisation.

The full set of three-loop massless $2\to2$ amplitudes is now complete~\cite{Caola:2021rqz,Caola:2021izf,Caola:2022dfa}.
Together with the \NNLOQCD $3j$ calculation, all amplitude building blocks are available to tackle jet production at \NNNLOQCD; a major obstacle here is to devise a formalism that can deal with the complex IR subtraction for this process at this order.

\medskip \noindent \textit{Experimental status}
Inclusive jets can be measured in both ATLAS and CMS with 5\% uncertainty in the cross sections (in the
precision range), a precision that requires \NNLOQCD cross sections.  Global PDF fits require \NNLOQCD calculations of double and even triple differential observables,  requiring the use of full colour predictions. The measurements extend to jet transverse momenta of the order of 3--5\,TeV,  necessitating the precise calculation of EW corrections as well. Eventually, PDFs will be determined at the \NNNLOQCD level, requiring the use of \NNNLOQCD predictions for the input processes, including inclusive jet production, necessitating the calculation of di-jet production to this order.

\subsubsection{$\geq$3j}

\textit{LH21 status} \NNLOQCD corrections for 3-jet with the double-virtual corrections treated in the leading-colour approximation~\cite{Czakon:2021mjy}.
\NLOQCD corrections for 4-jet~\cite{Bern:2011ep,Badger:2012pf} and 5-jet~\cite{Badger:2013yda} known.
Full \NLOSM calculation for 3-jet production was performed using \Sherpa interfaced to \Recola in Ref.~\cite{Reyer:2019obz}.
\medskip

Three-jet observables provide a better description of jet shapes, and have the potential for the determination of the strong coupling constant over an extended dynamic range.

\subsection{Vector-boson associated processes}

An overview of the status of vector-boson associated processes is given in Table~\ref{tab:SM_wishlist:wlV}.
If not stated explicitly, the leptonic decays are assumed.
In the same way, the off-shell description is the default one.
Finally, in some cases for $VV+2j$, the full NLO corrections are not known, and in these cases we indicate to which underlying Born contribution the corrections refer.

\begin{table}
  \renewcommand{\arraystretch}{1.5}
\setlength{\tabcolsep}{5pt}
  \centering
  \begin{tabular}{lll}
    \hline
    \multicolumn{1}{c}{process} & \multicolumn{1}{c}{known} & \multicolumn{1}{c}{desired} \\
    \hline
    $pp\to V$ &
    \begin{tabular}{l}
      \NNNLOQCD\!+\,\NLOQE11 \\
      \NLOEW
    \end{tabular} &
    \begin{tabular}{l}
      \NLOE2
    \end{tabular} \\
    \hline
    $pp\to VV'$ &
    \begin{tabular}{l}
      \NNLOQCD\!+\,\NLOEW{ }\wleptdecays{} \\
      \!+\, Full \NLOQCD{ } ($gg \to ZZ$), \\
      approx.\ \NLOQCD{} ($gg \to WW$) \wleptdecays{} \\
    \end{tabular} &
    \begin{tabular}{l}
      Full \NLOQCD{ } \\($gg$ channel, w/ massive loops) \\
      \NLOQE11
    \end{tabular} \\
    \hline
    $pp\to V+j$ &
    \begin{tabular}{l}
      \NNLOQCD\!+\,\NLOEW{ }\wleptdecays{} \\
    \end{tabular} &
    \begin{tabular}{l}
      hadronic decays
    \end{tabular} \\
    \hline
    $pp\to V+2j$ &
    \begin{tabular}{l}
      \NLOQCD\!+\,\NLOEW (QCD component) \\
      \NLOQCD\!+\,\NLOEW (EW component)
    \end{tabular} &
    \begin{tabular}{l}
      \NNLOQCD \wdecays{} \\
    \end{tabular}\\
    \hline
    $pp\to V+b\bar{b}$ &
    \begin{tabular}{l}
      \NLOQCD{ }\wleptdecays{} \\
    \end{tabular} &
    \begin{tabular}{l}
      \NNLOQCD \!+\,\NLOEW{ }\wdecays{} \\
    \end{tabular} \\
    \cline{2-2}
    $pp\to W+b\bar{b}$ &
    \begin{tabular}{l}
      \NNLOQCD \\
    \end{tabular} &
    \begin{tabular}{l}
      \\
    \end{tabular} \\
    \hline
    $pp\to VV'+1j$ &
    \begin{tabular}{l}
      \NLOQCD\!+\,\NLOEW{ }\wdecays{}
    \end{tabular} &
    \begin{tabular}{l}
      \NNLOQCD \\
    \end{tabular} \\
    \hline
    $pp\to VV'+2j$ &
    \begin{tabular}{l}
      \NLOQCD \wleptdecays{} (QCD component) \\
      \NLOQCD\!+\,\NLOEW{ }\wleptdecays{} (EW component)
    \end{tabular} &
    \begin{tabular}{l}
      Full \NLOQCD\!+\,\NLOEW{ }\wdecays{} \\
    \end{tabular} \\
    \cline{2-2}
    $pp\to W^+W^++2j$ &
    \begin{tabular}{l}
      Full \NLOQCD\!+\,\NLOEW{ }\wleptdecays{} \\
    \end{tabular} &
    \begin{tabular}{l}
      \\
    \end{tabular} \\
    \cline{2-2}
    $pp\to W^+W^-+2j$ &
    \begin{tabular}{l}
      \NLOQCD\!+\,\NLOEW{ }\wleptdecays{} (EW component)\\
    \end{tabular} &
    \begin{tabular}{l}
      \\
    \end{tabular} \\
    \cline{2-2}
    $pp\to W^+Z+2j$ &
    \begin{tabular}{l}
      \NLOQCD\!+\,\NLOEW{ }\wleptdecays{} (EW component)\\
    \end{tabular} &
    \begin{tabular}{l}
      \\
    \end{tabular} \\
    \cline{2-2}
    $pp\to ZZ+2j$ &
    \begin{tabular}{l}
      Full \NLOQCD\!+\,\NLOEW{ }\wleptdecays{} \\
    \end{tabular} &
    \begin{tabular}{l}
      \\
    \end{tabular} \\
    \hline
   $pp\to VV'V''$ &
    \begin{tabular}{l}
      \NLOQCD+\NLOEW (w/ decays)
    \end{tabular} &
    \begin{tabular}{l}
      \NLOQCD\!+\,\NLOEW (off-shell) \\
    \end{tabular} \\
    \cline{2-2}
   $pp\to WWW$  &
    \begin{tabular}{l}
      \NLOQCD + \NLOEW{ } (off-shell)
    \end{tabular} &
    \begin{tabular}{l} \\
    \end{tabular} \\
    \cline{2-2}
   $pp\to W^+W^+(V\to jj)$  &
    \begin{tabular}{l}
      \NLOQCD + \NLOEW{ } (off-shell)
    \end{tabular} &
    \begin{tabular}{l}
      \\
    \end{tabular} \\
    \cline{2-2}
   $pp\to WZ(V\to jj)$  &
    \begin{tabular}{l}
      \NLOQCD + \NLOEW{ } (off-shell)
    \end{tabular} &
    \begin{tabular}{l}
      \\
    \end{tabular} \\
    \hline
    $pp\to \gamma\gamma$ &
    \begin{tabular}{l}
      \NNLOQCD\!+\,\NLOEW
    \end{tabular} &
    \begin{tabular}{l}
      \NNNLOQCD \\
    \end{tabular} \\
    \hline
    $pp\to \gamma+j$ &
    \begin{tabular}{l}
      \NNLOQCD\!+\,\NLOEW
    \end{tabular} &
    \begin{tabular}{l}
      \NNNLOQCD \\
    \end{tabular} \\
    \hline
    $pp\to \gamma\gamma+j$ &
    \begin{tabular}{l}
      \NNLOQCD\!+\,\NLOEW \\
      \!+\,\NLOQCD{ }($gg$ channel)
    \end{tabular} &
    \begin{tabular}{cl}

    \end{tabular} \\
    \hline
    $pp\to \gamma\gamma\gamma$ &
    \begin{tabular}{l}
      \NNLOQCD
    \end{tabular} &
    \begin{tabular}{cl}
      \NLOEW \\
    \end{tabular} \\
    \hline
  \end{tabular}
  \caption{Precision wish list: vector boson final
    states. $V=W,Z$ and $V',V''=W,Z,\gamma$.
    Full leptonic decays are understood if not stated otherwise.}
  \label{tab:SM_wishlist:wlV}
  \renewcommand{\arraystretch}{1.0}
\end{table}

\subsubsection{$V$}

\textit{LH21 status}
\NNNLOQCD to the inclusive neutral-current Drell-Yan process~\cite{Duhr:2021vwj} and to the lepton-pair rapidity distribution in the photon-mediated Drell-Yan~\cite{Chen:2021vtu};
\NNNLOQCD to the inclusive charged-current Drell-Yan process~\cite{Duhr:2020sdp} and to the rapidity, transverse mass, and the charge asymmetry \cite{Chen:2022lwc};
\NLOEW corrections known for many years see \eg~Ref.~\cite{Alioli:2016fum} and references therein;
corrections at $\mathcal{O}(\alphas \alpha)$ (\NLOQE11) known for the off-shell neutral process~\cite{Bonciani:2021zzf,Buccioni:2022kgy} and the charged process up to the finite two-loop remainder~\cite{Buonocore:2021rxx};
Several results for on-shell $W$ or parts of the off-shell calculation for the charged process~\cite{Dittmaier:2015rxo,Behring:2020cqi,Behring:2021adr};
\\
\NNLOQCD computations matched to parton shower available using the
MiNLO method~\cite{Karlberg:2014qua}, SCET resummation \cite{Alioli:2015toa}, the UN${}^2$LOPS technique \cite{Hoche:2014uhw}, and the MINNLO${}_\text{PS}$ method~\cite{Monni:2019whf};
\NNNLOQCD\!+\,\NNNLL accuracy~\cite{Camarda:2021ict,Chen:2022cgv}.
\medskip

\noindent \textit{Progress}
In Ref.~\cite{Alekhin:2024mrq}, a comparative study at \NNLOQCD accuracy between several codes has been conducted.
Agreement has been found provided linear power corrections induced by the fiducial cuts are included for programs relying on phase-space slicing subtraction schemes.
It is shown that symmetric experimental event selection render unstable the fixed-order predictions unless they are supplemented by resummation. Recommendations for future experimental measurements are made.

In Ref.~\cite{Campbell:2023lcy}, a new  calculation for $W$ production at \NNNLOQCD has been presented, supplemented with transverse momentum resummation.
The authors present results for the total cross section and differential distributions.
In Ref.~\cite{Gehrmann-DeRidder:2023urf}, a triple-differential analysis has been carried out at \NNLOQCD\!+\,\NLOEW.
In addition, partial \NNNLOQCD as well as higher-order EW corrections are supplemented where appropriate.

Beyond developments for QCD corrections, mixed QCD--EW corrections have been investigated further.
The two-loop mixed QCD--EW amplitude to the charged~current Drell--Yan has been computed~\cite{Armadillo:2024nwk}.
The computation is done with massive leptons, thereby regularizing the associated collinear singularities.
The results can be used in terms of numerical grids.
In Ref.~\cite{Dittmaier:2024row}, the corrections of initial--initial type were computed, thus completing the full mixed QCD--EW in the pole approximation.
Various differential distributions are discussed along the forward--backward asymmetry.
Following on this, the full calculation, valid over the full range for the neutral current has been studied again in Ref.~\cite{Armadillo:2024ncf}.
In particular, a study of bare muons is presented there.

In Ref.~\cite{Buonocore:2024xmy}, resummation at NLL accuracy for both EW and mixed QCD-EW corrections has been presented.
These corrections are then combined to N3LL corrections to provide state-of-the-art differential predictions for the neutral and charged process.
In the same spirit, Ref.~\cite{Autieri:2023xme} presented the $q_T$ resummation of \NLL{}accuracy in QED, LL accuracy for mixed QCD-EW effects, and \NNLL accuracy in QCD.
In Ref.~\cite{Isaacson:2023iui}, the {\sc{ResBos}} programm has been promoted to \NNLOQCD\!+\,\NNNLL accuracy.
Threshold logarithms for this process were considered at \NNNLL, matched to the \NNNLOQCD results of Ref.~\cite{Baglio:2022wzu}, and found to be phenomenologically negligible~\cite{Das:2022zie}.
Along the same line, a new independent calculation at \NNLOQCD\!+\,N4LL' accuracy has presented~\cite{Neumann:2022lft}.
Finally, Ref.~\cite{Billis:2024dqq} also presented results at N3LL' accuracy and approximate N4LL in resummed perturbation theory, matched to the available $\mathcal{O}(\alphas^3)$ fixed-order results.
In addition, parametric uncertainties associated to $\alphas$, the collinear parton distribution functions, and the non-perturbative transverse momentum-dependent (TMD) dynamics are discussed.

With the recent advance in theoretical predictions, several phenomenological studies have been carried out.
For example, in Ref.~\cite{Amoroso:2023uux} the idea of probing the running of the weak-mixing angle has been explored, with a focus on the High-Luminosity phase of the LHC.
To that end, a new version of the {\sc{Powheg}} implementation of EW corrections has been released where both on-shell and $\overline{\rm MS}$ renormalisation scheme can be used.
New ideas on how to extract the W-boson mass at hadron colliders have also been promoted~\cite{Rottoli:2023xdc}, where also higher-order corrections are discussed.
Finally, the sensitivity of theoretical predictions to PDFs has been discussed in Ref.~\cite{Ball:2022qtp} by focusing on the forward-backward asymmetry.

\medskip \noindent \textit{Experimental status}
The Drell-Yan process (W and Z production) is arguably one of the best measured processes at the LHC, given its large cross section and simple final state. 
As a result, it is crucial for the determination of parton distribution functions.
The systematic uncertainties are dominated by that of the luminosity uncertainty, with other systematic uncertainties at the percent level or smaller. 
It is worth mentioning that luminosity uncertainty can be eliminated by considering normalised distributions, in which case an experimental uncertainty well below the percent level can be achieved in the $p_T$ distribution of the Z boson.
The relative precision between the measured $W$ and $Z$ boson cross sections achieved at 7\,TeV by ATLAS~\cite{ATLAS:2016nqi} resulted in an increase of the strange quark distribution in PDF fits using that data set.
Measurements at $\sqrt{s}=5.02$ and 13 TeV are available from CMS~\cite{CMS:2024myi}, and a measurement based on partial Run 3 data at $\sqrt{s}=13.6$ TeV has been performed by ATLAS~\cite{ATLAS:2024irg}.
It is worth pointing out that pure electroweak and mixed electroweak-QCD corrections are critical to match the experimental precision.

\subsubsection{$V/\gamma+j$}

\textit{LH21 status}
$Z+j$~\cite{Gehrmann-DeRidder:2015wbt,Boughezal:2015ded,Boughezal:2016isb,Boughezal:2016yfp,Gehrmann-DeRidder:2017mvr},
$W+j$~\cite{Boughezal:2015dva,Boughezal:2016dtm,Boughezal:2016yfp,Gehrmann-DeRidder:2017mvr}, and $\gamma+j$~\cite{Campbell:2016lzl,Chen:2019zmr}
completed through \NNLOQCD including leptonic decays;
all processes of this class, and in particular their ratios, investigated in great
detail in Ref.~\cite{Lindert:2017olm}, combining \NNLOQCD predictions with full NLO EW and
leading \NNLOEW effects in the Sudakov approximation, including also approximations for leading
\NLOQE11 effects, devoting particular attention to error estimates and
correlations between the processes.
Subleading EW corrections known for $Z+j$~\cite{Denner:2019zfp};
\NNLOQCD known for polarised $W+j$~\cite{Pellen:2021vpi};
\NNLOQCD known for $Z+b$ \cite{Gauld:2020deh} and $W+c$ \cite{Czakon:2020coa};\\
\NLOQCD with parton-shower corrections for mass charm for $W+c$ known~\cite{Bevilacqua:2021ovq}.
\medskip

\noindent \textit{Progress}
Processes featuring a vector boson produced in association with a flavoured jet have become of increasing interest in the last few years.
One of these is $W+c$ production, which provides sensitivity to the strange-quark content of the proton.
To that end, \NNLOQCD corrections retaining full CKM-matrix dependence have been computed, along with \NLOEW corrections~\cite{Czakon:2022khx}.
In addition, the influence of flavored jet algorithms and the experimental definition of the process has been investigated.
This work has been followed by a comparison with CMS measurement~\cite{CMS:2023aim} which showed good agreement.
In Ref.~\cite{Gehrmann-DeRidder:2023gdl}, the same calculation was performed, providing in addition the breakdown of the partonic channels, which is particularly useful in order to get a handle on the strange-quark parton-distribution functions.
For the same process, Ref.~\cite{FerrarioRavasio:2023jck} provided theoretical predictions at \NLOQCD accuracy matched to parton shower with massive charm quarks.
Particle-level results were presented while comparing several parton showers.
In addition, hadronisation and underlying-event effects were investigated.

In Ref.~\cite{Gauld:2023zlv}, \NNLOQCD predictions have been presented for $Z+c$ production in the fiducial region of the LHCb measurement.
It is important to notice that the authors refrain from comparing their predictions to the LHCb measurement~\cite{LHCb:2021stx} due to the difficulty of comparing theoretical predictions and experimental measurements involving flavour jets on an equal footing, as well as the large effects due to multiple-particle interactions observed in this set-up.

In Ref.~\cite{Caletti:2024xaw}, \NLOQCD predictions have been provided for Z production in association with light charged hadrons inside a jet or the production of a W along with a charm hadron.
Results are shown for several fragmentation functions and are compared to LHCb and ATLAS measurements at $13$TeV.

In Ref.~\cite{Guzzi:2024can}, a mass variable-flavor number scheme has been designed for Z+heavy quark. The authors illustrated their work by looking at $Z+b$ production at the LHC.

Advances have also been seen for the calculation of mixed QCD-EW corrections for Z+jet production.
In particular, in Ref.~\cite{Bargiela:2023npj}, the bosonic contribution to the two-loop mixed QCD-electroweak scattering amplitudes has been obtained.
The amplitudes have been evaluated on a two-dimensional grid in the rapidity and transverse momentum of the Z boson.

Pushing theoretical accuracy even further, there has been impressive progress  towards \NNNLOQCD  accuracy for V+jet production.
For example, the planar three-loop QCD helicity amplitudes for V+jet production have been obtained~\cite{Gehrmann:2023jyv}.
Along the same line, two-loop helicity amplitudes including axial-vector couplings~\cite{Gehrmann:2022vuk} to higher orders in $\epsilon$~\cite{Gehrmann:2023zpz} have been obtained.

Finally, N3LL resummation of one-jettiness for $Z+j$ production has been presented in Ref.~\cite{Alioli:2023rxx}.
The calculation has been matched to the corresponding fixed-order predictions, hence making these predictions also applicable to phase-space regions with extra hard jets.

\medskip \noindent \textit{Experimental status}
There are a number of kinematic variables related to $V$+jet production that probe the QCD dynamics of the hard scatter, most simply the transverse momenta of the boson and of the lead jet. At 13\,TeV, the boson and jet transverse momenta have been measured up to the order of  2\,TeV~\cite{ATLAS:2022nrp,CMS:2022ilp}. Better agreement with the data is obtained at NNLO than at NLO. Electroweak corrections are especially important for the transverse momentum of the heavy gauge boson. The cross section with non-zero transverse momentum of the heavy gauge boson, in particular, can be measured very precisely, to the order of a few percent.
A review of past experimental and theoretical progress has been presented for vector-boson production in association with jet(s) in Ref.~\cite{Tricoli:2020uxr}.

\subsubsection{$V+\geq2j$}

\textit{LH21 status}
\NLOQCD computations known for $V+2j$ final states in QCD~\cite{Campbell:2002tg,Campbell:2003hd} and EW~\cite{Oleari:2003tc} production modes, for $V+3j$~\cite{Ellis:2009zw,Berger:2009zg,Ellis:2009zyy,Berger:2009ep,Melnikov:2009wh,Berger:2010vm}, for $V+4j$~\cite{Berger:2010zx,Ita:2011wn} and for $W+5j$~\cite{Bern:2013gka};
\NLOEW corrections known~\cite{Denner:2014ina,Lindert:2022ejn}, including merging and showering~\cite{Kallweit:2014xda,Kallweit:2015dum};
Multi-jet merged prediction up to 9 jets at LO~\cite{Hoche:2019flt}.
\medskip

\noindent \textit{Progress}
The \NNLOQCD calculation of the production of an isolated photon in association with a jet pair has been presented~\cite{Badger:2023mgf}.
The authors perform a comparison with ATLAS data~\cite{ATLAS:2019iaa} and find that the agreement with their new calculation is better than the one with parton-shower-matched and multi-jet-merged predictions generated for the ATLAS analysis using the \Sherpa Monte Carlo.
It is worth noting that this was the first $2\to 3$ calculation at \NNLOQCD accuracy not reverting to the leading-colour approximation.
Nonetheless, the effect of full colour in the two-loop virtual part has been found to be small with respect to the remaining theoretical uncertainties.

\subsubsection{$V+b\bb$}

 \textit{LH21 status} \NNLOQCD for $Wbb$ known~\cite{Hartanto:2022qhh} while \NLOQCD known for $Zbb$~\cite{FebresCordero:2009xzo};
 $Wb\bb$ with up to three jets computed at \NLOQCD in Ref.~\cite{Anger:2017glm}; matching to parton shower at \NLOQCD accuracy~\cite{Frederix:2011qg,Oleari:2011ey,Krauss:2016orf,Bagnaschi:2018dnh};
 \NLOQCD for $Wb\bb j$ calculated with parton shower matching~\cite{Luisoni:2015mpa};
 multi-jet merged simulation, combining five- and four-flavour calculations for $Z+b\bb$ production at the LHC~\cite{Hoche:2019ncc}.
 \medskip

\noindent \textit{Progress}
While the calculation presented in Ref.~\cite{Hartanto:2022qhh} used the flavour $k_t$ algorithm, Ref.~\cite{Hartanto:2022ypo} presented results using this time the flavour anti-$k_t$ algorithm~\cite{Czakon:2022wam}.
To investigate the parametric freedom in the flavour anti-$k_t$ algorithm, the authors performed a comparison to CMS data~\cite{CMS:2016eha} which shows good agreement.

While Refs.~\cite{Badger:2021nhg} assumed massless bottom quarks, a new computation, this time with massive bottom quarks, has been presented in Ref.~\cite{Buonocore:2022pqq}.
The authors argue that using massive bottom quarks in their calculation avoids the ambiguities regarding flavour assignment that arises in massless calculations.

\NNLOQCD corrections matched to parton shower for the production of a Z boson in association with a bottom-quark pair are presented in Ref.~\cite{Mazzitelli:2024ura}.
Assuming a four-flavour scheme, the authors find that the \NNLOQCD corrections resolve previously-observed tensions between lower-order predictions in four- and five-flavour schemes.
These state-of-the-art predictions are compared to a CMS measurement~\cite{CMS:2021pcj}, showing good agreement.

\subsubsection{$VV'$}

\textit{LH21 status} \NNLOQCD publicly available for all vector-boson
pair production processes with full leptonic decays, namely
$WW$~\cite{Gehrmann:2014fva,Grazzini:2016ctr},
$ZZ$~\cite{Cascioli:2014yka,Grazzini:2015hta,Heinrich:2017bvg,Kallweit:2018nyv},
$WZ$~\cite{Grazzini:2016swo,Grazzini:2017ckn},
$Z\gamma$~\cite{Grazzini:2013bna,Grazzini:2015nwa,Campbell:2017aul},
$W\gamma$~\cite{Grazzini:2015nwa};
\NLOQCD corrections to the loop-induced $gg$ channels
computed for $ZZ$~\cite{Caola:2015psa,Grazzini:2018owa} and
$WW$~\cite{Caola:2015rqy,Grazzini:2020stb} involving full off-shell leptonic dacays;
interference effects with off-shell Higgs contributions known~\cite{Caola:2016trd,Campbell:2016ivq};
NLO EW corrections known for
all vector-boson pair production processes including full leptonic
decays~\cite{Denner:2014bna,Denner:2015fca,Biedermann:2016yvs,Biedermann:2016guo, Biedermann:2016lvg, Biedermann:2017oae,Biedermann:2017yoi,Kallweit:2017khh,Chiesa:2018lcs};
Polarised predictions at \NLOQCD $WZ$~\cite{Denner:2020eck}, at \NLOQCD+\NLOEW for $ZZ$~\cite{Denner:2021csi} and $WZ$~\cite{Le:2022lrp}, and at \NNLOQCD for $WW$~\cite{Poncelet:2021jmj};
combination of \NNLOQCD and \NLOEW corrections to all massive diboson processes known~\cite{Grazzini:2019jkl};\\
\NNLOQCD matched to a parton shower for $WW$~\cite{Re:2018vac,Lombardi:2021rvg}, $Z\gamma$~\cite{Lombardi:2020wju}, $W\gamma$~\cite{Cridge:2021hfr}, $ZZ$~\cite{Buonocore:2021fnj,Alioli:2021egp} production;
\NNNLL resummation for transverse momentum of $WW$ matched with \NNLOQCD~\cite{Kallweit:2020gva};
\NLOQCD matched to parton showers for the gluon--gluon loop-induced channel~\cite{Alioli:2021wpn,Grazzini:2021iae};
\NLOQCD+\NLOEW matched to parton shower~\cite{Chiesa:2020ttl,Brauer:2020kfv,Bothmann:2021led}.

\medskip

\noindent \textit{Progress}
In Ref.~\cite{Agarwal:2024pod}, the full \NLOQCD corrections to the loop induced process $gg\to ZZ$ have been obtained.
The crucial two-loop virtual contribution was obtained by combining analytic results for the massless, Higgs-mediated, and one-loop factorisable amplitudes with numerically computed amplitudes containing the top-quark mass.
The authors have found that the NLO corrections give a sizable impact at the third order in perturbative QCD (meaning the \NNNLOQCD  predictions).
In Ref.~\cite{Degrassi:2024fye}, the top-quark loops of the double virtual contribution have been obtained by an expansion in small transverse momentum. The results have then been combined with a high-energy expansion in order to provide analytic results valid over the whole phase space.

In Ref.~\cite{Gavardi:2023aco}, predictions at \NNLOQCD accuracy matched to parton shower were presented for $WW$ production.
An important point in this implementation is that, since the resummation is performed for the hardest jet transverse momentum, the matching ensures that no large logarithms appear when applying jet vetoes.
The predictions are compared to experimental measurements of both ATLAS and CMS~\cite{ATLAS:2019rob,CMS:2020mxy} and are found to be in good agreement.
In Ref.~\cite{Banerjee:2024xdh}, threshold resummation for $ZZ$ production at \NNLOQCD+NNLL have been presented.
The effect of resummation has been found to be at the level of few per cent.
In Ref.~\cite{Lindert:2022qdd}, \NNLOQCD matched to parton shower have been combined consistently with \NLOEW corrections for $WZ$ production.
This was the first time that such accuracy is achieved for a public event generator.

Polarised predictions with higher-order corrections have been of particular interest for diboson production in the last few years.
In particular, \NLOQCD corrections have been obtained for $WZ$ production in final states with two charged leptons and jets~\cite{Denner:2022riz}.
Furthermore, \NLOEW corrections have also been computed for $WW$~\cite{Denner:2023ehn,Dao:2023kwc} and $WZ$~\cite{Le:2022ppa} production (in the latter case, first results were already provided in Ref.~\cite{Le:2022lrp}).
In addition to strictly fixed order predictions, a first step toward matching higher orders with parton shower has been achieved.
In particular, in Ref.~\cite{Pelliccioli:2023zpd}, \NLOQCD corrections were matched to parton shower for all production mechanisms.
In addition, results on how to enhance doubly-longitudinal polarised states and study the radiation amplitude zero effect in $WZ$ production have been made public~\cite{Dao:2023pkl}.
In Ref.~\cite{Dao:2024ffg}, a study of the bottom-quark contribution at NLO QCD+EW  accuracy has been presented.
In Ref.~\cite{Javurkova:2024bwa}, polarised predictions for ZZ pairs in gluon fusion and in vector-boson fusion have been presented.

In Ref.~\cite{He:2024iqg}, two-loop planar master integrals for the massive \NNLOQCD corrections for $WW$ production have been obtained.
In addition, in Ref.~\cite{Long:2024bmi}, analytical results for three-loop ladder diagrams with two off-shell legs have been presented.
These contributions constitute relevant ingredients for the computation of \NNNLOQCD corrections to equal-mass diboson production.
Finally, in Ref.~\cite{Canko:2024ara} several three-loop master integrals relevant for the computation of \NNNLOQCD corrections for the production of two off-shell vector bosons with different masses have been presented.

\medskip \noindent \textit{Experimental status}
As illustration, it is interesting to discuss a measurement from the CMS collaboration~\cite{CMS:2020gtj}.
The total cross section reads $\sigma_\text{tot}(pp \to ZZ) = 17.4 \pm 0.3 (\text{stat}) \pm 0.5 (\text{syst}) \pm 0.4 (\text{theo}) \pm 0.3 (\text{lumi}) \text{pb}$.
With the upcoming high-luminosity phase of the LHC, the statistical uncertainty will significantly reduce while it is also expected that systematic uncertainty will shrink.
It means that theory uncertainty will most liekly become the dominant uncertainty.
The first source of theory uncertainty is related to the one originating from the strong coupling and the PDFS.
The use of NNLO QCD + PS predictions, combined with EW predictions are therefore crucial for future data-theory comparisons.
In addition, given the importance of EW corrections in tails of distributions for diboson production, mixed QCD-EW corrections are likely to become relevant for data description in the future.

\subsubsection{$VV'+j$}

\textit{LH21 status}
\NLOQCD corrections known for many years~\cite{Dittmaier:2007th,Campbell:2007ev,Dittmaier:2009un,Binoth:2009wk,Campanario:2010hp,Campanario:2009um,Campbell:2012ft,Campbell:2015hya};
Full \NLOEW corrections available \cite{Brauer:2020kfv,Bothmann:2021led}
along with matching with parton shower with approximate EW corrections.

\medskip \noindent \textit{Experimental status}
On the experimental side, for the $WWj$ channel, both ATLAS and CMS have measured the process~\cite{ATLAS:2016agv,ATLAS:2021jgw,CMS:2020mxy}.
The experimental errors are at the level of $10\%$ or below and are dominated for now by systematic uncertainty.
In the long term, statistical uncertainties will become negligible, while systematic uncertainties are likewise expected to shrink. The resulting total experimental uncertainty will therefore be of the same order as, or smaller than, the current theoretical uncertainty, calling for advances beyond the present state of the art to meet the precision of forthcoming high-luminosity data.

\subsubsection{$VV'+\geq2j$}

\textit{LH21 status}
Full \NLOSM corrections (\NLOQCD, \NLOEW and mixed \NLOgen) available for $W^+W^+jj$~\cite{Biedermann:2016yds,Biedermann:2017bss} and $ZZjj$~\cite{Denner:2020zit,Denner:2021hsa};
\NLOQCD+\NLOEW known for $WZjj$~\cite{Denner:2019tmn} and $W^+W^-jj$~\cite{Denner:2022pwc};
\NLOQCD corrections known for the EW production for all leptonic signatures in the vector-boson scattering approximation~\cite{Jager:2006zc,Jager:2006cp,Bozzi:2007ur,Jager:2009xx,Denner:2012dz,Campanario:2013eta,Campanario:2017ffz};
Same holds true for the QCD production modes~\cite{Melia:2010bm,Melia:2011dw,Greiner:2012im,Campanario:2013qba,Campanario:2013gea,Campanario:2014ioa,Campanario:2014dpa,Campanario:2014wga}; \NLOQCD calculated for $WW+3j$~\cite{FebresCordero:2015kfc};\\
All above computations matched to parton shower~\cite{Arnold:2008rz,Baglio:2011juf,Melia:2011gk,Jager:2011ms,Jager:2013iza,Jager:2013mu,Baglio:2014uba,Rauch:2016upa,Jager:2018cyo} (in the VBS approximation for EW production);
\NLOEW to same-sign $WW$ matched to parton/photon shower~\cite{Chiesa:2019ulk}.
Comparative study at \NLOQCD and with parton-shower corrections for same-sign $WW$~\cite{Ballestrero:2018anz}.
\medskip

\noindent \textit{Progress}
In Ref.~\cite{Jager:2024eet}, the final state $W^\pm W^\pm jjj$ has been computed at \NLOQCD matched to parton shower.
This allows a better description of observables using the third jet (in addition to the two tagging jets), e.g.\ for jet veto in the central region in experimental analyses.

So far, most work in VBS has been focused on leptonic channels, but
new results are becoming available for semi-hadronic and fully hadronic signatures.
For example, the implementation of the \NLOQCD corrections matched to parton shower for $WZjj$~\cite{Jager:2018cyo} in \Powheg has been extended to allow to consider the semi-leptonic and fully hadronic channels~\cite{Jager:2024sij}.
Results are shown for current and future possible hadron colliders up to $100$~TeV.
The spin-correlations and off-shell effects are also studied.
Along the same line, Ref.~\cite{Denner:2024xul} presented results at LO for the VBS production of $\ell\nu jjjj$  using a double-pole approximation.

In Ref.~\cite{Dittmaier:2023nac}, full \NLOQCD+EW corrections have been presented for the $W^+W^+jj$, confirming the results of Ref.~\cite{Biedermann:2016yds,Biedermann:2017bss}.
In addition, the \NLOQCD corrections have been compared to those in the double-pole and  VBS approximations.

In Ref.~\cite{Abreu:2024yit}, presented results for several planar master integrals which contribute to the \NNLOQCD virtual corrections in vector-boson pair production in association with a jet.

One of the key properties of the VBS is that it is particularly sensitive to the longitudinal polarisations of heavy gauge bosons.
In an effort to provide reliable predictions to experimental collaborations for the extraction of polarisation fractions, \NLOQCD+EW corrections have been computed for the same-sign $WW$ channel for definite polarisation~\cite{Denner:2024tlu}.
The extraction of polarisation fractions is expected to be one of the highlights of the Run~3.

\medskip \noindent \textit{Experimental status}
Several prospective studies regarding the measurement of VBS at the high-luminosity phase of the LHC have been made public~\cite{CMS:2016rcn,CMS:2018zxa,ATLAS:2018uld,CMS:2021uvc}.
They claim that at the end of the Run 3, the cross sections of VBS processes will be measurable with a total uncertainty of a few percent.
This will therefore allow the extraction of the longitudinal polarisations of the gauge boson in such processes with a significance of few sigma~\cite{ATLAS:2018uld,CMS:2021uvc}.
In that respect, first polarisation extraction by both ATLAS and CMS have been performed~\cite{CMS:2020etf,ATLAS:2025wuw}.
While still suffering from large uncertainties, these are expected shrink as more data will become available, making therefore precise theory predictions critical.

Note that in Ref.~\cite{Covarelli:2021gyz}, a review of past experimental and theoretical progress have been presented for vector-boson scattering at the LHC.

\subsubsection{$VV'V''$}

\textit{LH21 status}
\NLOQCD corrections known for many years~\cite{Hankele:2007sb,Binoth:2008kt,Campanario:2008yg,Bozzi:2009ig,Bozzi:2010sj,Bozzi:2011wwa,Bozzi:2011en,Campbell:2012ft},
also in case of $W\gamma\gamma j$~\cite{Campanario:2011ud};
\NLOEW corrections with full off-shell effects for $WWW$ production with leptonic decays \cite{Schonherr:2018jva,Dittmaier:2019twg};
\NLOEW corrections available for the
on-shell processes involving
three~\cite{Nhung:2013jta,Shen:2015cwj,Wang:2016fvj}
and two~\cite{Wang:2017wsy,Cheng:2021gbx} massive vector bosons; $V\gamma\gamma$ processes with full leptonic decays calculated
at \NLOQCD and \NLOEW accuracy~\cite{Greiner:2017mft,Zhu:2020ous}.
\medskip

\noindent \textit{Progress}
While up to now, calculations including the decays of the heavy gauge bosons have focused on the leptonic channels, several recent computations have  considered also the hadronic case.
In Ref.~\cite{Denner:2024ufg}, the case of $WW(V\to jj)$ has been considered.
Beyond the fixed-order results at full \NLOQCD+\NLOEW accuracy (i.e.\ for the EW and QCD production), results for the matching of QCD corrections to parton shower along with approximate EW corrections have  been presented.
Along the same line, full \NLOQCD+\NLOEW predictions have been presented for the $WZ(V\to jj)$ channel~\cite{Denner:2024ndl}.
Interestingly, the \NLOEW corrections to the EW production have been found to be at the level of $14\%$ i.e.\ twice as large as typical EW corrections for the triboson production.

In Ref.~\cite{Rosario:2024aht}, NLO QCD corrections matched to parton shower have been presented for the process $pp\to e^+ \nu_e \mu^- \bar \nu_\mu \gamma$.
To that end, {\sc Herwig} is used as the Monte Carlo event generator, while the amplitudes are tkane from {\sc Vbfnlo}.
Parton-shower effects are studied in detail.
In particular, the parton-shower corrections can reach $10\%$ in some distributions, potentially beyond the naive scale variation.

Finally, two-loop QCD amplitudes to $W\gamma\gamma$ have been presented~\cite{Badger:2024sqv}.
The results are available in the form of analytical results for the leading colour while they are available only numerically for the full colour case.
In Ref.~\cite{Kermanschah:2024utt}, $N_f$-contributions at the two-loop level in QCD have been obtained for a number of processes featuring the production of two and three vector bosons.
These results are relevant for the computation of \NNLOQCD for triboson production.

\medskip \noindent \textit{Experimental status}
The measurements of the triple-production of massive bosons has only started a few years ago and is therefore statistically limited~\cite{ATLAS:2016jeu,CMS:2020hjs,ATLAS:2022xnu,ATLAS:2024nab}.
For example, the WWW inclusive production has a $12\%$ statistical uncertainty and a $10\%$ systematic uncertainty.
Also, it is worth pointing out that such measurements typically have as irreducible background $VH$ production, making it therefore challending to single out the triple-gauge boson production.
For the future, having in mind the high-luminosity phase of the LHC, at least \NLOQCD+\NLOEW accuracy will be needed to match the upcoming experimental precision.

\subsubsection{$\gamma\gamma$}

\textit{LH21 status}
\NNLOQCD corrections known~\cite{Cieri:2015rqa,Catani:2018krb,Campbell:2016yrh,Grazzini:2017mhc,Gehrmann:2020oec};
\NLOQCD corrections including top-quark mass effects to loop-induced $gg$ channel known \cite{Maltoni:2018zvp,Chen:2019fla};
$q_T$ resummation computed at \NNLL~\cite{Cieri:2015rqa};
\NLOEW corrections available~\cite{Bierweiler:2013dja,Chiesa:2017gqx};\\
\NNLL+\NNLOQCD accuracy achieved~\cite{Alioli:2020qrd} as well as \NNLOQCD+PS~\cite{Gavardi:2022ixt}.
\medskip

\noindent \textit{Progress}
While \NLOQCD corrections are known for the loop-induced $gg$ channel, the EW ones are still unknown.
In that respect, the main bottleneck is the computation of the two-loop virtual corrections.
A key ingredient for this calculation is the availability of the relevant master integrals.
In Ref.~\cite{Fiore:2023myh}, these integrals for the light-quark contributions  were obtained.

In Ref.~\cite{Becchetti:2023yat}, the full top-mass dependence for both the $gg$ and $q\bar q$ channels has been assessed by explicit calculation, relying on the newly obtained two-loop form factors~\cite{Becchetti:2023wev}.
The work shows that the effect of top-mass contributions for the full calculation is below $1\%$.

Finally, Ref.~\cite{Neumann:2021zkb} presented a $q_T$ resummation at  \NNNLLp + NNLO QCD accuracy.
In addition to discussing the impact of newly implemented contributions, photon isolation prescriptions are also studied in this work, and a comparison to ATLAS data at 8 TeV~~\cite{ATLAS:2017cvh} is presented.

\subsubsection{$\gamma\gamma+\ge1j$}

\textit{LH21 status}
\NNLOQCD corrections known~\cite{Chawdhry:2021hkp} (at leading colour for the two-loop part) as well as \NLOQCD to the loop-induced process~\cite{Badger:2021ohm};
\NLOQCD known for $\gamma\gamma+2j$~\cite{Gehrmann:2013bga,Badger:2013ava,Bern:2014vza,Fah:2017wlf} and $\gamma\gamma+3j$~\cite{Badger:2013ava};
photon isolation effects studied at \NLOQCD~\cite{Gehrmann:2013aga};
\NLOQCD corrections for the EW production of $\gamma\gamma+2j$ \cite{Campanario:2020xaf};
\NLOEW corrections available for $\gamma\gamma j(j)$~\cite{Chiesa:2017gqx};
\medskip

\noindent \textit{Progress}
A new calculation of the production of prompt photons in association with two jets to \NLOQCD matched to parton showers within \Powheg has been presented~\cite{Jezo:2024wsc}.
In this work, a comparison with ATLAS data~\cite{ATLAS:2019iaa} is presented using two parton-shower programs (\Pythia and \Herwig).
Both variants provide a good description of the data.

\subsubsection{$\gamma\gamma\gamma$}

\textit{LH21 status}
\NNLOQCD corrections in the leading-colour approximation known \cite{Chawdhry:2019bji,Kallweit:2020gcp}.
\medskip

In Ref.\cite{Abreu:2023bdp}, the two-loop QCD corrections to three-photon production beyond the leading-colour approximation have been obtained.
This work constitutes the last missing piece of the full \NNLOQCD calculation with full colour.
The authors have estimated that the full-colour effect in the two-loop virtual corrections will decrease  the total cross section by a few percent with respect to the case with leading-color approximation in the two-loop virtual.

\subsection{Top-quark associated processes}

An overview of the status of top quark associated processes is given in Table~\ref{tab:SM_wishlist:wlTJ}.
\begin{table}
  \renewcommand{\arraystretch}{1.5}
\setlength{\tabcolsep}{5pt}
  \centering
  \begin{tabular}{lll}
    \hline
    \multicolumn{1}{c}{process} & \multicolumn{1}{c}{known} &
    \multicolumn{1}{c}{desired} \\
    \hline
    $pp\to t\tb$ &
    \begin{tabular}{l}
      \NNLOQCD\!+\,\NLOEW (w/o decays) \\
      \NLOQCD\!+\,\NLOEW{ }(off-shell) \\
      \NNLOQCD{ }(w/ decays)
    \end{tabular} &
    \begin{tabular}{l}
      \NNNLOQCD
    \end{tabular} \\
    \hline
    $pp\to t\tb+j$ &
    \begin{tabular}{l}
      \NLOQCD{ }(off-shell effects) \\
      \NLOEW (w/o decays)
    \end{tabular} &
    \begin{tabular}{l}
      \NNLOQCD\!+\,\NLOEW{ }(w decays)
    \end{tabular} \\
    \hline
    $pp\to t\tb+2j$ &
    \begin{tabular}{l}
      \NLOQCD{ }(w/o decays)
    \end{tabular} &
    \begin{tabular}{l}
      \NLOQCD\!+\,\NLOEW{ }(w decays)
    \end{tabular} \\
    \hline
    $pp\to t\tb+V'$ &
    \begin{tabular}{l}
      \NLOQCD\!+\,\NLOEW{ }(w decays)
    \end{tabular}
    &
    \NNLOQCD\!+\,\NLOEW{ }(w decays)
    \\
    $pp\to t\tb+\gamma$ &
    \begin{tabular}{l}
      \hline
      \NLOQCD{ }(off-shell)
    \end{tabular} & \\
    $pp\to t\tb+Z$ &
    \begin{tabular}{l}
      \hline
      \NLOQCD\!+\,\NLOEW{ }(off-shell)
    \end{tabular} & \\
    $pp\to t\tb+W$ &
    \begin{tabular}{l}
    \hline
    \NLOQCD\!+\,\NLOEW{ }(off-shell) \\
    \end{tabular} &
    \begin{tabular}{l}

    \end{tabular} \\
    \hline
    $pp\to t/\tb$ &
    \begin{tabular}{l}
      \NNLOQCD{*}(w decays) \\
      \NLOEW{ }(w/o decays)
    \end{tabular} &
    \begin{tabular}{l}
      \NNLOQCD\!+\,\NLOEW{ }(w decays)
    \end{tabular} \\
    \hline
    $pp\to tZj$ &
    \begin{tabular}{l}
      \NLOQCD\!+\,\NLOEW{ }(off shell)
    \end{tabular} &
    \begin{tabular}{l}
      \NNLOQCD\!+\,\NLOEW{ } (w/o decays)
    \end{tabular} \\
    \hline
    $pp\to t\tb t\tb$ &
    \begin{tabular}{l}
      \NLOQCD (w decay) \\
      \NLOEW{ }(w/o decays)
    \end{tabular} &
    \begin{tabular}{l}
      \NLOQCD\!+\,\NLOEW{ }(off-shell) \\
      \NNLOQCD
    \end{tabular} \\
    \hline
  \end{tabular}
  \caption{Precision wish list: top quark  final states. \NNLOQCD$^{*}$ means a
   calculation using the structure function approximation. $V'=W,Z,\gamma$.}
  \label{tab:SM_wishlist:wlTJ}
  \renewcommand{\arraystretch}{1.0}
\end{table}

\subsubsection{$t\tb$}

\textit{LH21 status}
Fully differential \NNLOQCD computed for on-shell top-quark pair production~\cite{Czakon:2015owf,Czakon:2016ckf,Czakon:2016dgf,Catani:2019hip,Catani:2020tko}, also available as {\tt fastNLO} tables~\cite{Czakon:2017dip};
polarised two-loop amplitudes known~\cite{Chen:2017jvi} ;
combination of \NNLOQCD and \NLOEW corrections performed~\cite{Czakon:2017wor};top quark decays known at \NNLOQCD~\cite{Gao:2012ja,Brucherseifer:2013iv};
Complete set of \NNLOQCD corrections to top-pair production and decay in the NWA for intermediate top quarks and $W$ bosons~\cite{Behring:2019iiv,Czakon:2020qbd}, including B-hadron production~\cite{Czakon:2021ohs}; $W^+W^- b\bar{b}$ production with full off-shell effects calculated
at \NLOQCD~\cite{Denner:2010jp,Denner:2012yc,Bevilacqua:2010qb,Heinrich:2013qaa}
including leptonic $W$ decays, and in the lepton plus jets channel~\cite{Denner:2017kzu};
full \NLOEW corrections for leptonic final state available~\cite{Denner:2016jyo};
calculations with massive bottom quarks available at
\NLOQCD~\cite{Frederix:2013gra,Cascioli:2013wga};\\
$b \bb 4\ell$ at \NLOQCD matched to a parton shower in the \Powheg framework retaining all off-shell and non-resonant contributions~\cite{Jezo:2016ujg};
\NNLOQCD matched to parton shower for on-shell tops~\cite{Mazzitelli:2020jio,Mazzitelli:2021mmm};
multi-jet merged predictions for up to 2 jets in \Sherpa~\cite{Hoeche:2014qda} and \Herwig\,7.1~\cite{Bellm:2017idv}; with \NLOEW corrections available~\cite{Gutschow:2018tuk};
resummation effects up to \NNLL computed~\cite{Beneke:2011mq,Cacciari:2011hy,Ferroglia:2013awa,Broggio:2014yca,Kidonakis:2015dla,Pecjak:2016nee,Alioli:2021ggd};
\NNLOQCD\!+\,\NNLL for (boosted) top-quark pair production~\cite{Czakon:2018nun}. Analytic results for leading-colour two-loop amplitudes for $gg \to t\bar t$ known~\cite{Badger:2021owl}.
\medskip

\noindent \textit{Progress}
In Ref.~\cite{Chen:2023dsi}, the analytic \NNNLOQCD corrections at leading colour to the semi-leptonic decay, which includes light-quark loop contribution, have been computed.
It was found that at \NNLOQCD, these contributions amount to $95\%$ of the total corrections, giving confidence it is a reliable approximation.
The third-order QCD corrections are at the level of  $-0.667\%$ with respect to the LO, and the scale uncertainty is reduced by half compared to the \NNLOQCD predictions.
Along the same line,
in Ref.~\cite{Chen:2023osm}, the top-quark decay width has been computed at \NNNLOQCD.
The authors found that the value of the decay width is decreased by about $0.8\%$, exceeding the scale variation at \NNLOQCD.
In addition, it was found the the \NNNLOQCD corrections to the polarisation fractions are much smaller.

In Ref.~\cite{Bernreuther:2024ltu}, a new analysis of spin correlation and polarization effect at the LHC has been conducted at \NLOQCD including electroweak effects.
Potential new physics effects parametrized through an effective field theory are also investigated.
The authors of Ref.~\cite{Mandal:2022vju} have reported the first analytical calculation of the two-loop amplitude for the production of a heavy quark pair via light-quark annihilation.
Reference~\cite{Campbell:2023fjg} has provided compact analytical expressions for the production of a pair of top quarks in association with up to two jets at tree level.
These amplitudes can be used for \NNLOQCD calculations to top-pair production.

The scale dependence of top-pair production has been investigated in several renormalisation schemes in Ref.~\cite{Makela:2023xnt}. This is particularly important for the experimental extraction of the top quark mass, especially in the low top-pair invariant mass regime.
Along the same line of research, a study~\cite{Garzelli:2023rvx} of the top-quark pole mass extraction has been carried out at \NNLOQCD accuracy, using total, single-, and double-differential cross sections.
In Ref.~\cite{Czakon:2022pyz}, non-perturbative fragmentation functions, for B-hadrons, J/$\Psi$ and muons resulting from semileptonic B decays have been derived at \NNLOQCD.
These fragmentation functions are then used to study the production of top-quark pairs with these final states.

In Ref.~\cite{Jezo:2023rht}, the first event generator at \NLOQCD accuracy matched to parton shower for the off-shell production of a top-quark pair in the lepton+jets channel has been presented.
This implementation also allows the separation between $tW$ and $t\bar t$ production mechanisms.

In Refs.~\cite{Ju:2022wia,Ju:2024xhd}, the differential transverse momentum and azimuthal decorrelation of the top-quark pair have been computed with degrees of fixed-order accuracy  combined with resummation, including \NNLL+\NNLOQCD accuracy.
Particular emphasis has been put on the interplay between soft-collinear resummation and Coulomb singularities.
In Ref.~\cite{Catani:2023tby}, the computation of the soft-parton contributions at low transverse momentum of the top-quark pair up to \NNLOQCD has been presented.
This is the final ingredient for the implementation of the $q_T$ subtraction formalism at \NNLOQCD for top-quark production.

In Ref.~\cite{Makarov:2023uet}, linear power corrections have been computed using renormalon techniques for the top-quark pair production, in the quark-antiquark partonic channel.
It is shown that for the total cross section, linear power corrections vanish, provided a short-distance scheme is used for the top-quark mass. In general, the effects computed are relatively small.

\medskip \noindent \textit{Experimental status}
The production of a pair of top quarks a critical process to be included in global PDF fits as it offers additional information on the gluon distribution, especially at higher $x$. To date, all decay channels (fully leptonic, semi-leptonic, and fully hadronic) have been measured. One advantage of this process is that it offers multiple observables that can be used in PDF fits, with statistical correlations  provided by the experiments that prevent double-counting. Measurements cover a very wide kinematic range with the top-quark pair mass currently being measured up to 4\,TeV~\cite{ATLAS:2022mlu}, which will extend to 7\,TeV at the high-luminosity LHC.
Electroweak corrections become very important at higher masses. 
Also, typically, both resolved and boosted topologies are measured (see e.g.~\cite{ATLAS:2017cez}) and the latter is particularly important for the higher mass range.

\subsubsection{$t\tb\,j$}

\textit{LH21 status}
\NLOQCD corrections calculated for on-shell top quarks~\cite{Dittmaier:2007wz,Melnikov:2010iu,Melnikov:2011qx},
full off-shell decays included at \NLOQCD~\cite{Bevilacqua:2015qha,Bevilacqua:2016jfk};
\NLOEW corrections known~\cite{Gutschow:2018tuk} for on-shell top quarks;\\
matching to parton showers~\cite{Kardos:2011qa,Alioli:2011as} for on-shell top quarks;
\medskip

\noindent \textit{Progress}
In order to compute \NNLOQCD corrections, the last missing piece is the two-loop virtual amplitude.
Therefore, much effort is being focused in this direction.
In Ref.~\cite{Badger:2022hno}, two-loop master integrals for one of the planar topologies contributing to the process have been presented.
In particular, it is the two-loop five-point pentagon-box integral configuration with one internal massive propagator.
Following this work, the differential equations for the two remaining integral topologies contributing to the leading colour two-loop amplitudes were also computed~\cite{Badger:2024fgb}.
Along the same line, one-loop QCD helicity amplitudes up to $\mathcal{O}(\eps^2)$ in the dimensional regularisation parameter, which are relevant for the calculation of \NNLOQCD corrections, have been presented~\cite{Badger:2022mrb}.
The amplitudes have been expressed in terms of a set of uniformly transcendental master integrals.
Finally in Ref.~\cite{Badger:2024gjs}, a numerical evaluation of the two-loop QCD helicity amplitudes for $gg\to t\tb g$ at leading colour has been presented.

In Ref.~\cite{Chargeishvili:2022ngl}, the one-loop soft anomalous dimension matrices have been presented.
It is a key ingredient for resumming logarithms associated to soft-gluon emissions in $t\bar t + j$ production.

\subsubsection{$t\tb+\ge2j$:}

\textit{LH21 status}
\NLOQCD corrections to $t\tb jj$ known for many years~\cite{Bevilacqua:2010ve,Bevilacqua:2011aa};
$t\tb jjj$ at \NLOQCD calculated~\cite{Hoche:2016elu}.
\medskip

\noindent \textit{Progress}
In Ref.~\cite{Bevilacqua:2022ozv}, resonant top quarks are considered in the NWA and \NLOQCD corrections have been computed for both the production and the decay part of the process,  retaining all spin information.

\subsubsection{$t\tb+b\bb$}

\textit{LH21 status}
\NLOQCD corrections to $t\tb b\bb$ with massless bottom quarks known
for off-shell top quarks~\cite{Denner:2020orv,Bevilacqua:2021cit,Bevilacqua:2022twl};
\NLOQCD corrections for $t\bar{t}b\bar{b}$ production in association with a
light jet~\cite{Buccioni:2019plc} for on-shell top quarks;\\
\NLOQCD with massive bottom quarks and matching to a parton shower investigated~\cite{Cascioli:2013era,Jezo:2018yaf} for on-shell top quarks.
\medskip

\noindent \textit{Progress}
In Ref.~\cite{Ferencz:2024pay}, \NLOQCD predictions for $t\tb+b\bb$ production with $b$-quark mass effects have been matched to a $t\bar t + {\rm jets}$ simulation in a variable flavor number scheme.
In Ref.~\cite{Frederix:2024sfi}, \NLOQCD corrections have been matched to parton shower in the five-flavour scheme.

\subsubsection{$t\tb t\tb$}

\textit{LH21 status}
\NLOQCD known~\cite{Bevilacqua:2012em};
\NLOEW known~\cite{Frederix:2017wme};\\
matching of \NLOQCD corrections to parton shower known~\cite{Jezo:2021smh}.
\medskip

\noindent \textit{Progress}
In Ref.~\cite{Dimitrakopoulos:2024qib}, \NLOQCD corrections were presented for the four-lepton channel using the NWA approximation while retaining top-quark spin correlations. \NLOQCD corrections are considered for both the production and decays of the top quarks.
The authors conclude that the main theoretical uncertainties originate from missing higher-order corrections . The authors emphasize the need to include corrections in the top quark decay, which impact results at the $10\%$ level.
The same authors did a similar study on the 3-lepton channel~\cite{Dimitrakopoulos:2024yjm}.

In Ref.~\cite{vanBeekveld:2022hty}, threshold resummation for $t\tb t\tb$ have been performed \NLLp accuracy.
The calculation is matched to the \NLOQCD and \NLOEW corrections.
The \NLLp corrections are positive at the level of $15\%$ for the total production rate and reduce the size of the scale variation by a factor of 2, which brings the theoretical error well below the current experimental uncertainty.

\medskip \noindent \textit{Experimental status}
The production of four top quarks has been measured by both ATLAS~\cite{ATLAS:2020hpj,ATLAS:2021kqb} and CMS~\cite{CMS:2019jsc,CMS:2019rvj}, with the ATLAS measurement reaching a significance of 4.7 sigma. The uncertainties are evenly balanced between statistical and systematic sources, with each being of the same order as the current theory uncertainty (estimated through scale variation at NLO). Both experimental uncertainties will go down as more data is accumulated. It is worth noting that a sizeable fraction of the systematic error is related to the signal modelling, which could be reduced by improvements in the theoretical predictions.  At the moment, the calculation of four-top production at hadron collider at \NNLOQCD ($2\rightarrow4$ with a heavy mass scale) is not expected for the near future. Shorter-term improvements would include NLO top-quark decays with NLO spin-correlations which will not reduced the uncertainty of the predictions but instead improve the modelling of the process.

\subsubsection{$t\tb V^\prime$}

\textit{LH21 status}

\NLOQCD for off-shell $t\tb Z$~\cite{Bevilacqua:2019cvp,Bevilacqua:2022nrm};
\NLOQCD for off-shell $t\tb W$~\cite{Bevilacqua:2020pzy,Denner:2020hgg,Bevilacqua:2020srb} as well as \NLOQCD+EW for QCD production and \NLOQCD for QCD production~\cite{Denner:2021hqi};
\NLOQCD for off-shell $t\tb \gamma$~\cite{Bevilacqua:2018woc} and \NLOEW for on-shell top quarks~\cite{Duan:2016qlc};
Full \NLOSM corrections for $t\tb W$ and $t\tb t\tb$ production~\cite{Frederix:2017wme}, $t\tb Z$~\cite{Frixione:2015zaa} as well as for $t\tb\gamma$, $t\tb\gamma\gamma$, and $t\gamma j$~\cite{Pagani:2021iwa};\\
\NLOQCD corrections matched to parton shower for on-shell top quarks for $t\tb \ell^+\ell^-$~\cite{Ghezzi:2021rpc} and $t\tb W$~\cite{FebresCordero:2021kcc};
Merged prediction for $t\tb W$~\cite{Frederix:2021agh};
\NLOQCD corrections to $t\tb\gamma\gamma$ production matched to parton shower~\cite{vanDeurzen:2015cga};
Resummed calculations up to \NNLL to $t\tb W$~\cite{Broggio:2016zgg,Kulesza:2018tqz}
and $t\tb Z$~\cite{Broggio:2017kzi,Kulesza:2018tqz} production;
Combination of these corrections with \NLOEW~\cite{Broggio:2019ewu} for $t\tb Z/W/H$;
\NNLL+\NLOQCD corrections for $t\tb W/Z/h$~\cite{Kulesza:2020nfh}.
\medskip

\noindent \textit{Progress}
The tensions observed in $t\tb W$ measurements for several years~\cite{ATLAS:2023gon,CMS:2022tkv} have triggered several theory studies.
For example, in Ref.~\cite{Buonocore:2023ljm}, the first \NNLOQCD corrections for $t\tb W$ have been computed.
The computation is exact apart from the finite part of the two-loop virtual amplitude, which is estimated through a soft-$W$ approximation and a massification procedure.
The \NNLOQCD corrections increase the cross section by $15\%$, and significantly reduce the perturbative uncertainty. Nevertheless,
the authors have found that the tensions with ATLAS and CMS measurements remain at the level of $1-2$ sigma.
Along the same line, it has been speculated that these tensions could be alleviated by including also $t\tb W j$ predictions in the theory predictions.
In Ref.~\cite{Bi:2023ucp}, \NLOQCD corrections to this process have been calculated, considering off-shell top quarks, and the $W$ boson in the NWA.

Full \NLOSM predictions for $t\tb Z$ for off-shell top quarks have been computed~\cite{Denner:2023eti}.
The authors highlight that, although a calculation with on-shell top quarks captures the majority of the effects across phase space (in particular the non-trivial hierarchy between the various orders in perturbation theory), fully off-shell calculations are vital, especially when considering stringent experimental cuts.

Full \NLOSM corrections have been calculated for $t\tb \gamma$ where the top quarks have been described in the NWA~\cite{Stremmer:2024ecl} and with full off-shell effects~\cite{Stremmer:2024zhd}. The residual perturbative uncertainty is between $5\%-8\%$.

\subsubsection{$t$/$\tb$}

\textit{LH21 status}
Fully differential \NNLOQCD corrections for the dominant $t$-channel production process completed
in the structure function approximation, for stable top quarks~\cite{Brucherseifer:2014ama} and
later including top-quark decays to \NNLOQCD accuracy in the NWA~\cite{Berger:2016oht,Berger:2017zof,Campbell:2020fhf};
\NLOEW corrections known~\cite{Frederix:2019ubd};
Non-factorisable contributions from the two-loop helicity amplitude for $t$-channel ~\cite{Bronnum-Hansen:2021pqc} and including all \NNLOQCD corrections~\cite{Bronnum-Hansen:2022tmr};
\NNLOQCD corrections for the $s$-channel and related decay,
neglecting the colour correlation between the light and heavy quark lines and applying the NWA~\cite{Liu:2018gxa};
\NLOQCD correction for stable $tW$ production~\cite{Giele:1995kr,Zhu:2002uj,Cao:2008af,Kant:2014oha}, including decays~\cite{Campbell:2005bb} and \NLOEW corrections~\cite{Beccaria:2007tc};
\NLOQCD corrections to $t$-channel electroweak $W+bj$ production available
within MG5\_aMC@NLO~\cite{Papanastasiou:2013dta,Frixione:2005vw};
\NLOQCD for single top-quark production in association with two jets \cite{Molbitz:2019uqz};\\
\NLOQCD corrections matched to parton shower for single-top production in the $t, s$, and $tW$ channels available~\cite{Frixione:2008yi,Alioli:2009je,Re:2010bp,Jezo:2016ujg,Bothmann:2017jfv,Frederix:2019ubd};
\NLOQCD matched to parton shower for single top-quark production in association with a jet in the \MiNLO method~\cite{Carrazza:2018mix};
Soft-gluon resummation at \NLLone for single-top production in the $t$-channel~\cite{Cao:2019uor} and the $s$-channel modes~\cite{Sun:2018byn}.
\medskip

\noindent \textit{Progress}
For single top production in the $t$ channel, linear power corrections to the $t$-channel production~\cite{Makarov:2023ttq} and including decays in the NWA~\cite{Makarov:2024ijn} have been obtained using renormalon calculus.
Beyond the phenomenological relevance of their work, the authors have shown that there are no linear power corrections to the total production cross section, provided that it is expressed in terms of a short-distance top-quark mass.
When the top quark decay is included,  linear corrections do impact the total cross section, as well as polarization observables and generic kinematic distributions of leptons originating from top-quark decays.

In Ref.~\cite{Wu:2023fyo}, a systematic computation of master integrals for the two-loop virtual amplitude for the non-factorizable corrections to $t$-channel single-top production at \NNLOQCD has been presented.
The results are expressed in the form of Goncharov polylogarithm functions.

Single top production can also be used to constrain PDF fits~\cite{Campbell:2021qgd}. In this study, the authors have shown that t-channel single-top-quark production can provide stringent constraints for $b$-quark PDF. They also conclude that the $b$-quark mass uncertainty is the dominant theory uncertainty for this process.

The $tW$ process is actually part of the off-shell $tt$ process~\cite{Jezo:2016ujg}; nonetheless the process is sometimes singled out in experimental analyses.
In order to compute \NNLOQCD corrections, the last missing piece is the two-loop contributions.
Following Ref.~\cite{Chen:2021gjv}, further two-loop master integrals have been obtained in Refs.~\cite{Long:2021vse,Wang:2022enl}.
This work then lead to the two-loop QCD amplitudes at leading colour with light fermion loop contributions~\cite{Chen:2022yni} and the full two-loop QCD amplitudes~\cite{Chen:2022pdw} for $tW$ production.

\subsubsection{$tZj$}

\textit{LH21 status} \NLOQCD+\NLOEW corrections known for off-shell top quarks and Z bosons \cite{Denner:2022fhu}.

\subsubsection{$tt\gamma\gamma$}

This process was not listed in the 2023 Les Houches wishlist.
With the increasing luminosity of the LHC experiments, it is justified to add it to the list of processes.
In particular, it is the main background for the measurement of Higgs production in association with a top-antitop pair where the Higgs boson decays into two photons.
This is actually one of the most sensitive channel for the measurement of the $ttH$ process~\cite{CMS:2020cga,ATLAS:2020ior}.
The \NLOQCD corrections and their matching to parton shower have been known for many years~\cite{Kardos:2014pba,Maltoni:2015ena,vanDeurzen:2015cga}
The \NLOEW corrections are also known~\cite{Pagani:2021iwa}.

In a recent study~\cite{Stremmer:2023kcd}, \NLOQCD corrections to both the $tt\gamma\gamma$ production process as well as the top decay have been computed in the NWA, while retaining spin correlations. Photon radiation is also considered from the top quark decay products.
 Results are presented for both the di-lepton and lepton+jet channel.
The authors found that the effects of photon bremsstrahlung are significant when two photons are emitted simultaneously in the production and decay of the $t\tb$ pair.

\subsection*{Acknowledgements}
We thank all of our colleagues who provided us with valuable input to update the wishlist.
This work is supported in part by the UK Science and Technology Facilities Council (STFC) through grant ST/T001011/1.
S.P.J.\ is supported by a Royal Society University Research Fellowship (Grant URF/R1/201268).
M.P.\ acknowledges support by the German Research Foundation (DFG) through the Research Training Group RTG2044.

\let\NLO\undefined
\let\NLL\undefined
\let\NLOH\undefined
\let\NLOQ\undefined
\let\NLOE\undefined
\let\NLOHone\undefined
\let\NLOQone\undefined
\let\NLOEone\undefined
\let\NLOQE\undefined
\let\LOQ\undefined
\let\NLOQonetb\undefined
\let\NLOQtb\undefined
\let\NLOQmtsix\undefined
\let\NLOQzzero\undefined
\let\NLOQoneVBF\undefined
\let\NLOQVBF\undefined
\let\NLOQoneDIS\undefined
\let\NLOQDIS\undefined
\let\NLOEoneVBF\undefined
\let\NLOQoneVBFstar\undefined
\let\NLOQVBFstar\undefined
\let\NLOEoneVBFstar\undefined
\let\NLOggHVtb\undefined

\let\xs\undefined
\let\tb\undefined
\let\bb\undefined
\let\qb\undefined
\let\VdkL\undefined
\let\VdkQ\undefined
\let\VdkLNWA\undefined
\let\VdkQNWA\undefined

\let\wodecay\undefined
\let\wdecay\undefined
\let\wodecays\undefined
\let\wdecays\undefined
\let\wleptdecays\undefined

\let\VdkALLNWA\undefined
\let\VdkALL\undefined
\let\tdk\undefined
\let\tdkNWA\undefined
\let\TVdkALLNWA\undefined

\let\MadgraphaMCatNLO\undefined
\let\Herwig\undefined
\let\Powheg\undefined
\let\Powhegboxres\undefined
\let\PowhegboxVtwo\undefined
\let\GoSam\undefined
\let\Recola\undefined
\let\OpenLoops\undefined
\let\MadLoop\undefined
\let\Matrix\undefined
\let\Munich\undefined
\let\Geneva\undefined
\let\Sherpa\undefined
\let\NNLOjet\undefined
\let\MiNLO\undefined
\let\NLOX\undefined


\clearpage
\bibliography{LH23}


\end{document}